\documentclass[useAMS,usenatbib]{mn2e}
\usepackage{epsfig,rotate,graphicx}
\usepackage[fleqn]{amsmath}
\usepackage{subfigure}
\usepackage{lscape}
\usepackage{bm}

\newcommand{\p}{\partial}
\newcommand{\mnras}{MNRAS}
\newcommand{\apj}{ApJ}
\newcommand{\aap}{A\&A}
\newcommand{\apjl}{ApJL}
\newcommand{\dd}{\delta}

\newcommand{\be}{\begin{equation}}
\newcommand{\ee}{\end{equation}}
\newcommand{\gtrsim}{\;\raisebox{-.8ex}{$\buildrel{\textstyle>}\over\sim$}\;}
\newcommand{\lesssim}{\; \raisebox{-.8ex}{$\buildrel{\textstyle<}\over\sim$}\;}

\newcommand{\araa}{{\it ARA\&A, }}
\newcommand{\apjs}{{\it ApJS, }}
\newcommand{\aaps}{{\it A\&AS, }}

\newcommand{\nat}{{\it Nature, }}

\newcommand{\sbar}{\bar{\sigma}}

\title[Vortex instabilities with self-gravity]{The effect of self-gravity on vortex instabilities in disc-planet
	interactions}

\author[Lin and Papaloizou]{Min-Kai Lin \thanks{E-mail: mkl23@cam.ac.uk} and John C. B. Papaloizou \thanks{E-mail: J.C.B.Papaloizou@damtp.cam.ac.uk}\\
Department of Applied Mathematics and Theoretical Physics,
University of Cambridge, Centre for Mathematical Sciences,\\
\  \ Wilberforce  Road, Cambridge, CB3 0WA, UK \\}

\begin{document}

\maketitle

\begin{abstract} \ \ \
We study the effect of disc self-gravity on instabilities associated with 
gaps opened by a giant Saturn mass 
 planet in a protoplanetary disc that lead
to the formation of vortices. We also study the nonlinear evolution
of the vortices when this kind of instability 
occurs in a self-gravitating disc  as well as  the potential
 effect on type III planetary migration due to angular momentum
exchange via coorbital flows.

 It is shown
analytically and is confirmed through linear
calculations that vortex forming  modes  with low  azimuthal
mode number $,m,$ are stabilised by the effect of self-gravity if 
the background  structure is assumed fixed.
However, the disc's self-gravity also affects
the background gap surface density
profile in  a way that destabilises modes with high $m.$
Linear calculations show that the combined effect of
self-gravity through its effect on   the background  structure and 
its direct effect on the linear modes 
 shifts the most rapidly growing  vortex mode to higher $m.$ 

Hydrodynamic simulations of gaps opened by a Saturn mass planet
 show more
vortices develop with increasing disc mass and therefore 
importance of self-gravity.
 For sufficiently large disc mass the
vortex instability is suppressed, consistent with
analytical expectations. In this case a new global instability
develops instead.

 In the non-linear regime, we found 
that vortex merging is in general increasingly delayed as  
the  disc mass increases 
and in some cases multiple vortices persist until
the end of simulations.
 For massive discs in which the vortices merge, the
post-merger vortex is localised in azimuth and has similar structure
to a Kida-like vortex. This is unlike the case without
 self-gravity where vortices
merge to form a larger vortex extended in azimuth. 

In order to study the properties of the vortex systems
without the influence of the planet,
we also performed a series of supplementary simulations of co-orbital 
Kida-like vortices. We found that self-gravity enables Kida-like vortices to 
execute horseshoe turns upon encountering each other.
 As a result vortex merging is avoided on
time-scales where it would occur without self-gravity.
Thus  we suggest that
mutual repulsion of self-gravitating vortices in a rotating flow is 
 responsible for the  delayed vortex merging seen in
 the disc-planet simulations. 

The effect of self-gravity on vortex-induced migration  in low
viscosity discs is briefly discussed.
 We found that when self-gravity is 
included and the disc mass is in the range where
 vortex forming instabilities occur,
 the vortex-induced type III migration  of \cite{lin10}
 is delayed. There are also  expected to be longer
periods of slow migration between the short  
bursts of rapid migration
compared to what occurs in a simulation without self-gravity.
 However, the extent
of induced rapid migration is unchanged and involves flow of vortex material across the gap, independent of whether
or not self-gravity is included. 
\end{abstract}\ \ \
\section{Introduction}

Since the first detection of an extrasolar giant planet
orbiting a main sequence star in a 4 day orbit by 
\cite{mayor95}, there have been many observations of hot Jupiters
\citep{marcy05,udry07}. One possible explanation for
their close proximity to their host stars is
orbital migration due to interaction with the gaseous
protoplanetary disc, from their site of formation to their current observed location. 
\citep{lin86}

Planetary migration was first studied by \cite{goldreich80}, and since then 
many analytical and numerical studies of disc-planet
interactions have been undertaken \citep[see][for a review]{papaloizou06}.
However, most of these adopt a low mass disc model for
which it may be assumed that self-gravity may be neglected.
This is even the case when rapid type III migration that requires
a massive disc is considered. It is the purpose of this paper
to investigate the effect of the disc self-gravity on such disc planet interactions and 
in particular to consider its effect on the stability
of the disc  to vortex formation, a phenomenon that has been found to occur
in discs with very low viscosity.

Astrophysical discs may be dynamically unstable for a variety of reasons. 
One example  is instability associated with 
steep surface density gradients or vortensity\footnote{The term vortensity
is used for the ratio of vorticity to surface density} extrema
 \citep{papaloizou85,papaloizou87,papaloizou89,papaloizou91,lovelace99,li00} that can 
lead to vortex formation in non-linear regime \citep{li01}. 
 The disc vortensity profile necessary for such instabilities to occur 
may be induced by  disc-planet interaction.
 For sufficiently large planetary masses and/or sufficiently low disc
viscosity, the tidal perturbation on the gaseous disc leads to 
a dip or gap in the surface density  \citep{lin86}. There have been a 
series of studies of the vortex instability associated with planetary
gaps \citep{koller03,li05,valborro07} and its consequences for planetary
migration \citep{ou07,li09,lin10,yu10}. However, the effect of
disc self-gravity  is still unclear.

In the context of disc-planet interactions, 
disc self-gravity is often  neglected.
 It  has thus far only been studied in a limited
number of works
\citep{nelson03a,nelson03b,zhang08,baruteau08}. \cite{li09} included 
self-gravity in their study of Type I migration in low viscosity
discs, where the vortex instability develops  for $\alpha$ viscosity parameters
$\leq 10^{-5}$, but the specific effect of self-gravity 
on the instability was not discussed analytically or explored numerically. 
In this paper we examine the effect of self-gravity on the vortex instability
associated with edges of gaps opened by a Saturn-mass planet and its
effect on the subsequent nonlinear evolution of vortices together
with a first calculation of the subsequent effects on orbital migration.

This paper is organised as follows. In \S\ref{governing} we present the
governing equations and models for our disc-planet systems. We discuss
the linearised stability problem in \S\ref{analytical}.
We show that vortex forming modes with long azimuthal wavelength  associated with vortensity
minima at gap edges,  induced by an embedded planet, are stabilised by self-gravity when the background model
is held fixed.
In \S\ref{linear} we perform linear mode calculations for disc
models of varying mass. We confirm the analytic discussion but show in addition
that changes to the gap structure that occur with increasing disc mass are destabilising, 
an effect that is stronger for modes with short wavelength in the azimuthal direction.
All of these effects cause the most unstable modes to shift 
to shorter azimuthal wavelength as the disc mass increases. 
However, in practice
these modes are eventually dominated by long azimuthal wavelength, global edge modes, 
which are not associated with localised vortex formation and  which are studied
in \cite{lin11}. 
Edge modes dominate our models when the disc-to-star mass ratio is $\gtrsim 0.047$, corresponding
to a Keplerian Toomre stability parameter $\lesssim 2$ at the outer disc boundary (see \S\ref{disc_planet_model} 
and Appendix \ref{Qo_Md}). As this paper is concerned with vortex formation, we focus on models with disc mass
below this threshold.

We go on to present nonlinear  
hydrodynamic simulations in \S\ref{hydro1}.
In accord with expectations from linear theory the instability 
produces more vortices of smaller scale as the disc mass increases. 
The vortices are found to survive for much longer against merging 
than in the non self-gravitating limit and themselves become self-gravitating 
with strong over surface densities. This is found to enable pairs of vortices
to undergo  co-orbital dynamics with horseshoe trajectories which aids to inhibit 
merging. In \S\ref{hydro2} we perform simulations of interacting Kida-like vortices
without an embedded planet in order to
clarify results obtained for vortices at gap
edges.  These show that the co-orbital dynamics occurs independently of the embedded planet.
In  \S\ref{migration} we give a preliminary investigation of the effects of vortex 
formation in low viscosity discs on orbital migration when
self-gravity is significant but not strong enough to prevent
effective vortex formation. As in the non self-gravitating case, 
episodes of rapid migration occur as vortices are scattered by the 
planet \citep{lin10}, but longer time intervals between them are expected on account of 
stabilisation by self-gravity and slowed vortex merging.  
Finally we summarise and conclude in \S\ref{conclusions}.

\section{Disc model and basic equations }\label{governing}
We consider a gaseous disc of mass $M_d$ orbiting a
central star of mass $M_*.$ We assume the disc to be razor thin
and thus work in the  the two dimensional flat disc
approximation.  We  adopt cylindrical polar co-ordinates $(r,\varphi)$
centred on the star and defined in the plane of the disc. 
The reference frame is non-rotating.   

The governing  hydrodynamic equations are the continuity
and momentum equations 
\begin{align}\label{continuity}
  &\frac{D\Sigma}{D t}=-\Sigma\nabla\cdot\mathbf{u}, \\
  &\frac{D\bm{u}}{D t} = -\frac{1}{\Sigma}\nabla p - \nabla\Phi +
  \frac{\bm{f}}{\Sigma},  
\end{align}
where $\Sigma$ is the surface density, $\bm{u}$ is the velocity field and
$p$ is the vertically  integrated pressure. Numerical calculations
adopt a locally isothermal equation of state, $p=c_s^2\Sigma$ where  
$c_s = H\Omega_k$ is the local sound speed and
$\Omega_k=\sqrt{GM_*/r^3}\equiv 2\pi/P.$ Here  $H=hr$ is the putative disc
semi-thickness. We fix the constant $h=0.05$.
The viscous force per unit area is  $\bm{f}$ being   characterised
by a uniform kinematic viscosity $\nu.$ 
 The gravitational potential  $\Phi$ includes the
indirect potential $\Phi_i$, the  potential  due to the central star,
$\Phi_* = -GM_*/r$, the  potential due to the disc 
$\Phi_d$ when  self-gravity is included 
and the potential due to an embedded planet,
$\Phi_p,$ for disc-planet systems.  
 Details of $\bm{f}$
are given in \cite{masset02}.  The potentials due to the
planet and  the disc  are   

\begin{align}
&\Phi_p = -\frac{GM_p}{\sqrt{r^2+r_p^2-2rr_p\cos{(\varphi-\varphi_p)+\epsilon_p^2}}} \\
&\Phi_d = -\int_{{\cal D}} \frac{G\Sigma(r^\prime,\varphi^\prime)}{\sqrt{r^2+r^{\prime 2} -
    2rr^\prime\cos{(\varphi-\varphi^\prime)}+ \epsilon_g^2}}r^\prime
dr^\prime d\varphi^\prime,
\end{align}
where the integral is taken over the domain of the disc ${\cal D}.$
The planet has mass $M_p$, position $(r_p,\varphi_p)$ and its potential
is softened by use of the softening length   
$\epsilon_p= 0.6H(r_p)$. Similarly  the disc potential is softened by 
use of the softening length $\epsilon_g= 0.3H(r^\prime).$ 
The indirect potential takes account of the the forces on the central star,
which is at the origin of a non inertial frame, due to the disc and planet.
It is given by
\begin{align}
\Phi_{i} = & r\int_{\cal{D}}
\frac{G\Sigma(r^\prime,\varphi^\prime)}{r^{\prime2}}\cos{(\varphi-\varphi^\prime)}r^\prime
dr^\prime d\varphi^\prime\notag\\
&+\frac{GM_p}{r_p^2}r\cos{(\varphi-\varphi_p)}.
\end{align}
 We adopt units such that $G=M_*=1$. The
Keplerian orbital period at $r=1$ is then $P(1)=2\pi$. 
As in our previous work, for disc-planet simulations we use a uniform viscosity $\nu=10^{-6}$ in code units.
This is an order of magnitude smaller than the typically adopted 
value of $\nu=10^{-5}$ that suppresses
the vortex instability \citep{valborro07,lin10}.

\subsection{Disc-planet model}\label{disc_planet_model}
Most of our discussion will be applied to disc-planet systems, so we
describe these models here.  Supplementary
simulations in \S\ref{hydro2} employ a different set up from that described below.

The disc occupies $r=[r_i,r_o]=[1,10]$. The initial surface density
profile is modified from \cite{armitage99}: 
\begin{align} 
 \Sigma(r) = \Sigma_0 r^{-3/2}\left(1-\sqrt{\frac{r_1}{r+H_1}}\right),
\end{align}
where $H_1=H(r_1)$ is introduced  to prevent a singular
pressure   force at the inner boundary because $\Sigma\to 0$ there. 
$\Sigma_0$ is chosen via the parameter $Q_o\equiv Q_\mathrm{Kep}(r_o)$ where   
\begin{align}\label{myQ} 
  Q_\mathrm{Kep}(r) = \frac{c_s\kappa}{\pi G\Sigma}=\frac{hM_*}{\pi r^2\Sigma(r)}
\end{align}
is the Toomre $Q$ parameter for  a thin Keplerian disc with a locally
isothermal equation of state  and 
 $\kappa^2 = 2\Omega r^{-1} d(r^2\Omega)/dr $ is the square of the  epicycle  frequency, with $\Omega$ being the disc angular velocity. We
use $Q_o$ to  label  disc models and also define 
$Q_p\equiv Q_\mathrm{Kep}(r_p)$. We note that specifying 
$Q_o$ also determines the disc-to-star mass ratio, $M_d/M_*$. The conversion between
$Q_o,\,Q_p$ and $M_d/M_*$ for the models used in disc-planet interactions is given by 
Table \ref{Qo_Md_conversion} in Appendix \ref{Qo_Md}.

The disc is initialised with  the azimuthal velocity  required  by 
hydrostatic equilibrium.  The initial
radial velocity is $u_r = 3\nu/r$, corresponding to the initial radial velocity
profile of a Keplerian disc with uniform kinematic viscosity and surface density
$\propto r^{-3/2}$. Note that $|u_r|\ll|u_\varphi|$. 


In most of this work we focus on stability of the gap induced by an embedded planet, 
accordingly  we fix the planet
on a circular orbit of radius $ r_p= 5$. We quote time in units
$P_0=2\pi/\Omega_k(r_p)$. The planet is introduced at $t=25P_0$ with
azimuthal velocity   that takes account the contribution from the 
 gravitational force due to the disc. The planet potential is ramped up
over a time period of $10P_0$.     

\section{Linear modes in discs with structured  surface density }\label{analytical}

In this section and the next we  study linear disturbances
associated with internal surface density depressions in a disc in which
 self-gravity is not neglected. We will
consider dips/gaps self-consistently opened by a giant planet. To
simplify the discussion, we ignore viscosity, indirect potentials and
the planet potential. 
The planet's role is then only to set up the
basic state, assumed to be axisymmetric and defined by
$\Sigma(r),\,u_\varphi(r)$ and $u_r = 0$. We 
begin with the linearised equations with a local isothermal
equation of state.

We consider perturbations to the disc
state variables with azimuthal and time dependence
through a factor  $\exp{i(\sigma t +m\varphi)}$ which is taken as read.
Here  $\sigma = \sigma_R -  i\gamma$ is a complex frequency
 with $\sigma_R$ and  $ -\gamma$ being the real and imaginary
 parts respectively. The azimuthal mode number $m$ is taken to be  a positive
integer. Denoting perturbations 
by a prime, the linearised equations of motion read 

\begin{align}
 u_r' &= -\frac{1}{D}
\left[i\bar{\sigma} \left(c_s^2\frac{dW}{dr} + \frac{d
      \Phi'}{dr}\right) + \frac{2im\Omega}{r}\left( c_s^2W +
    \Phi'\right)\right]\\  
 u_\varphi' &= \frac{1}{D}
\left[\frac{\kappa^2}{2\Omega}\left(c_s^2\frac{dW}{dr} +
    \frac{d\Phi'}{dr}\right) + \frac{m\bar{\sigma}}{r}\left( c_s^2 W
  + \Phi'\right)\right],
\end{align}
where $D = \kappa^2 - \sbar^2$,  $\bar{\sigma}\equiv \sigma
+m\Omega(r)$ is  the Doppler shifted frequency, $W=\Sigma'/\Sigma$ is  the
relative surface density perturbation and $\Phi'$ is the disc
gravitational potential perturbation which is given by the Poisson integral

\begin{align}\label{poisson}
&\Phi' = -G \int_{r_i}^{r_o} K_m(r,\xi)\Sigma(\xi)W(\xi)\xi d\xi ,\hspace{2mm} {\rm where}\\
&K_m(r,\xi) = \int_{0}^{2\pi} \frac{\cos(m\varphi)d\varphi}{\sqrt{r^2+\xi^2 -
    2r\xi\cos{(\varphi)}+ \epsilon_g^2(\xi)}}.
\end{align}
Using the linearised equations of motion to eliminate the velocity
component perturbations from  the linearised continuity equation 
\begin{align}
i\bar{\sigma} W = -\frac{1}{r\Sigma}\frac{d}{dr}\left(r\Sigma
  u_r'\right) - \frac{im}{r} u_\phi',
\end{align}
yields the  governing equation 
\begin{align}\label{sggoverning}
  &\frac{d}{dr}\left[\frac{r \Sigma}{D}\left(c_s^2\frac{dW}{dr} 
      + \frac{d\Phi'}{dr}\right)\right] +
  \left[ \frac{2m}{\bar{\sigma}} 
    \left(\frac{\Sigma\Omega}{D}\right)
    \frac{dc_s^2}{dr} - r\Sigma \right]W \notag\\
  &+ \left[\frac{2m}{\bar{\sigma}}\frac{d}{dr}
    \left(\frac{\Sigma\Omega}{D}\right)
    -\frac{m^2\Sigma}{rD}\right]
  \left(c_s^2 W +\Phi'\right)= 0.
\end{align}

\subsection{Modes leading to vortex formation: the effect
of  self-gravity}\label{perttheory0}
We shall be concerned with 
the vortex-forming instabilities commonly observed in
disc-planet interactions in discs with low viscosity
\citep{koller03,li05,li09,lin10}.
The unstable  modes  are found to be localised to gap edges
and  associated with minima in the
vortensity \citep{papaloizou89, papaloizou91, lovelace99}.

Here, we investigate the effect of self-gravity on these
modes and show somewhat paradoxically, in view of the fact
that increasing self-gravity in general destabilises discs, that
that it tends to stabilise the modes.
To do this we consider a  small change to the  strength of self-gravity 
and  apply perturbation theory to
determine the consequence  of the change  for  these modes.

Non-linear simulations show more vortices develop as self-gravity is increased. This is 
in fact consistent with the stabilisation effect demonstrated below, 
which is effective for small $m$. In other words, stabilisation by self-gravity 
discourages low-$m$ vortex modes, supporting the observation that only higher $m$ modes
develop in simulations.

To make the  analysis tractable,
 further simplifications are made. We take the fluid 
to be such that either  $c_s^2$
is constant (strictly isothermal) or more generally the equation
of state is  barotropic such that  $p=p(\Sigma)$. We do not expect a significant
difference between adopting a strictly isothermal,  or
 barotropic equation of state,  and the
fixed $c_s$ profile  as used in simulations. This is  because the modes
of interest are driven by local features and disturbances are
localised, whereas $c_s(r)$ for a locally isothermal equation
of state varies on a global scale.
For mathematical convenience  we also take a softening prescription
such that $K_m(r,\xi) = K_m(\xi, r)$ is symmetric, eg.
$\epsilon_g=\mathrm{constant}$. Although convenient mathematically,
this is not expected to lead to significant changes for the same reason
as that given above.

Introducing the variable $S$  
\begin{align}\label{s_def}
  S \equiv  c_s^2 W+ \Phi' = c_s^2 W - G\int_{r_i}^{r_o}
  K_m(r,\xi)\xi\Sigma(\xi) W(\xi)d\xi, 
\end{align}
the governing equation 
 derived from  equation (\ref{sggoverning})  simply by taking  $c_s$ to be
 constant is
\begin{align}\label{barotropic}
r\Sigma W = &\frac{d}{dr}\left[\frac{r \Sigma}{D}\left(\frac{dS}{dr}\right)\right] 
+ \left\{\frac{m}{\bar{\sigma}}\frac{d}{dr}
 \left[\frac{1}{\eta (1-\bar{\nu}^2)}\right]
 -\frac{m^2\Sigma}{rD}\right\}S\notag\\
& \equiv  rL (S).
\end{align}
where we have used the expression for vortensity $\eta =
\kappa^2/2\Omega\Sigma$ and $\bar{\nu}\equiv \bar{\sigma}/\kappa$.  
We remark that in fact these equations  also hold for a general barotropic 
equation of state.
\subsection{The limit of negligible self-gravity}
When self-gravity is negligible, we may set $G=0$ in (\ref{s_def})
from which we obtain $W=S/c_s^2.$
Substituting this into (\ref{barotropic}) gives the  second order
ordinary differential equation for $S$ that governs
stability in the limit of zero self-gravity  in the form
\begin{align}\label{NoselfG}
\frac{r\Sigma S}{c_s^2}  = &\frac{d}{dr}\left[\frac{r \Sigma}{D}\left(\frac{dS}{dr}\right)\right]
+ \left\{\frac{m}{\bar{\sigma}}\frac{d}{dr}
 \left[\frac{1}{\eta (1-\bar{\nu}^2)}\right]
 -\frac{m^2\Sigma}{rD}\right\}S.
\end{align}
This equation was studied by \citet{lin10} for gap profiles of the type 
we consider. They found that depending on parameters such as $m,$
there were neutral modes with co-rotation radius $,r_c,$
where  ${\bar \sigma}(r_c)=0$ coincident with a  vortensity minimum.
This could be associated with either the inner or outer gap edge.
Small variations of the disc parameters could then lead to an instability
associated with vortex formation. When unstable these modes retain
$r_c$ to be close to the vortensity minimum with which they can be considered
to be associated.

\subsection{Localised low $m$ modes}\label{Lowmm}
When $m$ is small the neutral modes described above are localised around co-rotation
and therefore insensitive to distant boundary conditions.
For these modes we may neglect $\bar{\nu}$ and set $D=\kappa^2$ in (\ref{NoselfG}) above
to get the simpler equation
\begin{align}\label{NoselfG1}
\frac{r\Sigma S}{c_s^2}  = &\frac{d}{dr}\left[\frac{r \Sigma}{\kappa^2}\left(\frac{dS}{dr}\right)\right]
+ \left\{\frac{m}{\bar{\sigma}}\frac{d}{dr}
 \left[\frac{1}{\eta}\right]
 -\frac{m^2\Sigma}{r\kappa^2}\right\}S.
\end{align}
Localisation occurs because the solutions of (\ref{NoselfG1}) can be seen to
decay exponentially away from the co-rotation region, where there are large background gradients,
on a length scale comparable to $H.$  However, for (\ref{NoselfG1}) to be a good approximation,
we require $ |\bar{\nu}^2|\ll 1$ in the region of 
localisation which is also comparable
to $H$ in extent. This in turn requires $H \ll 2r/(3m),$
 a condition which is satisfied for low $m.$
For larger $m$ the mode becomes de-localised with the excitation of density
waves that propagate into the extended disc \citep{lin10}.
In this case the boundary conditions can play a role. The analysis
below assumes localisation so that boundary conditions do not play a role.
It therefore only applies for low $m.$

\subsection{Evaluating the effect of small changes 
to low $m$ modes  using perturbation theory}\label{S3.4}
In order to investigate the effect of self-gravity
we return to equations (\ref{s_def}) and (\ref{barotropic})
and note that strengthening self-gravity  by scaling up the surface density
is equivalent to increasing
$G,$ provided that the background form remains fixed.
Thus although we increase the effective disc gravity, we
shall assume the background disc model remains unchanged.
In fact, in our case the background model is structured by a perturbing planet
and so its response to changing the disc gravity is non-trivial
to evaluate analytically. Calculations presented below in section \ref{SGRole}  show that  changes to the background 
 surface density profile
induced by incorporating the disc gravity tend to act to make the vortex modes
we consider  more unstable. 
However, the direct effect of self-gravity on the linear
response considered below turns out to be more important 
for localised modes with low $m,$ and acts to stabilise them.

Accordingly
we assume a solution $S$ corresponding to a neutral mode
with co-rotation radius, $r_c,$ located at a vortensity minimum, such that
 $d\eta/dr|_{r=r_c} =0.$
 As the associated  $\sigma$ is real,  for this value,
 the operator $L$ is real and regular everywhere. We then 
perturb this  solution by altering the strength of  self-gravity via   
\begin{align*}
  G \to G + \dd G
\end{align*}
so that $\dd G>0$ corresponds  to increasing the importance of
self-gravity and vice versa (note that the initial value, $G,$
could be zero). This induces  perturbations
\begin{align*}
   &S \to S + \dd S,\quad 
   \Sigma' \to \Sigma' + \dd\Sigma', \quad
   \sigma \to \sigma + \dd \sigma, \\
   &L \to L + \dd L ,
\end{align*} 
with
\begin{align*}
  &\dd \sigma = \dd \sigma_R - i\gamma\\
  &\dd L = \frac{\p L}{\p\sigma}\dd\sigma
\end{align*}
where $\dd \sigma_R$ and $\gamma$ are real. 

 Noting that $\delta$ denotes a small change and $\gamma$
is small, we linearise in terms of these quantities  about the
assumed original neutral mode and determine $\gamma.$
The governing equations lead to
\begin{align}
  &L(\dd S) + \dd L(S) = \dd\Sigma',\label{perttheory1}\\
  &\dd S = c_s^2 \frac{\dd\Sigma'}{\Sigma} - G\int_{r_i}^{r_o} 
  K_m(r,\xi)\xi\dd\Sigma'(\xi)d\xi \notag\\
  &\phantom{\dd S =} - \dd G\int_{r_i}^{r_o} 
  K_m(r,\xi)\xi\Sigma'(\xi)d\xi .\label{perttheory2}
\end{align}
Next, we define the inner product between two functions $U(r),\,V(r)$ as 
\begin{align}
  \langle U, V\rangle = \int_{r_i}^{r_o} rU^*(r)V(r) dr,
\end{align}
Then,  assuming localised functions
corresponding to localised modes so that  we can assume boundary values
 vanish  when integrating by parts, we have 
\begin{align}
  \langle U, L(V)\rangle =  \langle L(U), V\rangle,
\end{align}
which used the fact that $L$ corresponding to the neutral mode is
real, which makes $L$ self-adjoint. 
Equations \ref{perttheory1}---\ref{perttheory2}  can be
manipulated to yield
\begin{align}\label{perttheory3}
  \langle S, \dd L(S)\rangle &=  \dd
  G\int_{r_i}^{r_o}\int_{r_i}^{r_o}
  r\xi K_m(r,\xi)\Sigma'^*(\xi)\Sigma'(r) d\xi dr.\notag \\
  &\equiv \dd GE, 
\end{align}
where we have used the fact that the  kernel $K_m$ is
symmetric. Note that $E>0$ and is  proportional to the magnitude of
the gravitational energy of the mode. Following \cite{papaloizou89},
we separate out the contribution to the perturbed linear operator $\dd
L$  that is proportional to the vortensity gradient and  is potentially
singular at co-rotation and write 
\begin{align}\label{perttheory4}   
  \langle S, \dd L(S)\rangle \equiv \langle S, \dd L_1(S)\rangle +
  \langle S, \dd L_2(S)\rangle,
\end{align}
where $\dd L_2(S)$  contains  the potentially singular  contribution
and we have
\begin{align}
  \langle S, \dd L_2(S)\rangle &= -m
\dd\sigma \left[
\mathcal{P}\left(\int_{r_i}^{r_o}\frac{g(r)}{\bar{\sigma}}dr\right)
+ \int_{r_i}^{r_o}{\rm i}\pi\dd(\bar{\sigma})g(r)dr\right]\label{deltaL2}\\
 {\rm where} \hspace{2mm}  g(r)&\equiv\frac{|S|^2}{\bar{\sigma}
}\frac{d}{dr}\left(\frac{1}{\eta}\right).
\end{align}
Note that the usual Landau prescription for positive $\gamma$ has been used to deal
with the pole at co-rotation and that ${\cal P}$ outside the above integral  indicates
that the principal value is to be taken.
Note too  that  $\dd L_1$ accounts for  the remainder  of $\dd L$ and
is not singular at co-rotation. The contribution from 
 $\dd 
L_2$ is the only one  that can lead to an imaginary contribution
in the limit $\dd\sigma\to 0$ because of the pole at
co-rotation. In this regard we remark that one does not get
such contributions arising from Lindblad resonances where $D=0.$
 This is because, as is well known, these do not
constitute effective  singularities  in a gaseous disc
 \citep[eg.][]{papaloizou91}.

Recalling that  $\gamma$ is small,  equations \ref{perttheory3}- \ref{deltaL2}  can be combined to give
\begin{align}
  \dd GE &=\dd\sigma\left[\frac{\p\left\langle S, 
         L_1(S)\right\rangle}{\p \sigma} -m{\cal P}\left(\int_{r_i}^{r_o}
      \frac{g(r)}{\bar{\sigma}}dr\right) \right. \notag\\ 
    &\phantom{=}\left. - i m\pi\int_{r_i}^{r_o} 
    \dd(\bar{\sigma})g(r)dr \right]\notag\\  
  & \equiv (\mathcal{A} - im\chi)\dd\sigma
\end{align}
where $\mathcal{A}$ is the
contribution from the  $ L_1$ term plus the principle  value integral
 and  
\begin{align}
  \chi \equiv \left.\frac{\pi
      |S|^2}{m\Omega^\prime|
      m\Omega^\prime|}\frac{d^2}{dr^2}\left(\frac{1}{\eta}\right)
  \right|_{r=r_c}.  
\end{align}
In the limit $ \gamma\to 0_+$, $\mathcal{A}$ is real so we have
\begin{align}\label{stabilisation}
   \gamma = -\frac{m\chi}{\mathcal{A}^2 + m^2\chi^2}\dd GE.
\end{align}
 As we have remarked above, vortex forming modes are associated with vortensity
minima or at maxima of $\eta^{-1}.$  Consider a marginally stable mode with co-rotation at
$\mathrm{max}(\eta^{-1})$. Typical rotation profiles have
$\Omega^\prime < 0$, which means $\chi > 0$ for this mode.
For consistency with the assumption of $\gamma > 0$ in deriving
equation \ref{stabilisation}, we require $\dd G < 0$ since $E>0$. Accordingly in order 
to de-stabilise this mode, the strength of self-gravity needs to be
reduced. 

This leads to the conclusion that:
increasing self-gravity \emph{stabilises}  low $m$ modes with co-rotation at
a vortensity \emph{minimum} (as has been borne out by
our linear and nonlinear calculations presented below);
 while increasing self-gravity would
\emph{de-stabilise} modes which had  co-rotation at a vortensity
\emph{maximum}. This suggests that  for sufficiently strong self-gravity,
modes associated with vortensity maxima  should  be favoured. This has been
found to be the case but 
they  are not  modes leading to vortex formation \citep[see][]{lin11}.

Note $\chi\propto 1/m^2$, suggesting that $\gamma$ decreases
for large $m$, so the stabilisation/destabilisation effect of
self-gravity diminishes for increasing azimuthal wave-number. 
This is only speculative because there are implicit
dependencies on $m$ through terms in the integrals  and through the original
eigenfunctions  $S$. Nevertheless, a weakening effect of self-gravity through the potential perturbation is
anticipated because increasing $m$ decreases the magnitude of the  Poisson kernel
$K_m$.

\subsection{Association of localised low $m$  normal modes with vortensity minima
for the strength of self-gravity below a threshold}\label{S3.5}
We can also show that corotation radii for localised neutral modes with low $m$
must be at vortensity minima unless self-gravity (
or an appropriate mean value of $Q$) is above (below) a threshold level.
Let us consider a disc with localised steep surface density gradients and a
non axisymmetric disturbance with co-rotation radius  in this region. Multiplying
equation (\ref{barotropic})  by $S^*$ and integrating over the disc,  assuming 
most  of the contribution is from near corotation, as is expected for low $m$ modes
( see section \ref{Lowmm}),  so that the term
that is potentially singular at corotation
and is proportional to the  vortensity gradient is
dominant on the right hand side, we have the balance
\begin{align}
&\int_{r_i}^{r_o} r c_s^2\frac{ |\Sigma'|^2}{\Sigma}dr -
  G\int_{r_i}^{r_o} K_m(r,\xi)r\xi\Sigma'^*(\xi)\Sigma'(r)d\xi dr\notag\\
&\simeq  
  \int_{r_i}^{r_o}\frac{m|S|^2}{\sbar}\frac{d}{dr}\left(\frac{1}{\eta}\right)dr.
\end{align}
If the left hand side can be shown to always be positive 
then  co-rotation for a localised 
neutral mode  must lie at a  vortensity minimum, or
$\mathrm{max}(\eta^{-1}).$  This holds if the maximum possible value of $\Lambda$
for any $\Sigma'$ is less than unity, where
\begin{align}
\Lambda=\frac{ G\int_{r_i}^{r_o} K_m(r,\xi)r\xi\Sigma'^*(\xi)\Sigma'(r)d\xi dr}
{\int_{r_i}^{r_o} r c_s^2\frac{ |\Sigma'|^2}{\Sigma}dr}.
\end{align}
Now from the Cauchy-Schwartz inequality
it follows that
\begin{align}
\Lambda \le  G\sqrt{\int_{r_i}^{r_o} K^2_m(r,\xi)\frac{\Sigma^*(r)\Sigma(\xi)}
{c_s^2(r)c_s^2(\xi)} r\xi d\xi dr}.
 \end{align}
Thus if the right hand side of the above is less than unity,  corotation of a neutral mode
localised at a vortensity  extremum must be localised at a vortensity minimum.
This condition requires that the strength of self-gravity be below a threshold
and this implies a  lower  bound on an appropriate mean $Q$ value.
The fact that this fails for sufficiently strong self-gravity is consistent
with the discussion above that led to the conclusion that  increasing self-gravity tends
to stabilise the vortex forming instability.

Indeed when self-gravity increases further, instability is transferred
to modes with corotation associated with vortensity maxima.
These are different in character to vortex forming modes being more global
and are referred to as edge modes and they are studied 
in \cite{lin11}. 


\section{Linear calculations}\label{linear}

We now present numerical solutions to equation (\ref{sggoverning}) with a  local
isothermal equation of state for consistency with numerical simulations used to set up
the basic state. We regard the governing equation as   
the requirement that  an operator $\mathcal{L}$ acting on $W$ should be zero, thus
\begin{align}\label{linearode}
  \mathcal{L}(W) = 0. 
\end{align}
One form of $\mathcal{L}$ is 
\begin{align}\label{operator}
  \mathcal{L} =& rc_s^2\frac{d^2}{dr^2} + r\mathcal{I}_2 
 + \left[\frac{1}{r} + \frac{\Sigma^\prime}{\Sigma} -
  \frac{D^\prime }{D}\right]
\left(rc_s^2\frac{d}{dr} + r\mathcal{I}_1\right) \notag\\
& +\left\{\frac{2m}{\bar{\sigma}}\left[\frac{\Sigma^\prime\Omega}{\Sigma}
    + \Omega^\prime - \frac{\Omega D^\prime}{D}\right] -
  \frac{m^2}{r}\right\}\left(c_s^2 +\mathcal{I}_0\right)\notag\\
& + rc_s^{2\prime} \frac{d}{dr} 
 +\left[\frac{2m\Omega
    c_s^{2\prime}}{\bar{\sigma}} - rD\right].
\end{align}
It is understood that primes in equation \ref{operator} denote $d/dr$. 
The integro-differential operators $\mathcal{I}_n$ are such that
\begin{align}
\mathcal{I}_n(W) = \frac{d^n \Phi'}{dr^n}.
\end{align}
First and second derivatives of the perturbed potential are  performed by replacing
$K_m(r,\xi)$ by $\p K_m/\p r$ and $\p^2 K_m/\p r^2$ in the Poisson
integral respectively. 

\subsection{Quadratic approximation}
Vortex modes are localised about co-rotation at a  vortensity minimum. 
They extend over a region  characterised by small $|\bar{\sigma}|$ in the
neighbourhood of $r_c$ and are expected to be
 insensitive  to boundary conditions.    
If we multiply equation (\ref{operator})  by $\bar{\sigma}D$
and expand the resulting equation in powers of $\bar{\sigma}$
and neglect terms proportional to $\bar{\sigma}^3$ and higher powers, we can
derive an equation of the form  
\begin{align}\label{qepapprox}
  \left( \sigma^2 \mathcal{L}_2 + \sigma \mathcal{L}_1 +
    \mathcal{L}_0\right)W = 0, 
\end{align}
where the operators $\mathcal{L}_i$  are real and only depend on the basic
state. 
The eigenfrequency  now appears explicitly. We call this the quadratic
approximation.    
Although the above procedure is not expected to be  valid in general, it should
be valid for modes localised about their corotation radii
with the scale of localisation being much less than $r_c$ itself.
Modes can be found from (\ref{qepapprox}) using standard numerical methods.
 The eigenvalues are then used as  starting values
 in order to obtain an iterative solution for the eigenvalues for  the full equation 
( \ref{linearode}). 
The concept of the vortex forming mode as a localised
mode \emph{is confirmed} if the final solution of ( \ref{linearode})
is not significantly changed from the solution of (\ref{qepapprox}). 
 Note that since the operators $\mathcal{L}_i$  are real,
eigenfrequencies are either real or  occur in complex conjugate pairs.

\subsection{Distinction between including and excluding
  self-gravity}

Since we are concerned with effects of self-gravity, We need to
clearly define self-gravitating (SG) and non-self-gravitating cases (NSG). 
The
background state on which we perform linear stability analysis were obtained
from non-linear hydrodynamic simulations (see \S\ref{linear_basic}), which can be run 
with self-gravity (SGBG) or without self-gravity (NSGBG). 
In the linear calculations, self-gravity can be disabled 
 by setting $\Phi'=0$. Therefore we have two
disturbance types: a self-gravitating response (SGRSP) and a
non-self-gravitating response (NSGRSP). 

The analytical discussion in \S\ref{perttheory0} applies to the effect
of self-gravity through the linear response, assuming the form of the background
remains unchanged. In the context of hydrodynamic simulations,
including or neglecting self-gravity simply means whether or not the
disc gravitational potential is included.  If it is included, the
background state set up by simulations also depends on self-gravity. 
Hydrodynamic simulations therefore correspond either  to
the combination  SGBG+SGRSP or to the combination  NSGBG+NSGRSP. We call these fully
self-gravitating and fully non self-gravitating cases. 
The advantage of
performing linear calculations is that we can distinguish between
the effects of self-gravity  arising through its effects on  the background state 
and  the effects resulting from its influence on  the linear  response.

\subsection{Numerical approach}
 We discretised 
the linear operators  on a
grid that divides the radial range $[r_i, r_o]$ into  typically  $N_r=385$
 equally spaced grid points at  which  $W$ is evaluated as $W_j, j=1,2..N_r.$
The governing equation thus becomes a system of
$N_r$ simultaneous equations for the $W_j$ which determine an eigenvalue
problem for $\sigma.$ The system of equations incorporates the boundary conditions.
For simplicity we impose $dW/dr = 0$ at boundaries in discretised form. 
We remark that  it is known from other linear calculations and simulations 
that vortex-forming modes are localised and insensitive to
boundary conditions \citep{valborro07,lin10}. Tests have shown  that the  boundary
condition  does not influence  the essential effect of self-gravity on localised
modes. 

The discretised quadratic approximation obtained from  equation (\ref{qepapprox}) leads
to a  quadratic  eigenvalue problem that is equivalent to a $2N_r\times
2N_r$ standard linear matrix  eigenvalue problem.  This is  solved by
methods which  give all eigenvalues $\sigma.$  The most unstable
  eigenvalue is then used as a trial  solution for  the full  system which 
 yields a discriminant that is  then solved iteratively
for the actual eigenvalue $,\sigma,$ using the Newton Raphson method.
  We restrict attention to modes with eigenfrequencies such that
$|\sigma_R/m\Omega_e + 1| < 0.1$ and $|\gamma|< \Omega_e$, where
$\Omega_e$ is the rotation frequency of the vortensity minimum at the
outer gap edge.  
 
\subsection{Background  state}\label{linear_basic}
The background states used for linear calculations were set up by running
disc-planet simulations for the models described  in \S\ref{disc_planet_model}.
Numerical details will be  described in \S\ref{hydro1}. We allow the planet
to open a gap, then take azimuthal averages to obtain one-dimensional
profiles.  

\begin{figure}
\centering
\includegraphics[width=0.99\linewidth]{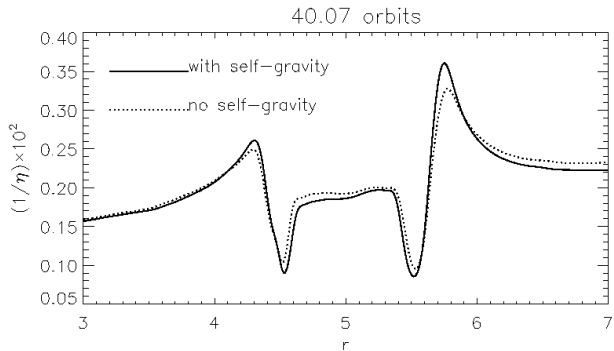}
\caption{
  Gap profiles  opened by a Saturn-mass planet in an initial
  disc model  with $Q_o=4.$ The inverse
  vortensity $\eta^{-1}$, obtained  with (solid) and without (dotted)
  self-gravity included is shown. Profiles were obtained from azimuthal
  averages  taken from disc-planet simulation outputs. The planet is at the
  fixed  location
  $r=5$. 
\label{basic}}
\end{figure}

Fig.\ref{basic} shows the gap profile, in terms of the inverse vortensity $1/\eta$,
 opened up by a Saturn-mass planet
in the $Q_o=4$ disc.  The
formation of 
vortensity peaks is due to the  generation of vortensity as material passes through  shocks
induced by the giant planet which
extend into its co-orbital region \citep{lin10}. 
We shall see  that in 
hydrodynamic simulations vortices predominantly develop at the
outer gap edge, so we focus on modes associated with the outer
vortensity extrema. The outer  vortensity maximum or $\mathrm{min}(\eta^{-1})$ is
located at $r\simeq 5.5$ while the outer vortensity minimum
or $\mathrm{max}(\eta^{-1})$ is located at $r\simeq 5.75.$  These
extrema are  separated by about $0.89H$. 
As we increase the strength of self-gravity,
the discussion in sections \ref{S3.4} and \ref{S3.5} 
indicates  that eventually vortex modes associated with vortensity minima
are suppressed and modes associated with vortensity maxima become favoured instead. 
This is consistent with self-gravity
stabilising low $m$ vortex modes through the linear response.

The gap profiles set up with and without self-gravity are  similar. 
When self-gravity is included, the peaks and troughs have slightly
larger amplitudes due to increased effective planet mass 
(see \S\ref{gapprofiles}). 

\subsection{Solution   in the quadratic approximation}
\begin{figure}
\centering
\includegraphics[width=0.99\linewidth]{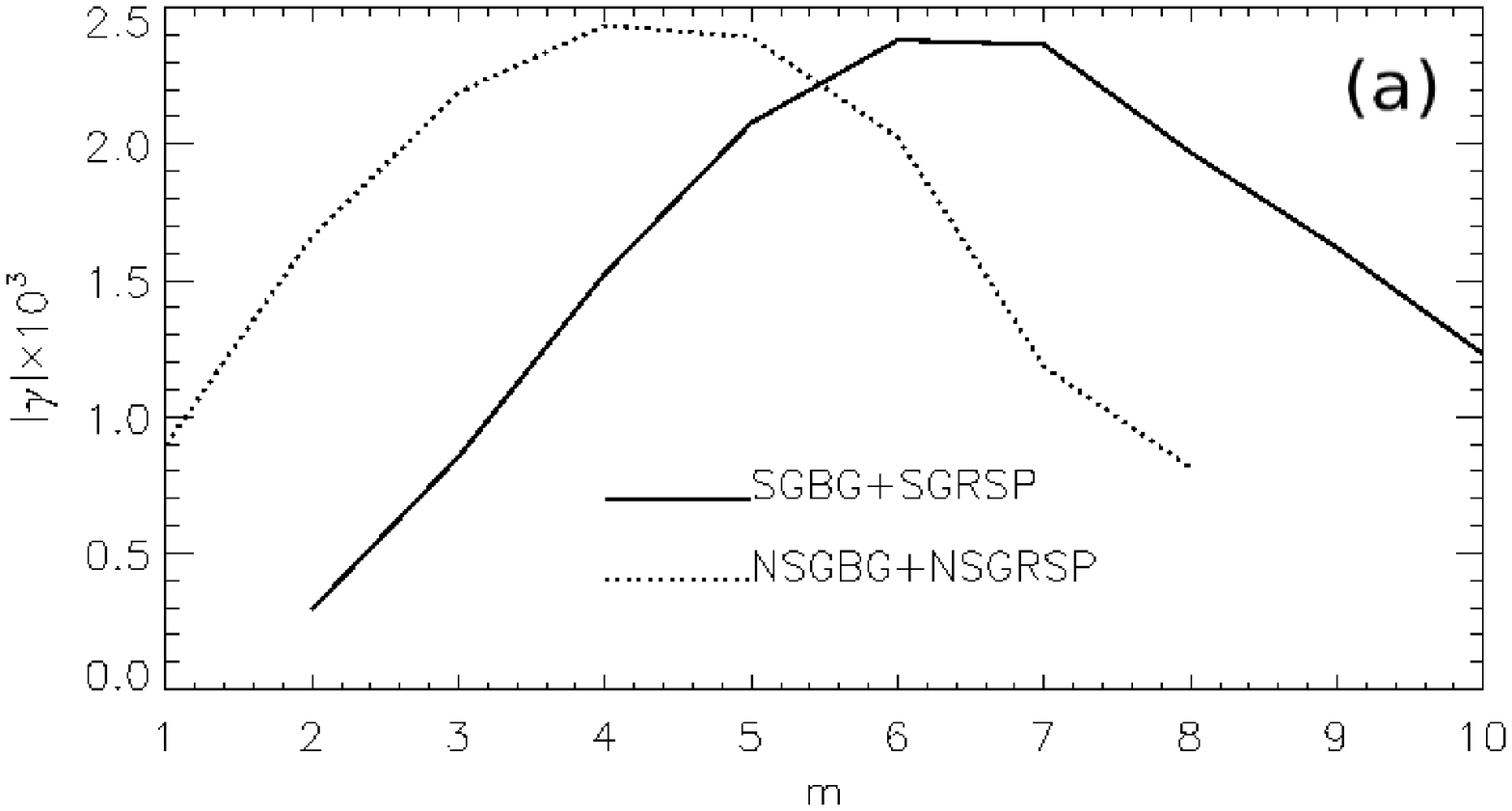}
\includegraphics[width=0.99\linewidth]{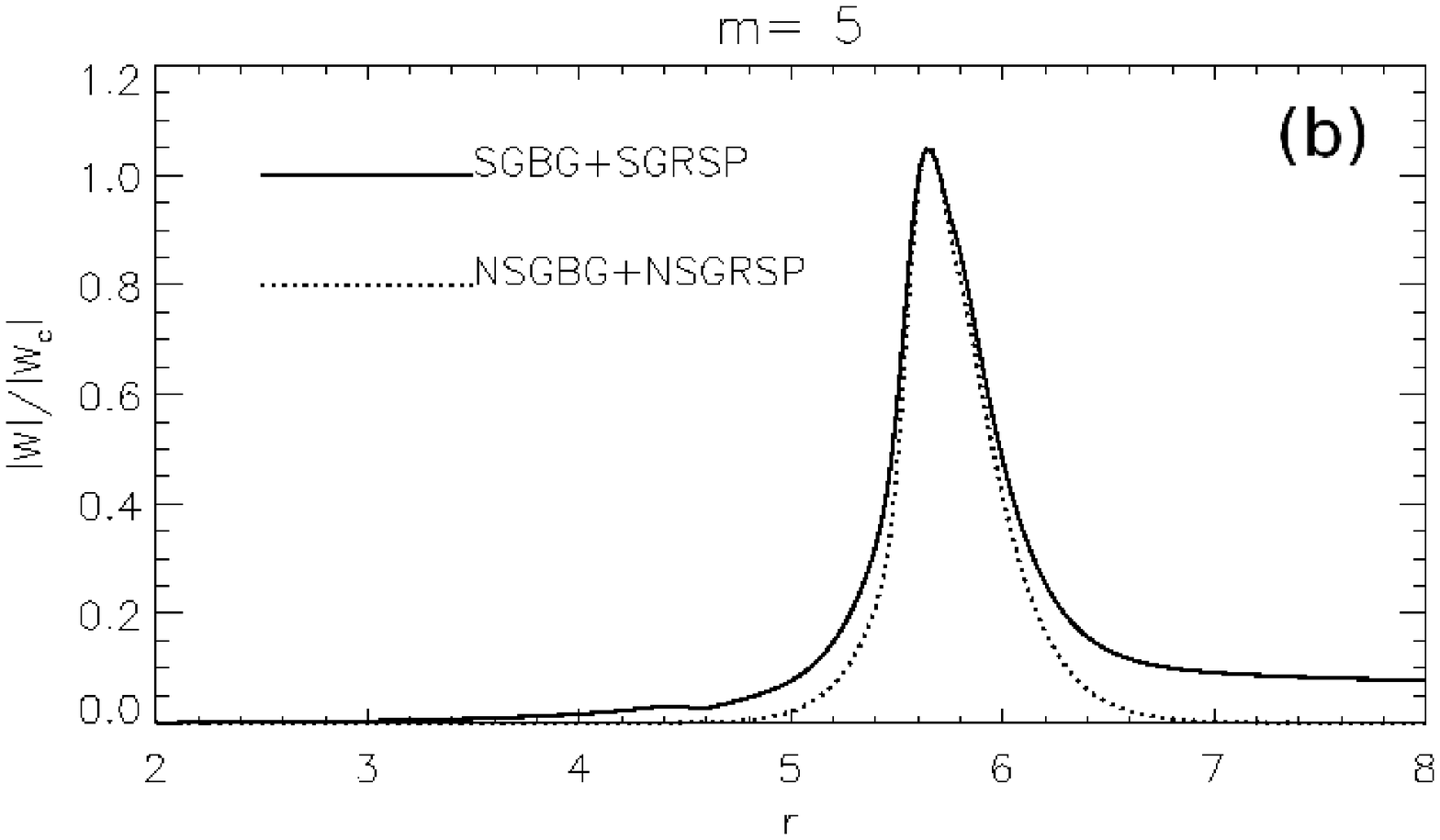}
  \caption{Growth rates (a) obtained from the quadratic approximation to the
    governing equation and the modulus of the $m=5$ eigenfunction
    (b). Here $W_c=W(r=5.7).$  The solid (dotted) lines correspond to the
    case with (without) self-gravity in both  the background  state and response.
    The initial disc model had  $Q_o=4.$
    \label{QEP_compare1}}
\end{figure}

We first present solutions for linear modes for  the $Q_o=4$ disc in the quadratic
approximation.  Fig. \ref{QEP_compare1} compares  the growth rates
$,\gamma,$ and the eigenfunctions,  $W,$ obtained for $m=5$  for  the fully
self-gravitating case  and the fully  non self-gravitating case.  
Growth rates  are such that the most unstable mode shifts
to higher $m$ when self-gravity  is included. Without self-gravity, 
the dominant mode has
$m=3,$ whereas with self-gravity,  the dominant mode has $m=6$---7.  The combination of
self-gravity acting through the background and the linear  response stabilises modes with
$m \leq 5$ and de-stabilises modes with $m\geq 6$.  Thus  higher $m$ vortex
modes are enabled by self-gravity.  

The eigenfunctions $W$  with and without self-gravity are similar, and are
for the most part  localised about co-rotation. This
suggests the nature of the instability is unchanged in this case.    
 However, the  eigenfunction amplitude in $r>6,\,r<5.2$
is larger when self-gravity is included.
 Note too that the amplitude is larger for
 $r>6$ than for  $r<5.2.$ This is not unexpected since the outer 
disc has smaller Toomre $Q$ values
it is expected to be more responsive.  


It is important to remember that the quadratic approximation assumes
modes are localised about co-rotation. The global nature of
some disturbances may invalidate this approximation. Increasing $m$  eventually 
quenches modes in the non self-gravitating case, thus we did not find 
modes for $m\geq 9$ that fit   the description of being localised
modes. For the full governing equation (\ref{sggoverning}), high $m$ modes are expected to have
increasing wave-like behaviour and be  less focused around corotation,
contradicting the quadratic approximation and
so it is not surprising that they are  not found here. 
Similarly, we did not find a localised $m=1$  vortex mode in the self-gravitating case  because
it had been stabilised by the inclusion of  self-gravity, according to earlier
analysis. However, there exists other types of low $m$ mode not
captured by the quadratic approximation. Modes with  extreme values of
$m$ are of less relevance since they are not observed to develop in
hydrodynamic simulations of interest, which typically show the number
of vortices in agreement with the most unstable $m.$       

\subsection{Solutions to the full governing linear  equation}
\begin{figure}
\includegraphics[width=0.99\linewidth]{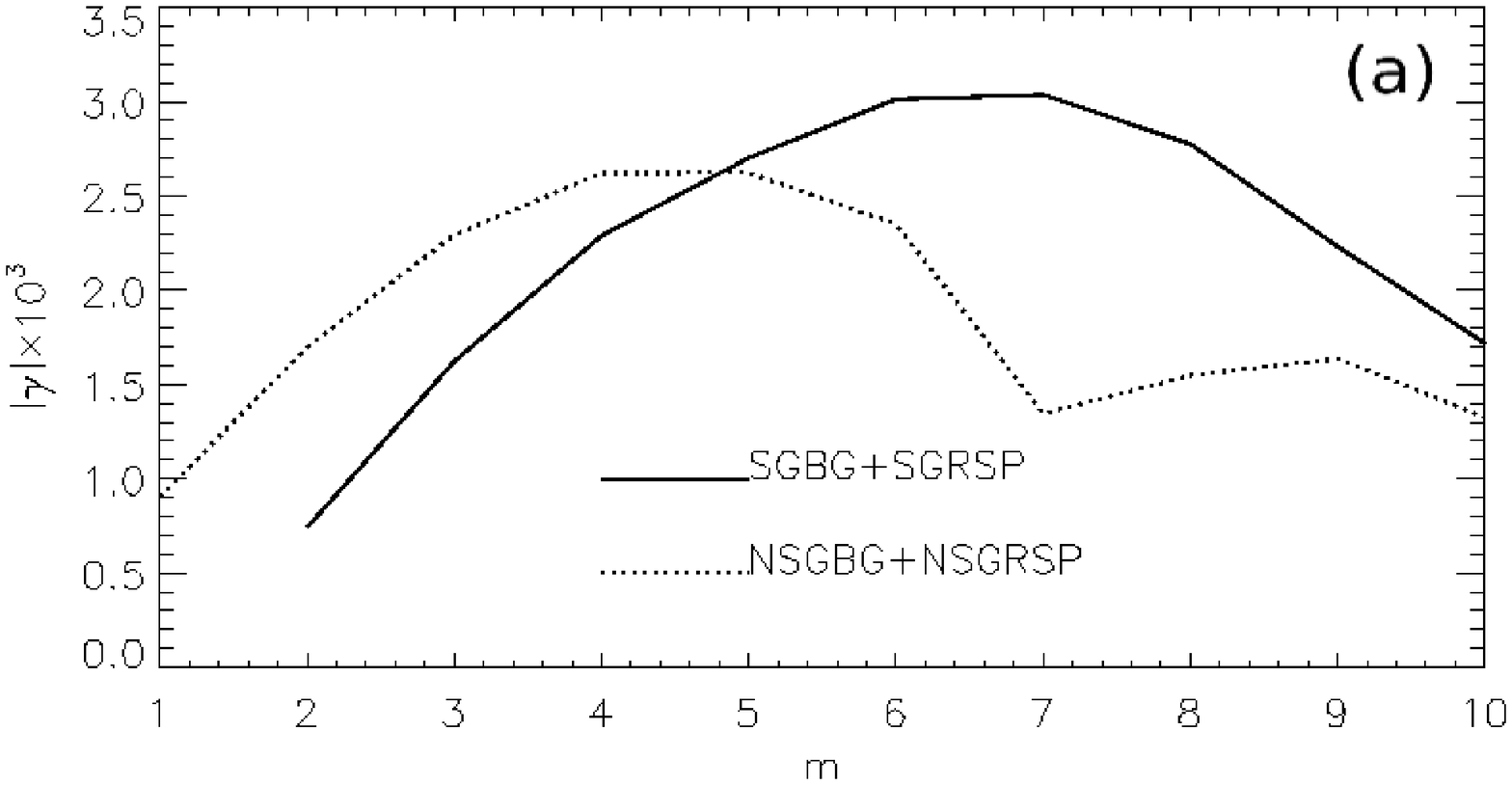}
\includegraphics[width=0.99\linewidth]{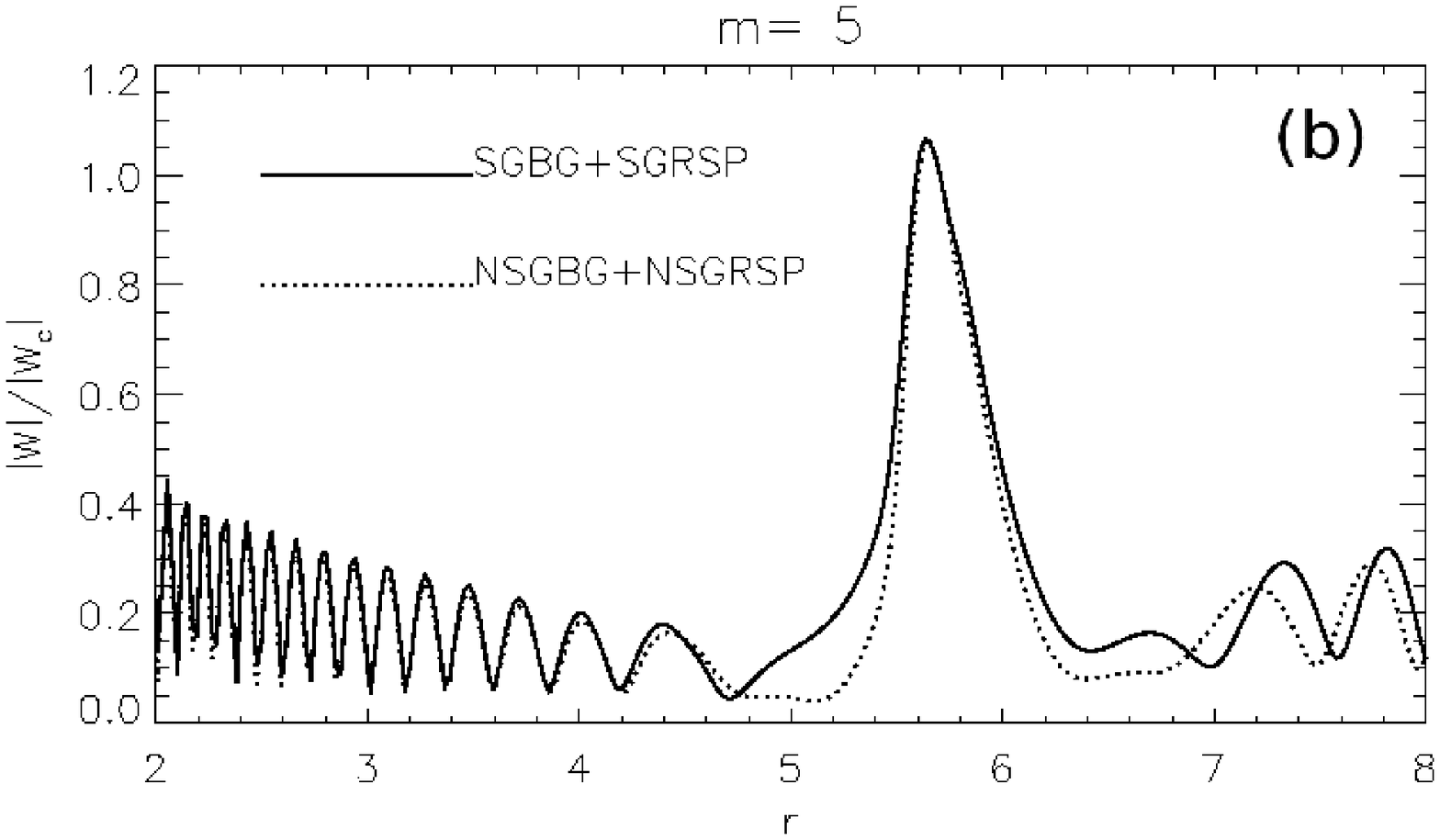}
\caption{Growth rates obtained  for the full governing equation (a) and the eigenfunction
   $|W|$ for $m=5$ (b). Here $W_c = W(r=5.7)$.  Solid lines are for the fully  
  self-gravitating case and dotted lines are for the fully
  non self-gravitating case. The initial disc had  $Q_o=4.$ In
    (a), the local maximum around $m=9$ in the dotted curve is most
    likely caused by a boundary condition 
    effect. Modes with
    $m\geq7$ without self-gravity are not seen in nonlinear
    simulations and are not relevant. 
\label{FULL_compare1}}
\end{figure}

Solutions to the full governing equation (\ref{sggoverning}), for
the fiducial cases  with $Q_o=4$  above are shown in Fig. \ref{FULL_compare1}.
In Fig. \ref{FULL_compare1}  we compare growth rates and the
$m=5$ eigenfunction.  The corotation radii for the
self-gravitating and non-self gravitating cases are  $r_c =
5.88,\,5.79$, respectively.  These radii are close to the local vortensity minimum or
$\mathrm{max}(\eta^{-1})$. In the non self-gravitating case, 
 co-rotation almost coincides with
the extremum. 


The non-self-gravitating case has maximum growth
rate for  $m=4$ and the self-gravitating case is most unstable for
$m=6$---7. Modes with $m<5$ are stabilised while $m>5$
are de-stabilised by self-gravity.
 This results in a shift to higher
$m$ modes  when self-gravity is included in nonlinear simulations of the model. 
Modes with the larger $m$ values  are destabilised, but this is 
not in contradiction with with earlier analysis which assumed small $m$
and did not take account of variations in the form of the background state.
It is shown in section \ref{SGRole} below that the destabilisation of higher $m$ modes
is in fact due to the change in the background state.

Results for $m < 5$ are in essential agreement with those obtained
in  the quadratic  approximation giving us
confidence in our view of the importance of the dominance of the corotation
region.  
As with the quadratic approximation, we did not find a $m=1$ mode with dominant disturbance about the gap edge
in the self-gravitating case.
This suggest such modes are the most easily   stabilised  by increasing self-gravity.
 In the non self-gravitating case, the  growth
rates for $m\geq 8$ are larger than $m = 7$ and do not follow the
trend of decreasing growth rates seen from $m=5\to7.$  
Notice in Fig. \ref{FULL_compare1}(a) (and Fig. \ref{sgrole}(a) below) the local maxima at $m=9$ for non-self-gravitating cases. 
The non-self-gravitating $m\geq8$  modes
are unlikely to be the same type of vortex modes as $m\leq7$
because the former  have significant wave-like regions in $W$ and are not
concentrated near co-rotation as for  $m \leq 7.$  In addition, the $m\geq
8$ region is also where the quadratic approximation  appears to 
fail. These wave-dominated modes demand radiative boundary conditions
rather than the simplistic conditions applied here, which are
appropriate for boundary condition  insensitive modes such as local vortex forming modes.
The non-self-gravitating $m\geq8$ modes identified here are thus likely to be artifacts of the boundary condition.  
Fortunately, these modes  
are irrelevant because they are not the
most unstable, nor are they observed in the corresponding
nonlinear simulations.
By considering  the behaviour for  $m\leq7,$ we can expect a cut-off for vortex modes
around $m=8.$ for this model.

The $m=5$ eigenfunctions in shown in  Fig. \ref{FULL_compare1} are similar. Both
have dominant disturbance around  co-rotation. Behaviour in the  region $r\leq4.6$ is 
essentially identical. The largest difference is found  in the region
$r\geq6.4$. The  disturbance around co-rotation is also somewhat
wider in the self-gravitating case. Comparing with Fig. \ref{QEP_compare1}, we see that
the quadratic approximation captures the main feature of the mode in 
the co-rotation region. However, it removes the
wave-like behaviour in the exterior disc.

\subsection{The role of self-gravity}\label{SGRole}
The calculations above can be  compared to hydrodynamic
simulations where self-gravity is either enabled or not.  
The linear problem allows one to examine separately the effect 
of self-gravity through its influence on the basic state and through
its influence on the
linear response. 

We continue with the disc model with  $Q_o=4$ and compare growth rates for
 a pair of models, one with and the other without
 self-gravity in setting up the basic state
(i.e. the   simulations were run with and without self-gravity enabled)
but both without self-gravity implemented in the
 linear mode calculation (i.e. setting
$\Phi' = 0$). In addition we compared a 
another pair of models,  one  with and the other without self-gravity
implemented  in the
linear response, but with  both having  the background 
state calculated with the disc self-gravity incorporated.

\begin{figure}
\centering
\includegraphics[width=0.99\linewidth]{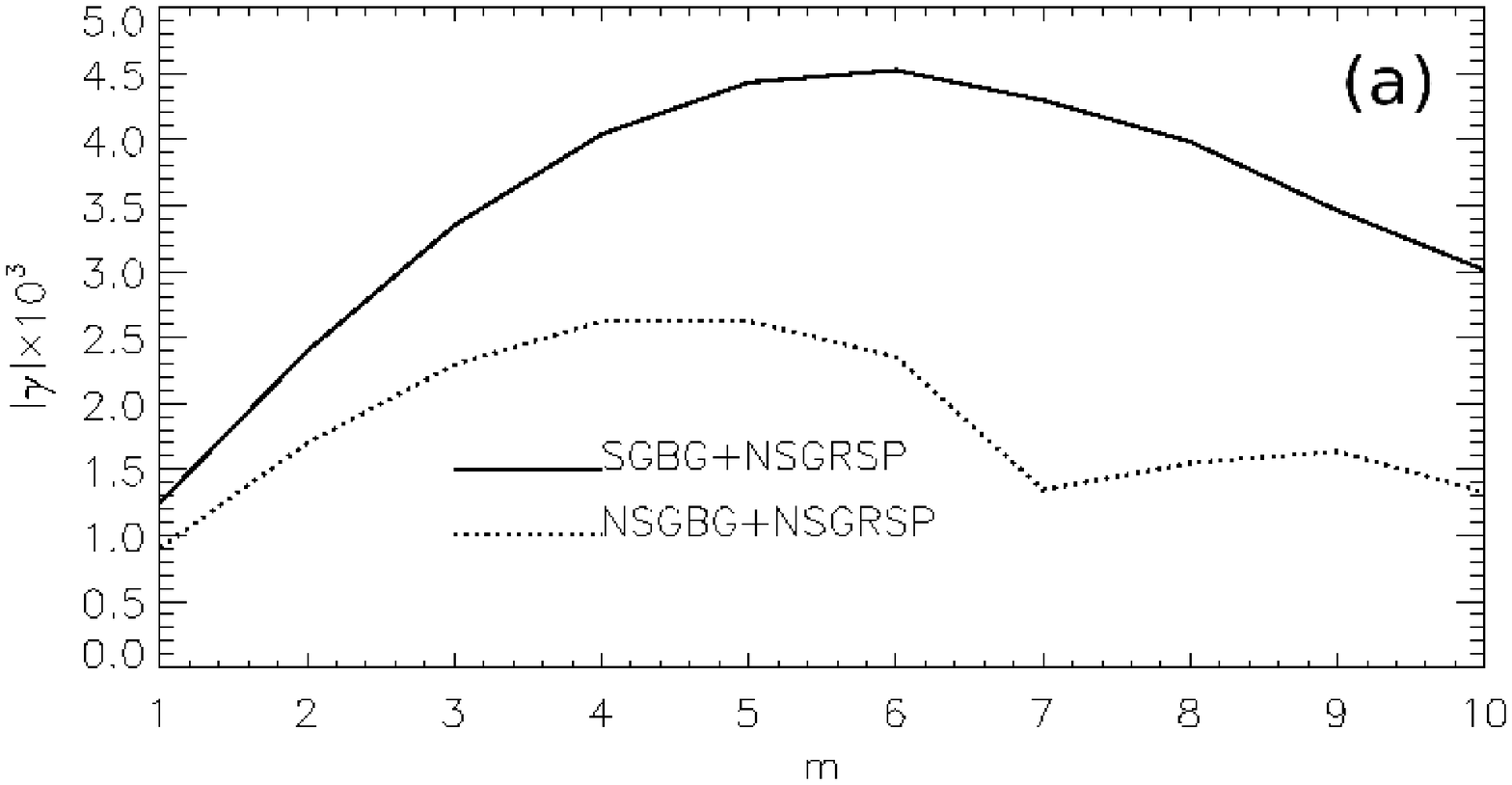}
\includegraphics[width=0.99\linewidth]{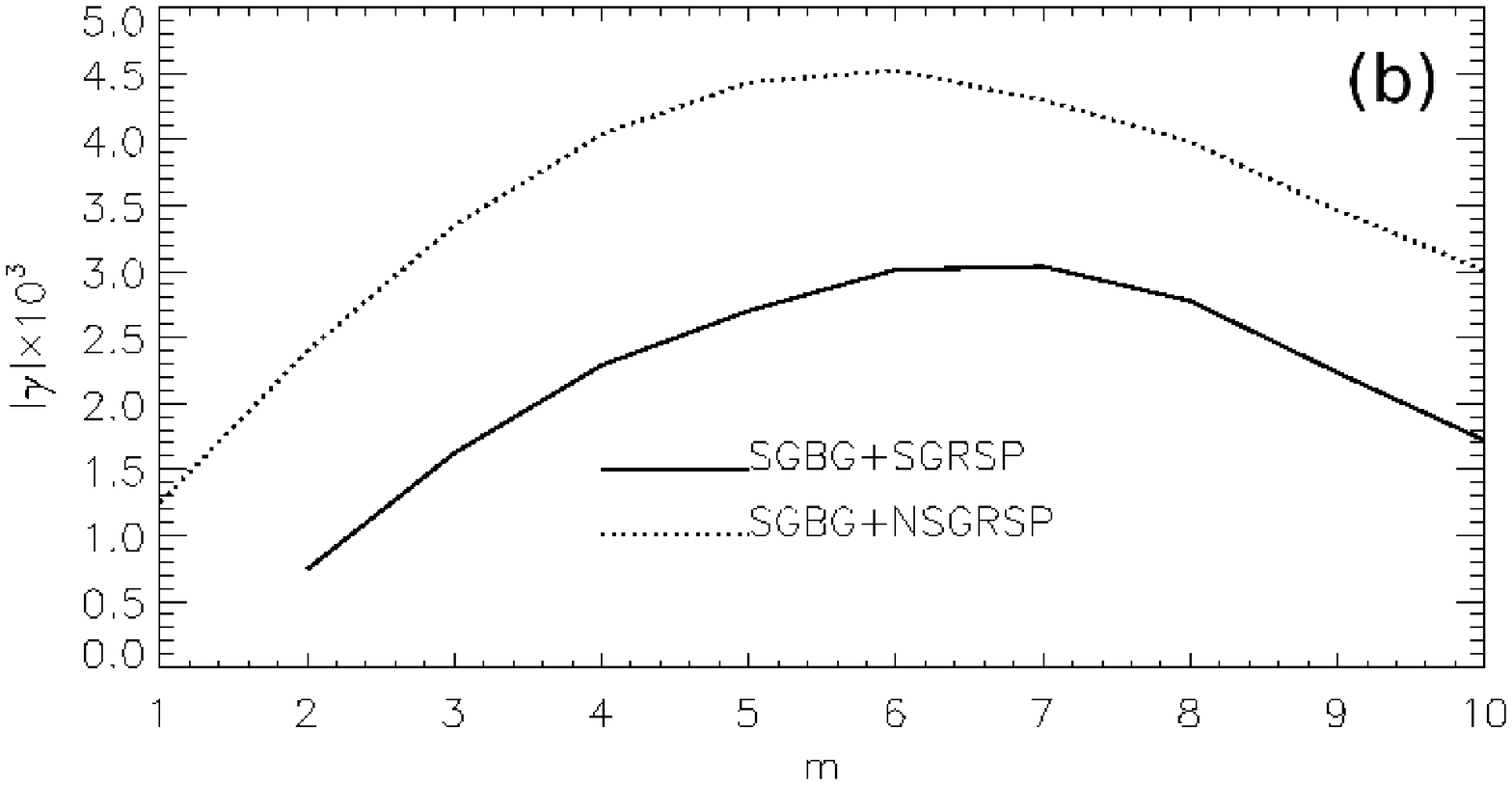}
\caption{The effect of self-gravity on growth rates (as a function of azimuthal wavenumber
$m$) of vortex-forming, gap edge instabilities through the background (a) and through the response (b).  
The disc model is $Q_o=4$. In
    (a), the local maximum around $m=9$ in the dotted curve is most
    likely caused by a boundary condition 
    effect. Modes with
    $m\geq7$ without self-gravity are not seen in nonlinear
    simulations and are not relevant. 
\label{sgrole}}
\end{figure}

Fig. \ref{sgrole}(a) examines the influence of self-gravity through its
modification of  the
background state. It shows that this modification  is de-stabilising.
This is not surprising because a deeper gap is set up when self-gravity is included in the simulation.
In going from the case without self-gravity
to the one with self-gravity, the most unstable mode shifts
from $m=4$---5 to $m=6.$ The vortex modes do not require self-gravity to
operate. However, the difference in growth rates decreases  as $m$ decreases so the
effect due to the modification of the background
is smallest for small $m.$ In other words,  high $m$ modes can be made more unstable  
by including self-gravity in the base state. 


Fig.\ref{sgrole}(b) compares the growth rates obtained from linear
calculations for the same background  state but
 with and without self-gravity
implemented in the response.  The analysis
in \S\ref{perttheory0} applies to this comparison.  Consistent with
that, enabling self-gravity in the response decreases
$|\gamma|$ and stabilises the system
against modes with co-rotation associated with a  vortensity
minimum for \emph{ any value of $m$}.
 Unlike the effect acting  through the background, 
the influence of self gravity  is more significant for low $m$
and can lead to stabilisation.
 The $m=1$ mode has been
stabilised. The most unstable mode  without self-gravity has  $m=5$  and
including self-gravity
 shifts it to $m=6$. 

The fact that self-gravity acting in the linear 
 response and background  state has
opposite effects on growth rates is consistent with the comparison between fully
self-gravitating and fully non-self-gravitating cases 
(Fig. \ref{FULL_compare1}). 
The effect of self-gravity through changes to the background
 and through direct influence on the linear response both
contribute to favouring higher $m$. However, the background effect is
achieved by increasing high $m$ growth rates, whereas the effect via
the response works by  stabilising low $m$ modes in accordance
with the discussion in section \ref{S3.4}.
 Thus  low $m$ vortex
modes become  disfavoured. Overall, we expect more vortices to form,
corresponding to increasing $m,$  with
increasing self-gravity.  

\subsection{Models with different $Q_o$ }
In the calculations presented below,
self-gravity is included both  in  setting up the background and 
in evaluating the linear response.
 We compare the fiducial calculation 
to  cases with $Q_o=3$ (a more massive disc) and $Q_o=8$ 
(a less massive disc). 
The $Q_o=8$ case has a  mass $M_d = 0.012M_*$ which is  usually
considered to be non-self-gravitating in disc-planet simulations. 

\begin{figure}
\centering
\includegraphics[width=0.99\linewidth]{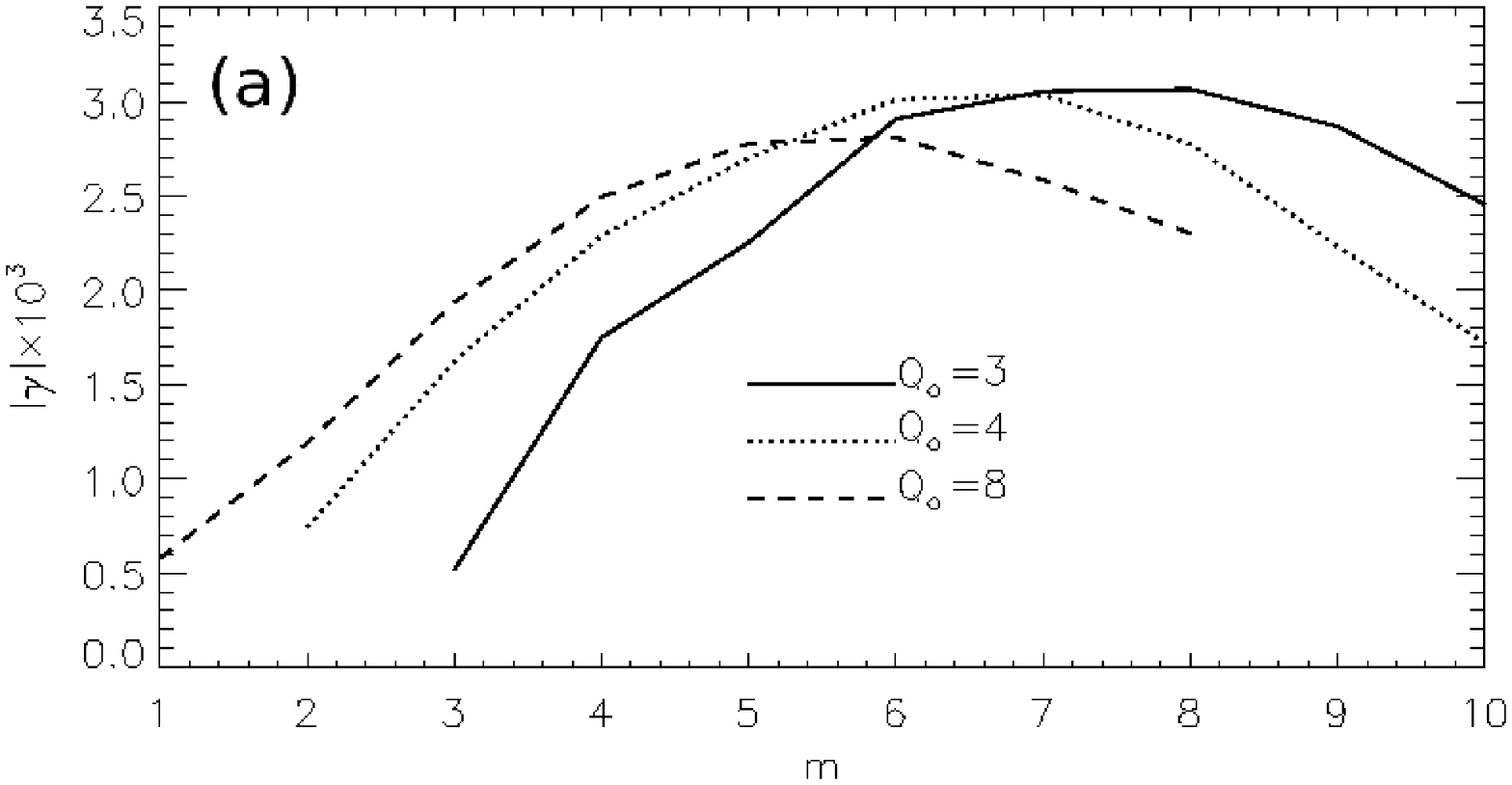}
    \includegraphics[width=0.99\linewidth]{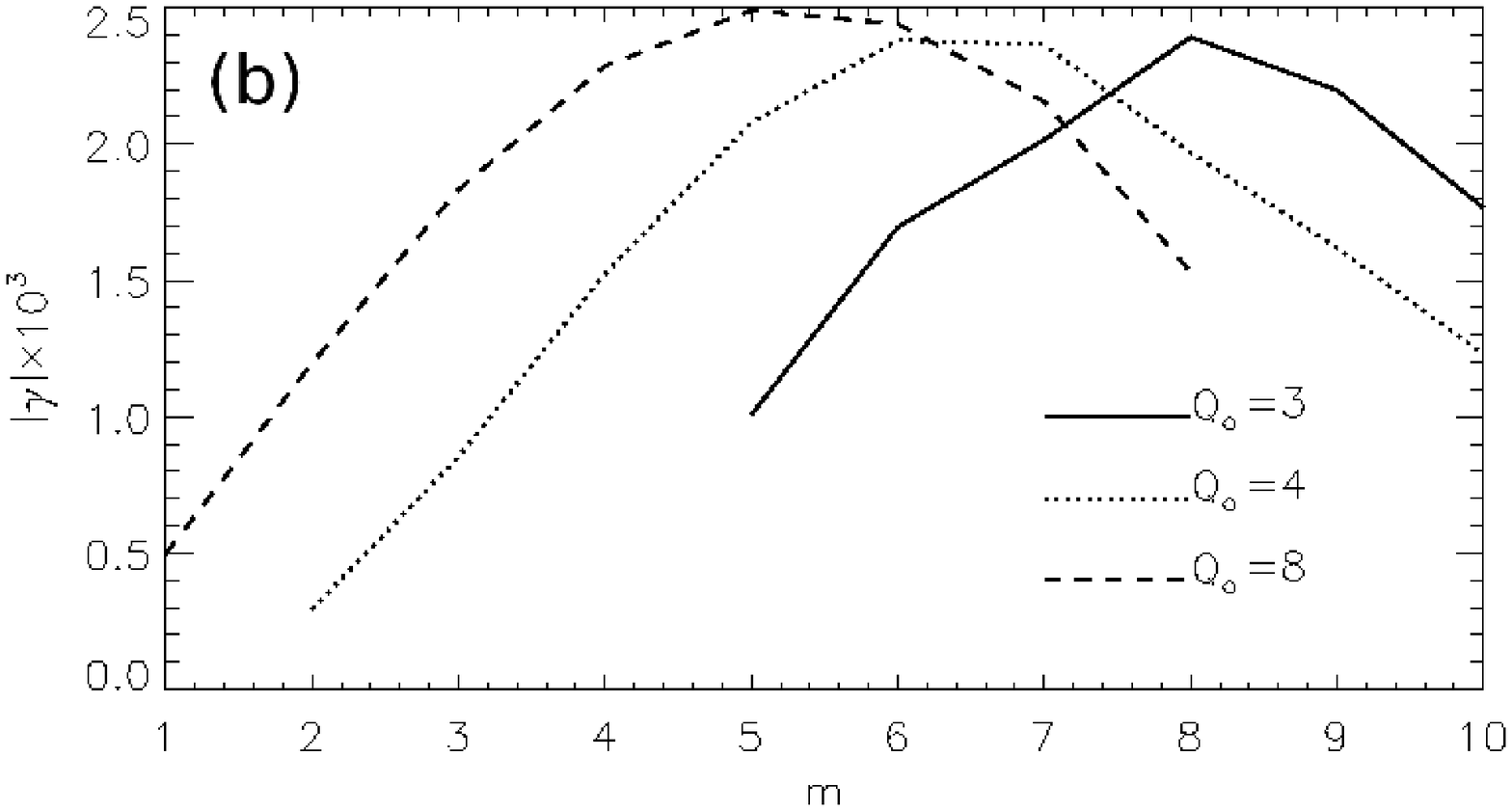}
\caption{Growth rates for unstable modes in background discs with
  $Q_o=3,\,4,\,8$.  (a) is obtained from solving the full linear equation, while (b)
	is from the quadratic approximation. Self-gravity is fully incorporated throughout. 
\label{compareQs}}  
\end{figure}

Growth rates are shown in Fig. \ref{compareQs}(a). The plot
demonstrates a clear shift of the most rapidly 
growing mode  to higher $m$ as $Q_o$ is lowered
(increasing disc mass). For $Q_o=8$ the most unstable mode has
$m=5-6$ and for $Q_o=3$ it shifts 
to $m=7-8$.  The shift is accompanied by the stabilisation (or loss) of low $m$
modes. For example, when the disc mass is halved  as $Q_o$ changes
from  $Q_o=4$ to $Q_o=8,$  the
$m=3$ growth rate decreases by a factor of 3. Modes with the  lowest $m$  are
stabilised by  strong self-gravity, as we  did not find $m=1,\,2$ modes
when  $Q_o=3$.

These calculations have also been performed in the quadratic
approximation.  Fig. \ref{compareQs}(b) shows a similar dependence 
of the  growth rates on  $Q_o$ in this case.  The shift to higher $m$ is more
apparent. For $Q_o=3$ we did not find modes with $m\leq 4.$   Since
the approximation reinforces the localised property of vortex-forming
modes, it means that localised modes for low $m$ become increasing
unlikely as we lower $Q_o$.  Hence, for massive discs only high $m$
vortex forming  modes can develop. The agreement between 
results obtained using the quadratic
approximation and the  full governing 
 equation indicates a lack of sensitivity to boundary conditions
and so is reassuring.

\section{Nonlinear hydrodynamic simulations of vortex
forming  instabilities}\label{hydro1}
We present hydrodynamic simulations of vortex formation and evolution
at edges of gaps opened by a giant planet for discs of varying masses
with self-gravity self-consistently included. 
The disc-planet  models have been already described in \S\ref{disc_planet_model}. 
We consider a Saturn-mass planet with $M_p=3\times10^{-4}M_*$.
{ The kinematic viscosity $\nu=10^{-6}$
corresponds to an $\alpha$
viscosity of $\alpha = \nu/(c_sH) = 1.8\times10^{-4}$ at the planet's
fixed orbital radius at  $r_p=5.$
}

Higher viscosities corresponding to  $\alpha = O(10^{-3})$, 
which is commonly assumed for protoplanetary discs, inhibit
vortex formation \citep{valborro07,lin10}.
If planetary masses appropriate for type I migration  are used,
then a much lower viscosity would be required for vortex formation to occur
\cite[e.g.][ where $\alpha \leq 10^{-5}$ was needed]{li09}. Since we consider
a giant planet, such low viscosity regimes are not needed in order 
to develop vortices.

\subsection{Numerical scheme}
The hydrodynamic equations are evolved using the FARGO code
\citep{masset00}. FARGO is an explicit finite-difference code similar
to ZEUS \citep{stone92} with second order accuracy in space. It
employs a modified azimuthal transport to achieved large time-steps,
otherwise limited by the Keplerian velocity   near the inner
boundary of the domain. Self-gravity for FARGO was implemented and
tested by \cite{baruteau08}. The gravitational acceleration due to the  disc
 is calculated directly using Fast Fourier Transforms in both
azimuth and radius. The latter requires the radial domain to be doubled
when calculating the self-gravity potential.

The disc is divided into $N_r\times N_\varphi = 768\times2304$ computational
grid cells  in  radius
and azimuth. The computational grid 
in the  radial direction is logarithmically spaced. We impose
an open boundary at $r=r_i$ and non-reflecting boundary as used by
\cite{zhang08} at $r=r_o$ \citep{godon96}.    
Since vortices are localised features, as
long as  gap edges are  far from boundaries, boundary conditions can only  have a
limited effect. We also performed  some  simulations with  damping
boundary conditions \citep{valborro07} and  open outer boundaries.  We
found similar results to those presented below.

\subsection{The effect of self-gravity}\label{sg_nsg}
We  compare  cases where disc gravitational potential is either  included or
excluded in the total potential calculation. This is the standard distinction
between self-gravitating and non-self-gravitating disc-planet
simulations \citep{nelson03a,zhang08}. We consider the $Q_o=4$ (
$Q_p=7$) disc. The disc mass is $M_d = 0.024M_*$.

\begin{figure}
\centering
\includegraphics[scale=.43,clip=true,trim=0cm 0cm 1.5cm 0cm]{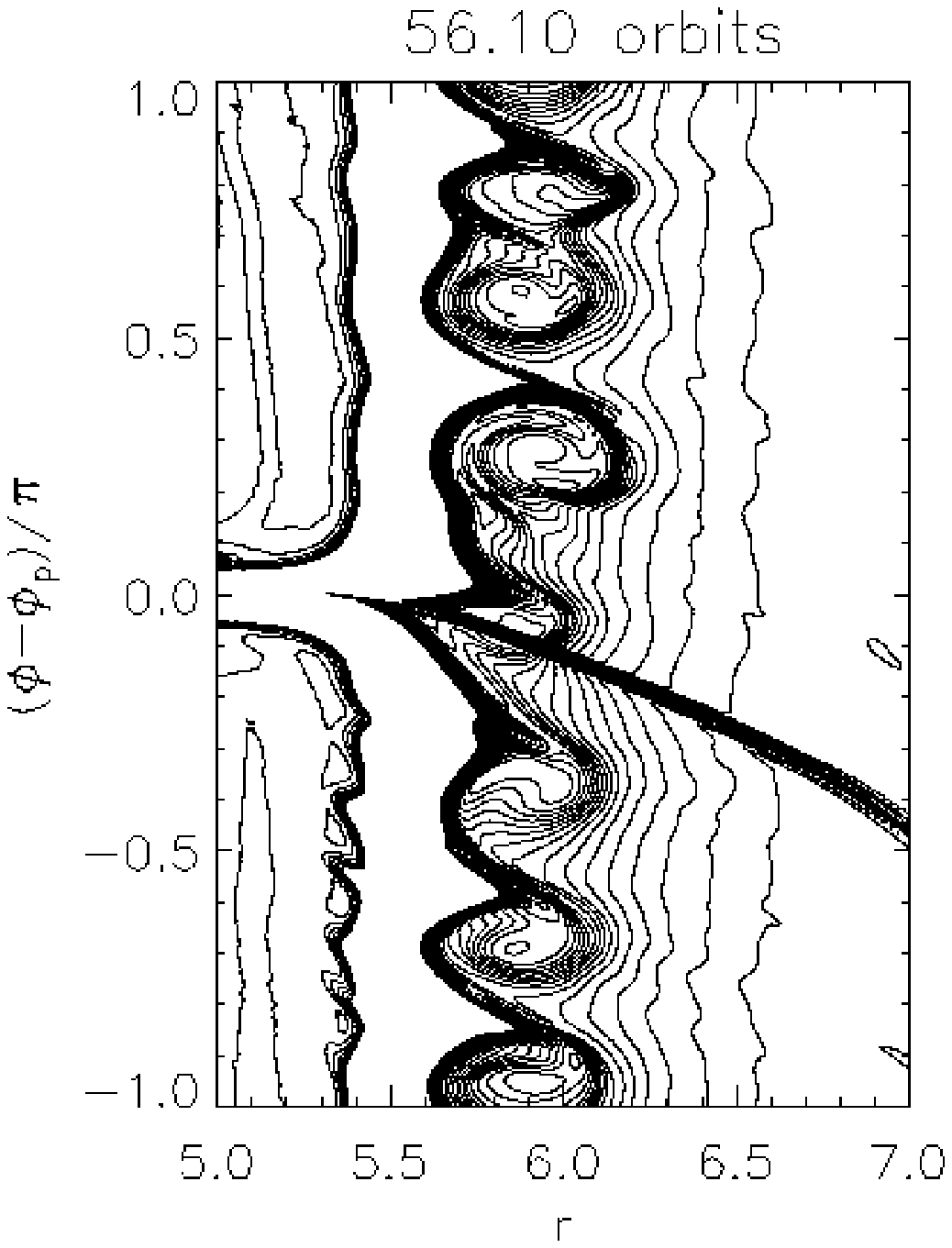}
\includegraphics[scale=.43,clip=true,trim=1.1cm 0cm 0cm 0cm]{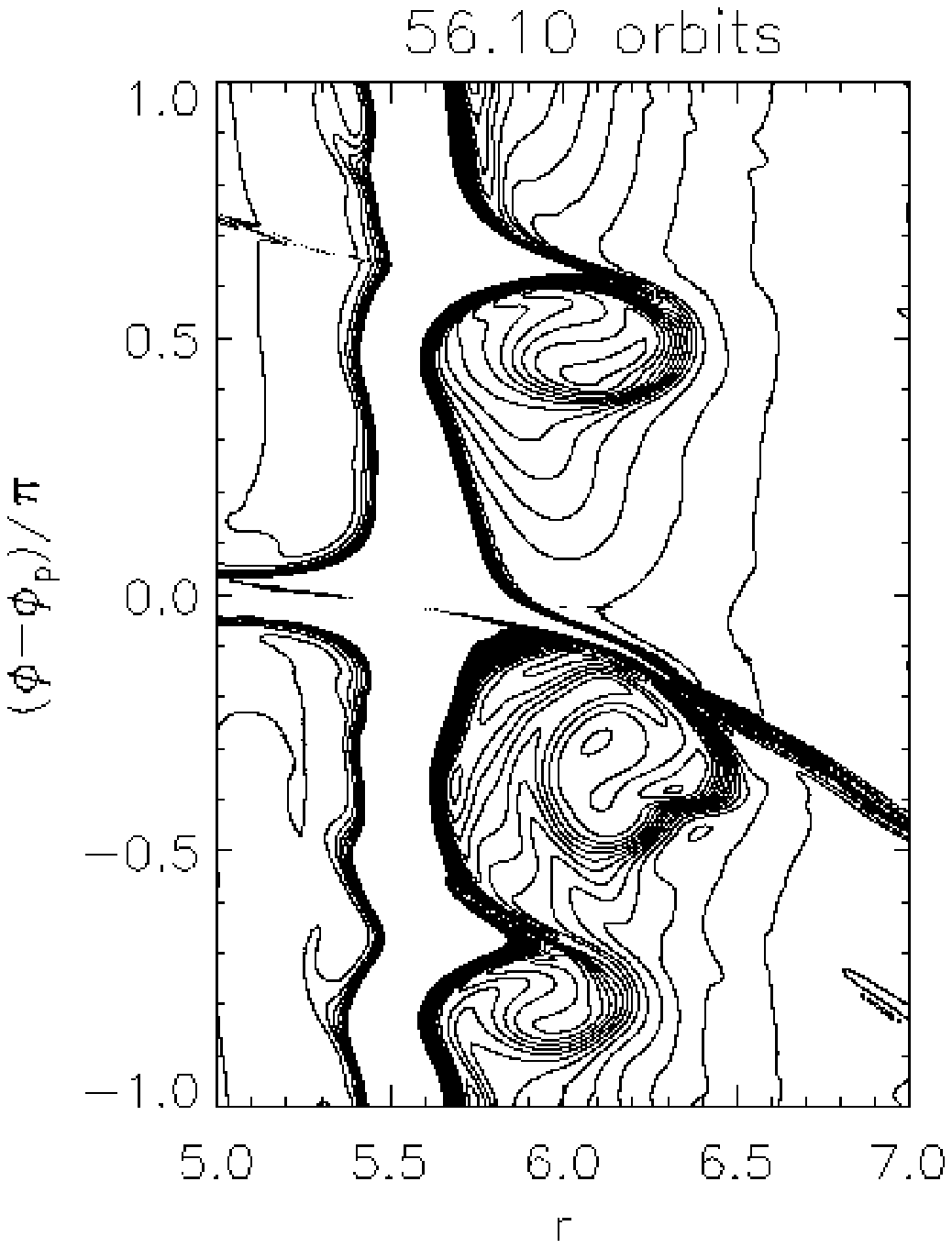}
\caption{Vortensity contours for $Q_o=4$ with (left) and without (right)
  self-gravity. The planet is located at $r = 5,\,\phi =   \phi_p$. The
  thick lines crossing the outer radial boundaries
   correspond to spiral shocks induced
  by the planet. 
\label{vortices_sg_nsg}}
\end{figure}

Fig. \ref{vortices_sg_nsg} shows vortensity
contours that illustrate  vortex formation at the outer gap edge.
There is a local
vortensity maximum at about $r=5.5$
that has been  produced  by flow through shocks induced by the planet.
 This maximum 
remains  largely undisturbed  indicating that instability 
is associated with structure outside it and associated
with  the vortensity minimum.
More
vortices are excited when self-gravity is included. In that case  the
$m=6$ vortex mode dominate whereas  the $m=3$ mode
dominates in the non self-gravitating case.
 This behaviour  is consistent with linear calculations. 

Fig. \ref{vortices_sg_nsg} also shows that vortices have radial
length-scales comparable to 
 the local scale-height ($\simeq H(6) = 0.3$). 
The  vortices in these cases  have radial sizes of  $\sim 2.2H(6)$ without self-gravity and
$\sim 1.3H(6)$ with self-gravity. The  
vortices are of smaller radial extent  when self-gravity is included
 because of the preference for
higher $m.$ A decrease in radial size can be expected from the
decrease in width of the WKBJ evanescent zone centred on corotation.
This is determined by the condition 
\begin{align}
  (\sigma + m\Omega)^2 = \kappa^2(1 - 1/Q^2). 
\end{align}
 Writing $\sigma = -m\Omega(r_c)$, it
is straight forward to show that the radial width of the evanescent zone 
approximately scales as $1/m$.  
Since
the preferred $m$ is approximately a factor of two smaller, we expect
vortices in the case without self-gravity to be double the size of those
in the case  where
self-gravity is included. 

In the case with self-gravity,
the  vortices are approximately centred  along the radius $r=5.9$, close to the
co-rotation radius expected from linear calculations  $r_c=5.88$. 
In the non self-gravitating case,
 linear calculation gives $r_c=5.81$ for $m=3$, but 
the  vortices are approximately centred along $r=6.$  
However, perturbations here are already in the non-linear regime
and interaction between vortices  or with the spiral shock
may shift the vortices around. 
Note too that in this regard  there is 
more variation in the radial locations of the vortices as compared
to the case with self-gravity.

\subsection{Varying disc mass: gap profiles}\label{gapprofiles} 
\begin{figure}
\centering
\includegraphics[width=0.99\linewidth]{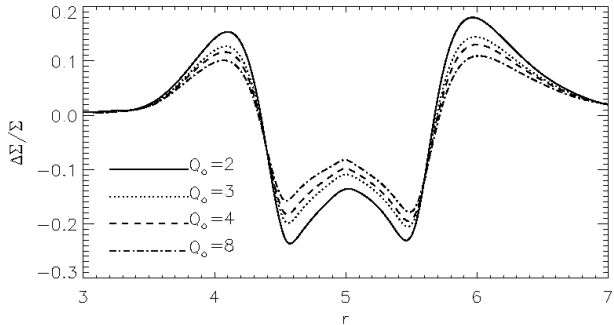}
\caption{Gap profiles opened by a Saturn mass planet in
  self-gravitating discs as a function of disc mass, parametrised by
  $Q_o$. 
  The azimuthally averaged relative surface density
  perturbation is shown. The planet located at $r = 5 $.  
\label{gap_profiles}}
\end{figure}

We present  simulations of discs models with $ 1.5 \leq Q_o \leq
8$, equivalently $ 2.6 \leq Q_p \leq 14$  or  $0.063 \geq M_d/M_* \geq
0.012$. The equilibrium gap profiles have a range of $Q$ values
with local extrema of $1.5\leq
\mathrm{min}(Q)\leq 9.5$ and $4.8\leq \mathrm{max}(Q)\leq 21.6$ near the outer gap edge. 
The gap
profiles opened by the planet are given  in Fig. \ref{gap_profiles}
for a range of $Q_o$ that develop vortices. Outside the plotted region
the curves are indistinguishable. 
The gap  deepens with decreasing $Q_o$ and gap edges become steeper. 
In going from $Q_o=8\to2$ the
gap depth $|\Delta\Sigma/\Sigma|$ increases by about 0.05---0.08, similarly the
bumps near gap edges increase by 0.05---0.06. On a global scale,
self-gravity becomes more important with increasing radius. However, we
observe no trend in the difference between gap profiles with respect to radius. 

Self-gravity affects gap structure on  a local scale by increasing
the effective planet mass so that $M_p \rightarrow M_p^\prime$.
 A straightforward estimate, based on the unperturbed disc model,
 of the expected  mass within  the Hill
radius $,r_H,$ of a
point mass planet with $M_p=3\times10^{-4}M_*$ is $M_H=0.047M_p$ for $Q_o=8$ and
$M_H=0.17M_p$ for $Q_o=2.$ It is likely that
 $M_H$, or at least some significant  fraction of it
adds to the effective mass of the planet acting  on the disc when
self-gravity is self-consistently included.
 Thus $M_p^\prime,$ and therefore also the gap depth, 
 is expected to  increase with disc
mass. Thus we note that  without carefully tuning $M_p$,
self-gravitating disc-planet calculations with different surface
density scales will always differ in $M_p^\prime$. Since the gap profiles
differ, as we have seen above,  stability is affected also.
 Note that the gap
width $w$ in Fig. \ref{gap_profiles} does not change greatly with $Q_o$, 
which is consistent with the scaling
$w\propto r_H\propto M_p^{\prime1/3}.$ However, the 
peaks/troughs for $Q_o=2$  do lie slightly closer to the orbital radius  $r_p$
 than other cases. This
is because shocks are induced closer to the planet due to increased 
$M_p^\prime$ \citep{lin10}.

\subsection{Varying disc mass: gap stability} 
\begin{figure*}
\centering
\includegraphics[scale=0.33]{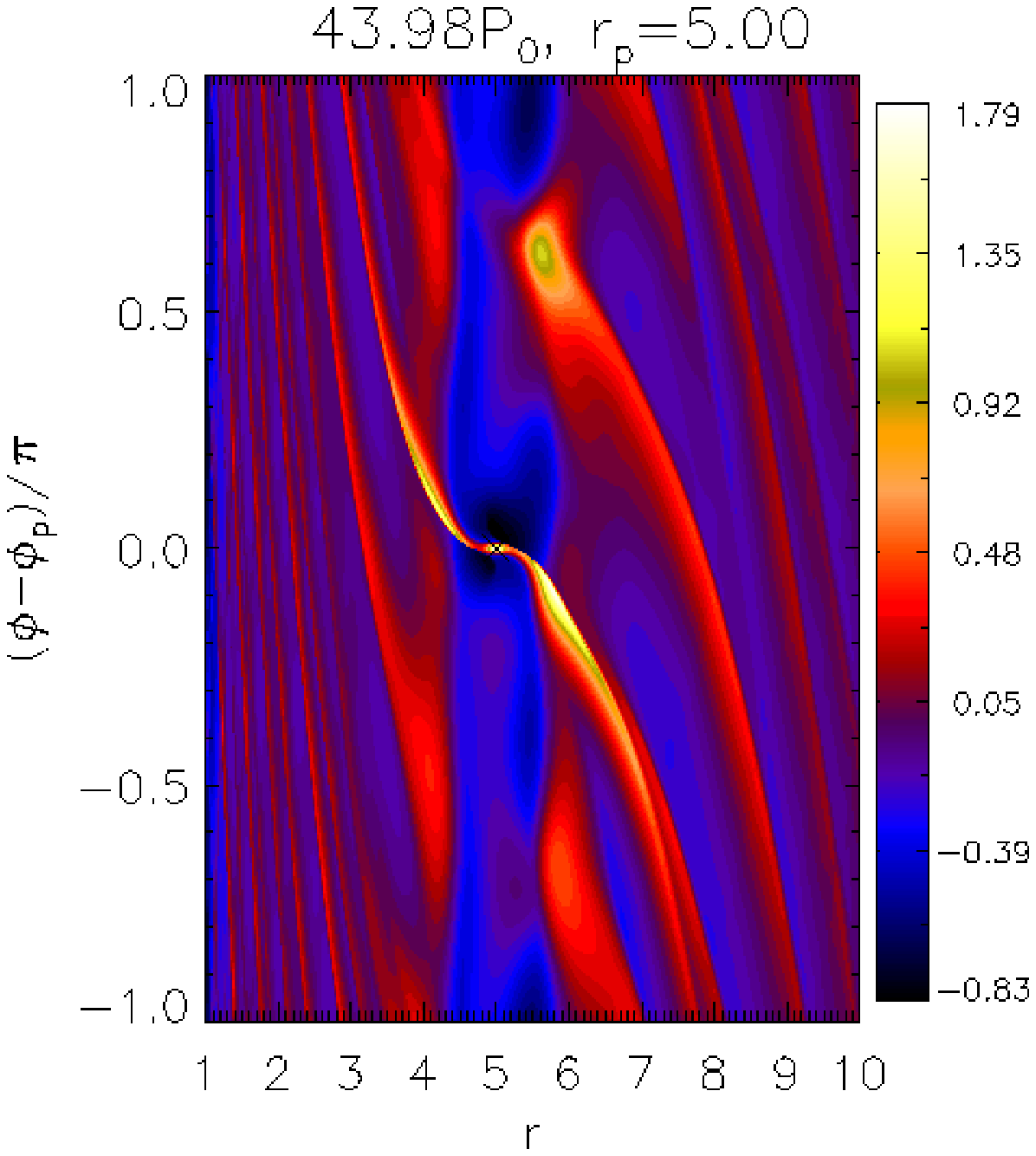}
\includegraphics[scale=0.33,clip=true,trim=2.3cm 0cm 0cm 0cm]{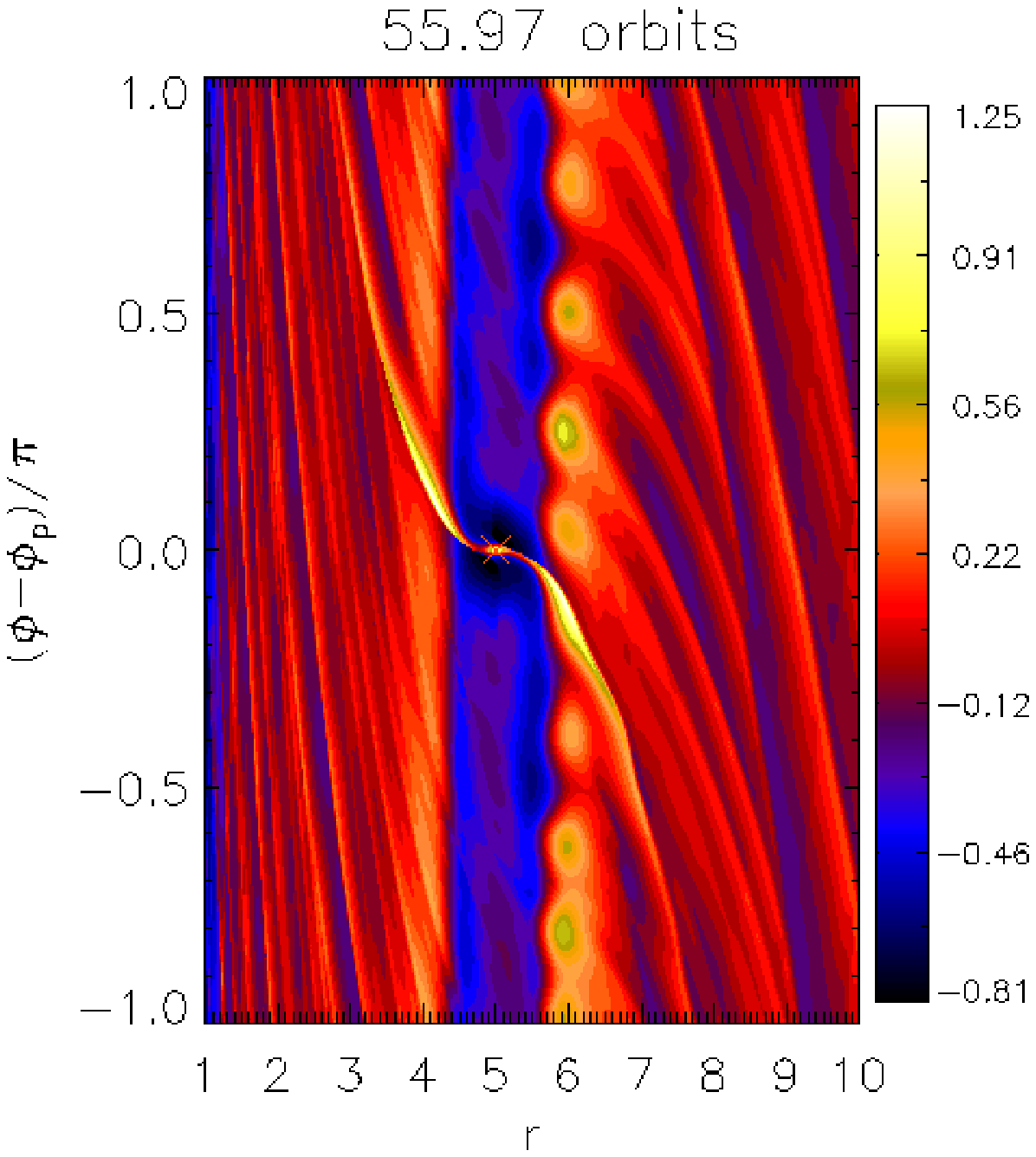}
\includegraphics[scale=0.33,clip=true,trim=2.3cm 0cm 0cm 0cm]{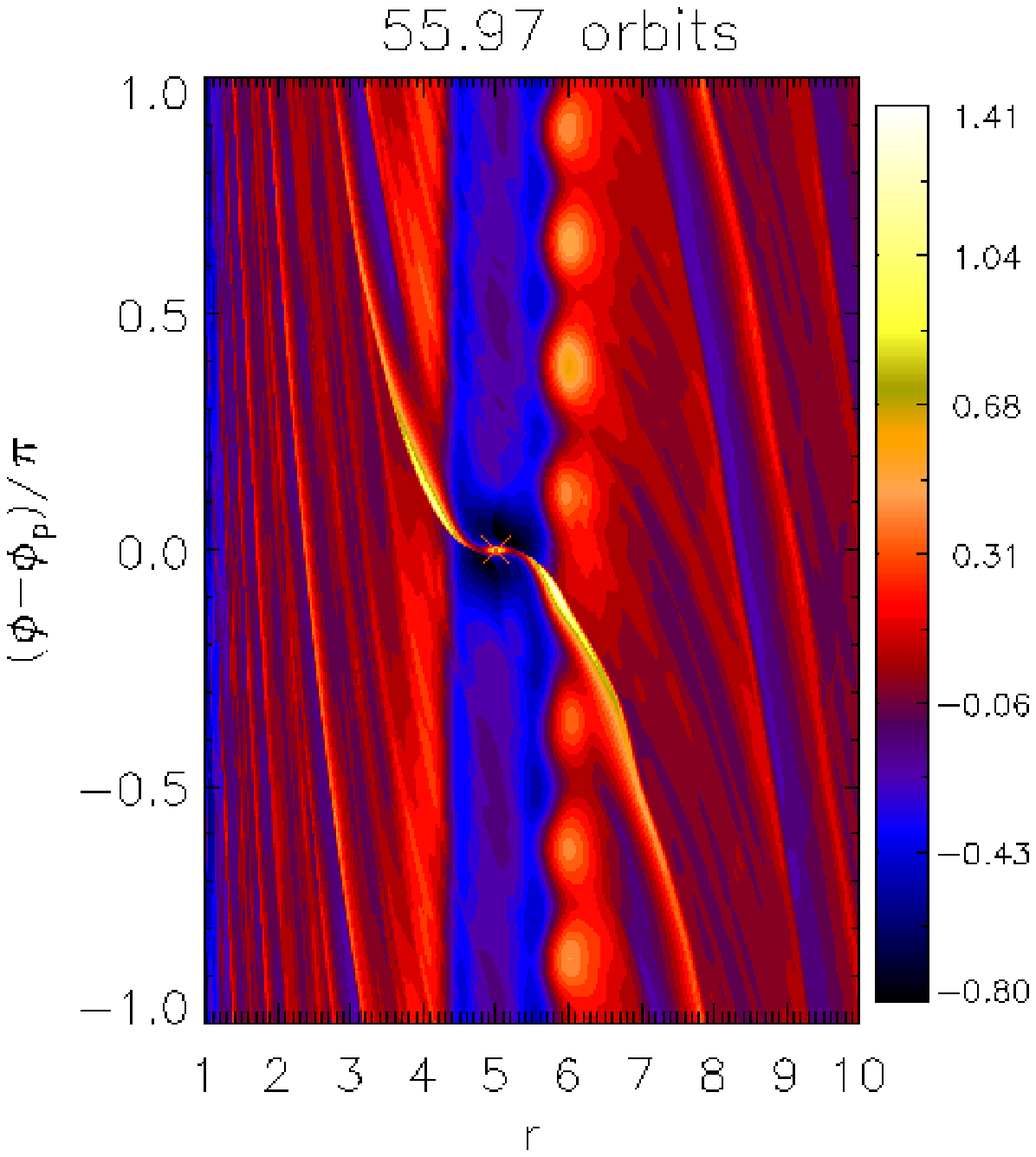}
\includegraphics[scale=0.33,clip=true,trim=2.3cm 0cm 0cm 0cm]{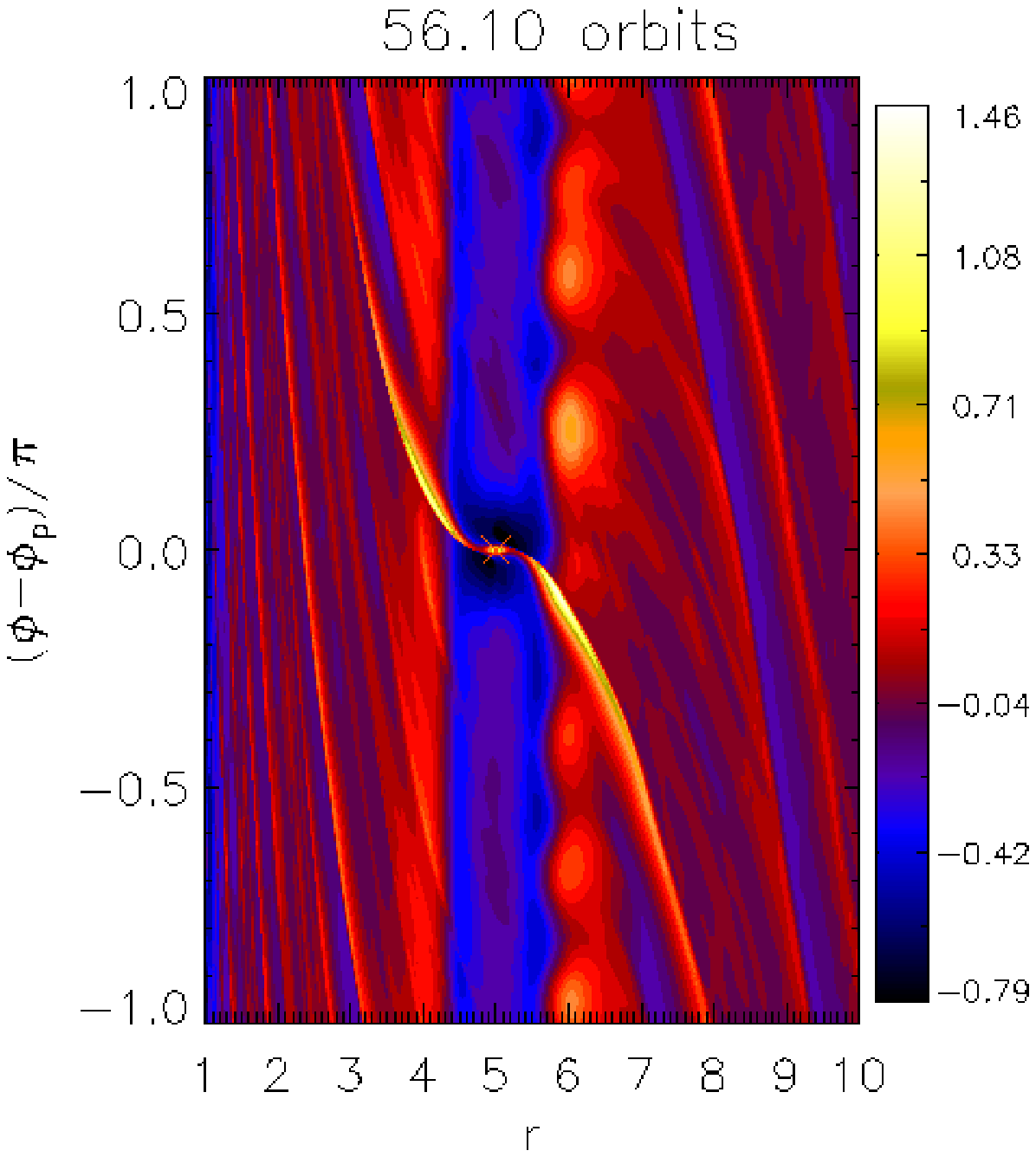}
\includegraphics[scale=0.33,clip=true,trim=2.3cm 0cm 0cm 0cm]{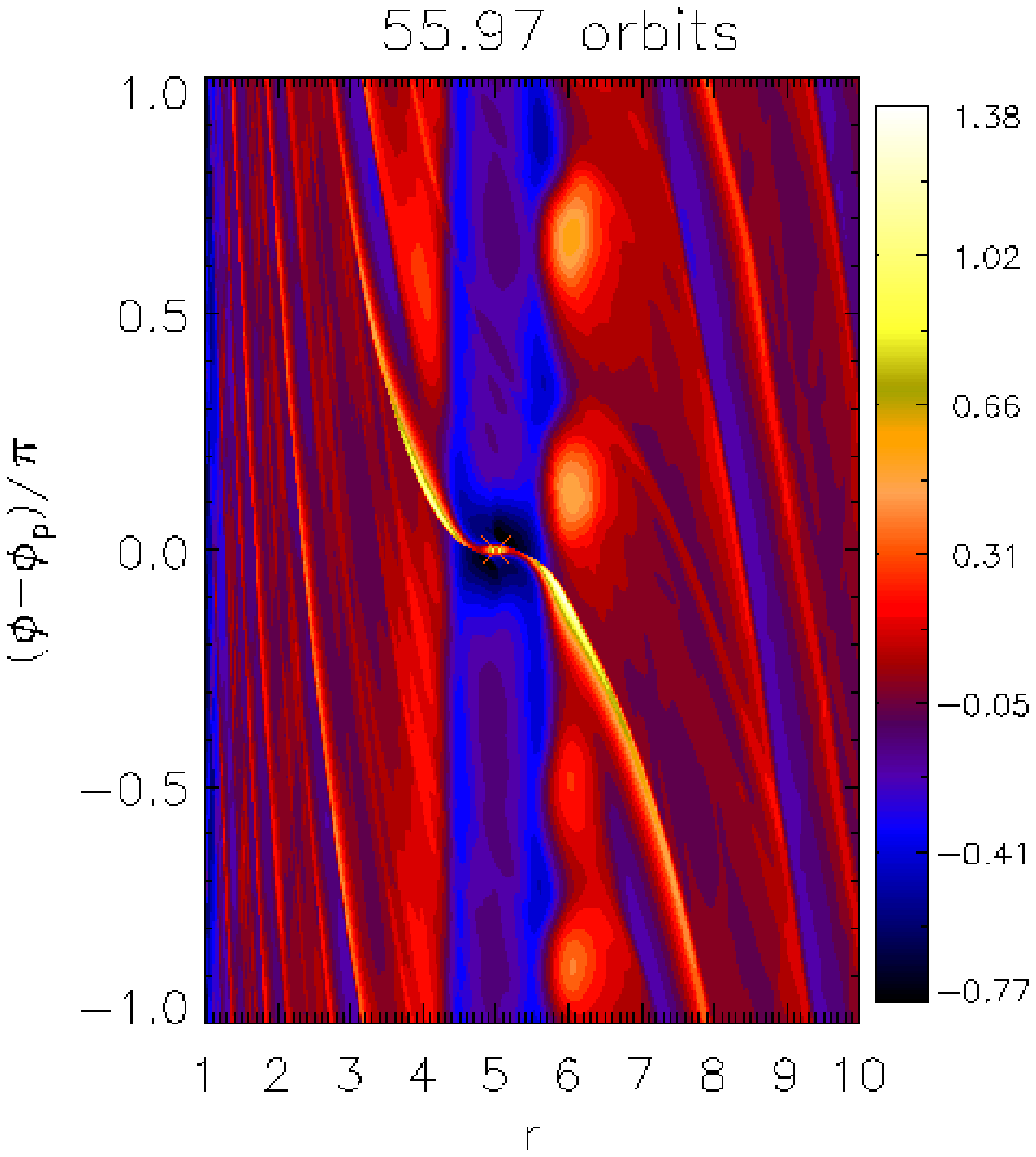}\\
\caption{Instability at the outer gap edge of a Saturn-mass planet, in
  a discs with minimum Toomre parameter, from left to right, of $Q_o = 1.5,\, 2,\, 3,\, 4,\, 8$.
  The relative surface density perturbation is shown.
\label{compare_v}}
\end{figure*}

Fig. \ref{compare_v} shows snapshots of the  relative surface density
perturbation as instability sets in. Consider  $Q_o\geq 2$ first.
The instability is associated with the outer gap edge while the inner
edge remains relatively stable. In the least massive
disc $Q_o=8$, 3---4  vortices form at the outer gap edge,
similar to what happens in the non-self-gravitating disc in \S\ref{sg_nsg}.
 As we
increase the disc mass, more vortices develop. By $Q_o=2$, 
8 surface density maxima can be identified. Note that a
vortex may be obscured if it coincides with the planetary wake. In
moving from $Q_o=8\to Q_o=2,$  vortices become smaller and their centres
move inwards. When  $Q_o=8 $ local surface density maxima lie just
outside $r=6$ while for $Q_o = 2 $ they  lie just interior to
$r=6$. 

Increasing $M_d$  makes the perturbation more
global. Vortices in massive discs can gravitationally perturb parts of
the disc further out, similar to a planet. Lowering $Q_o$, vortices
develop longer, more prominent trailing  spirals exterior to them. This
is most notable with $Q_o=2$, where the vortex spirals can have
comparable amplitudes as the planet wake. However, increasing
self-gravity even further,  we expect a global instability to eventually develop as
the vortices perturb the disc strongly via self-gravity.

The $Q_o=1.5$ disc does not develop vortices. This is
consistent with the stabilisation effect of self-gravity on the vortex
instability through the linear response. Instead, $Q_o=1.5$ develops
global spiral instabilities associated with the gap edge. We call
these edge modes.  They have been suggested to occur in
self-gravitating discs with surface density depression
\citep{meschiari08}. Here, we have shown that they can indeed develop
in gaps self-consistently opened by a giant planet. Detailed
discussions of edge modes are beyond the scope of this paper and are
presented in \cite{lin11}, but for reference
we note some important differences to the vortex modes. Edge modes 
here are associated with local vortensity \emph{maxima}, which lie closer to the planet than vortensity
minima.
 Perturbations at the inner gap edge are  also identified and correlate
with those  at the outer gap edge, where the over-density is largest
(and stronger than for vortex disturbances).  Communication between the two
sides of the gap is only possible with sufficient self-gravity.




\begin{figure}
  \centering
  \includegraphics[width=0.99\linewidth]{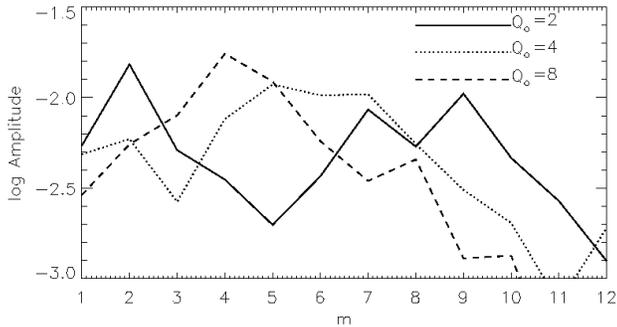} 
  \caption{Fourier amplitudes of the surface density in the outer disc
  $r\in[5,10],$ normalised by the $m=0$ component,
    for some of the  models illustrated in Fig. \ref{compare_v}. 
  \label{compare_amps}}
\end{figure}

The Fourier amplitudes of the surface density in $r\in[5,10]$,
as a function of $m,$
is shown in Fig. \ref{compare_amps} for
$Q_o=2,\,4,\,8$ at $t=56P_0$. Amplitudes have been normalised by the
axisymmetric component. The shift to higher $m$ vortex modes with
increasing disc mass is evident as expected from linear calculations.  For
$Q_o=8,4,2$ the preferred vortex modes have  respectively  $m=4,\, 5$---7
and 7---9, with average amplitudes that decrease with  $Q_o$. The
latter may reflect the stabilisation effect of self-gravity
on  linear modes with low $m.$  

An important region in Fig. \ref{compare_amps} is $m\leq
3$. There is a peak in amplitude at $m=2$ for $Q_o=2,\,4$. These are
the edge modes described above. They were not seen in linear
calculations because there we focused on finding vortex modes, which
have co-rotation at or close to local vortensity minima.  The loss of low $m$
vortex modes with increasing self-gravity observed in linear
calculations, is replaced by the increasing prominence of global edge
modes. Fig. \ref{compare_amps} shows the $m=2$ amplitude becomes more
significant with increasing self-gravity.  

In the $Q_o=2$ simulations we do see evidence of an $m=2$ disturbance
hindering vortex evolution. This transition case is difficult to
analyse since both types of instabilities
develop. Fig. \ref{compare_amps} shows the 
edge and vortex modes have comparable amplitudes in $r>5$. However, considering
the region $r\geq 7$ for $Q_o=2$ we found
$m=2$ is dominant, because the edge mode is global 
whereas vortex mode disturbances are localised to the
edge ($r\simeq6$). Increasing self-gravity further, we expect eventually edge
modes become dominant, this is seen in the $Q_o=1.5$ case in
Fig. \ref{compare_v}.  


\subsection{Evolution  and merging of self-gravitating vortices} 
\begin{figure*}
\centering
\includegraphics[scale=.33]{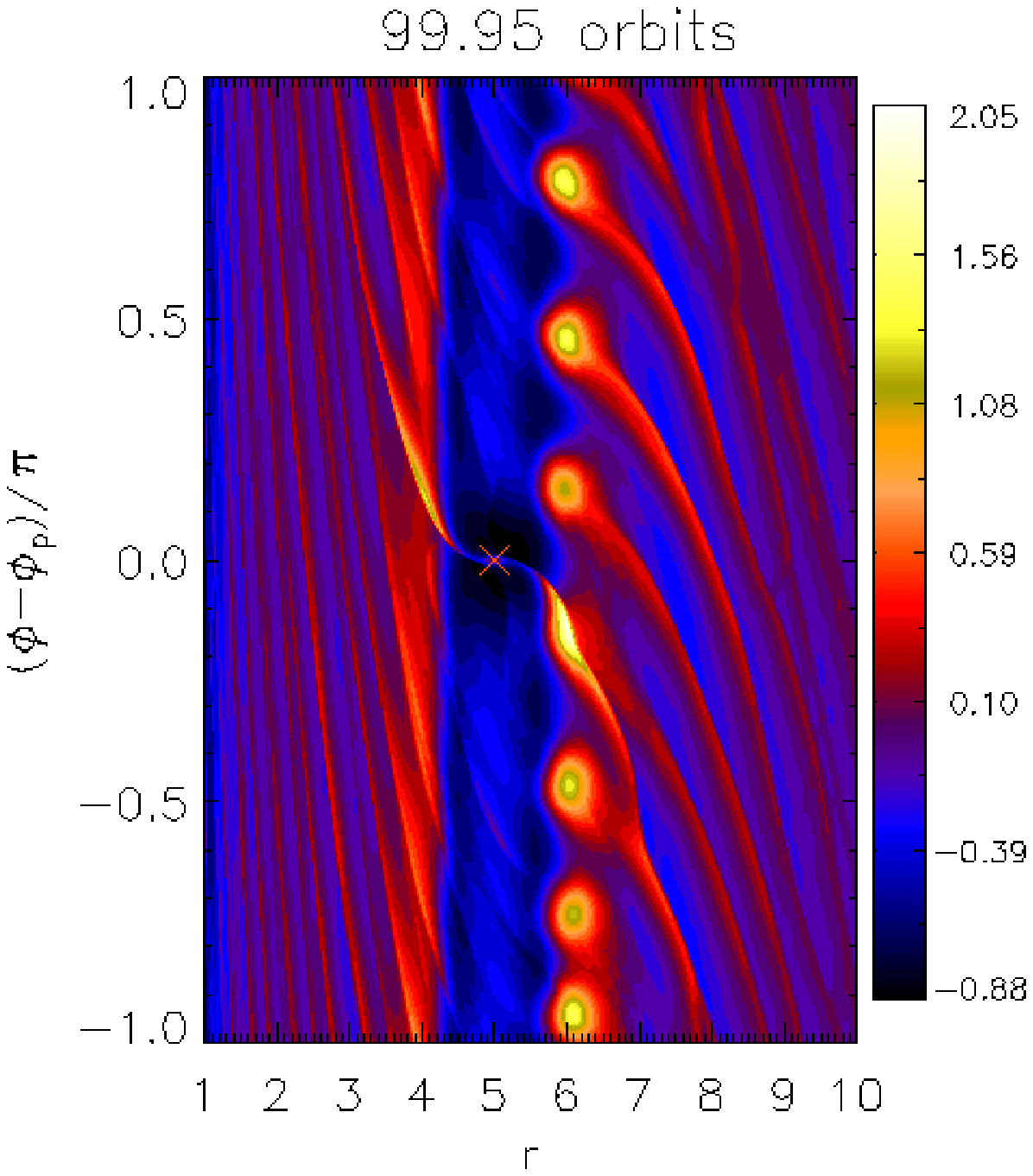}
\includegraphics[scale=.33,clip=true,trim=2.3cm 0cm 0cm 0cm]{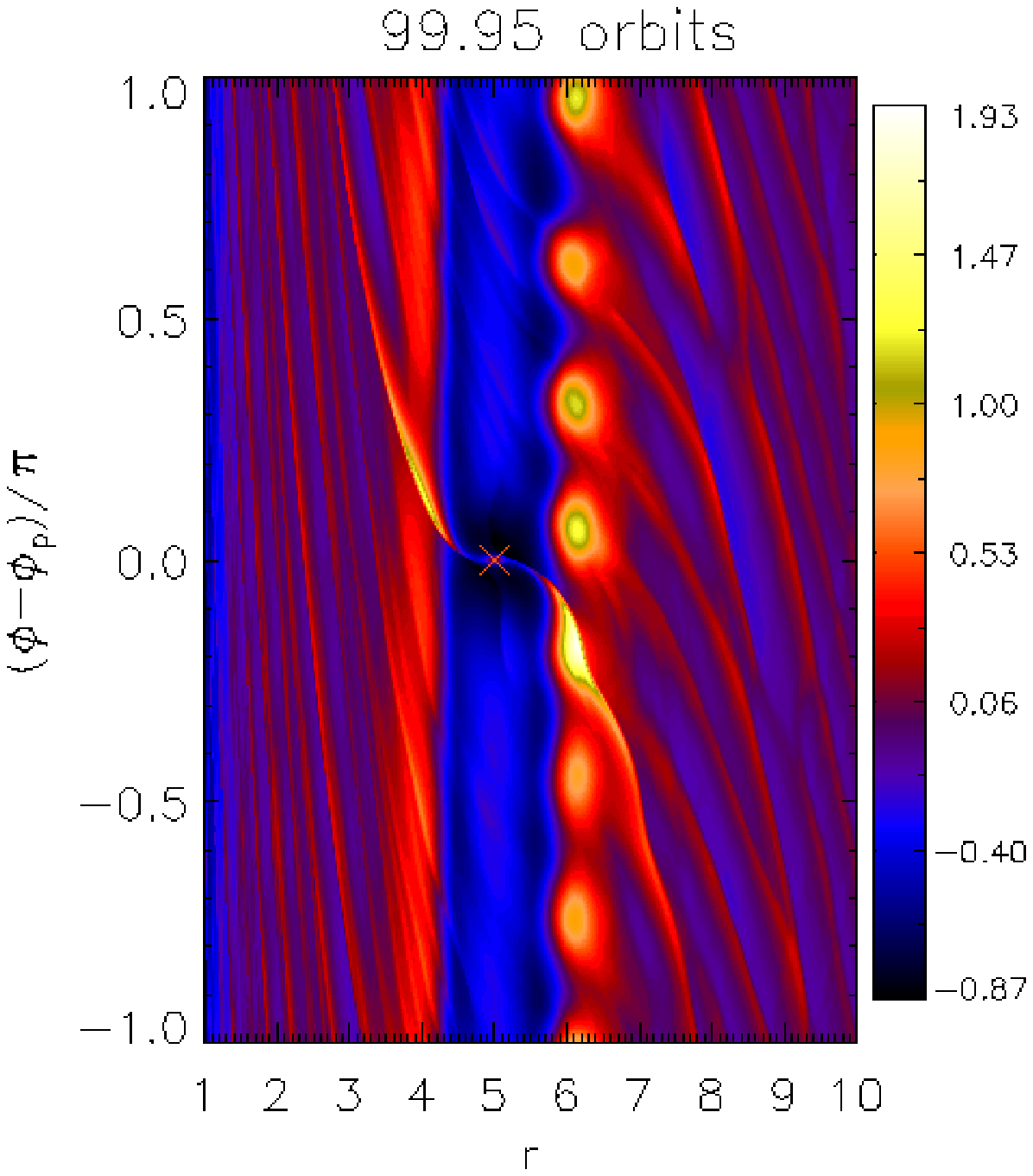}
\includegraphics[scale=.33,clip=true,trim=2.3cm 0cm 0cm 0cm]{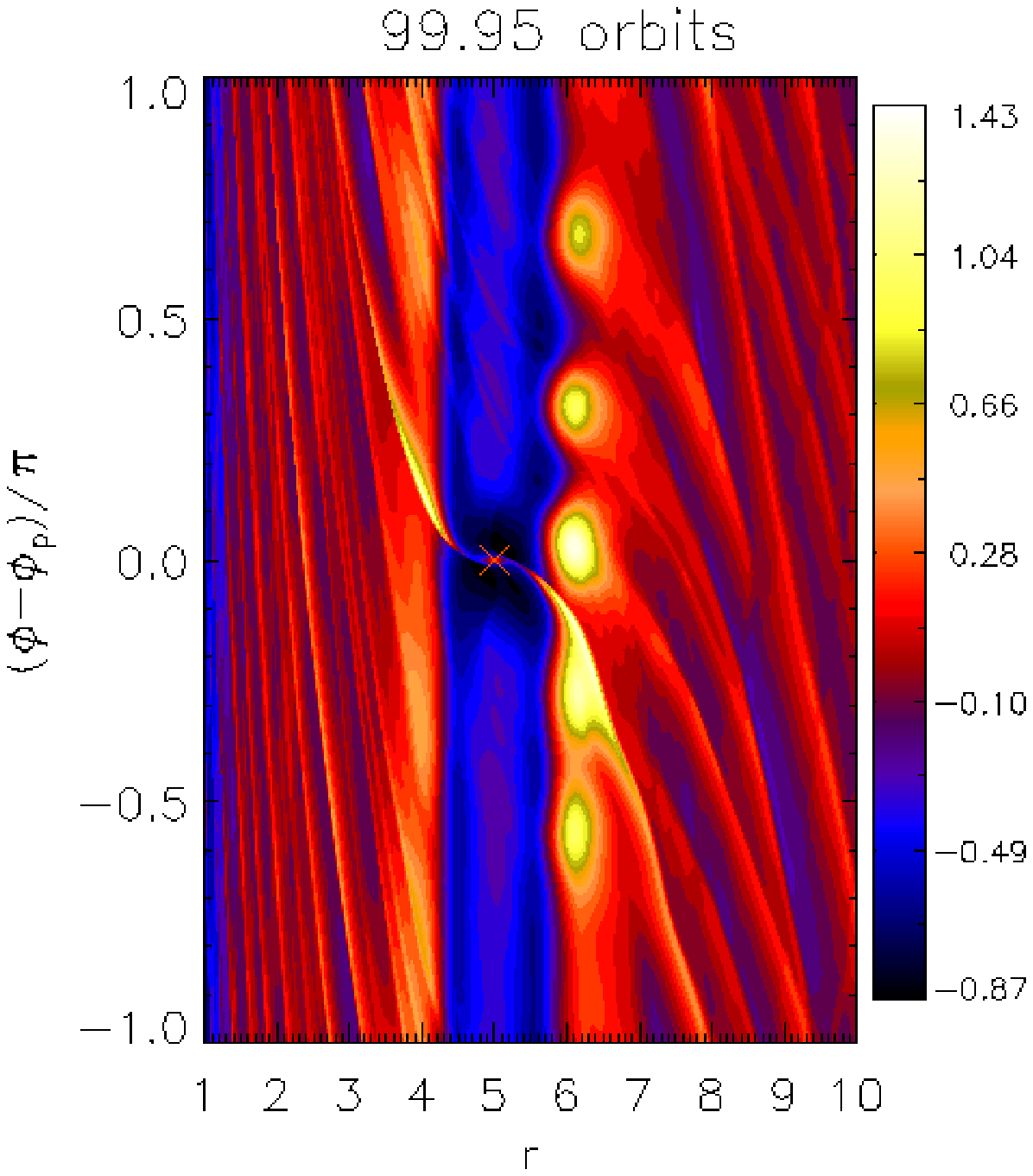}
\includegraphics[scale=.33,clip=true,trim=2.3cm 0cm 0cm 0cm]{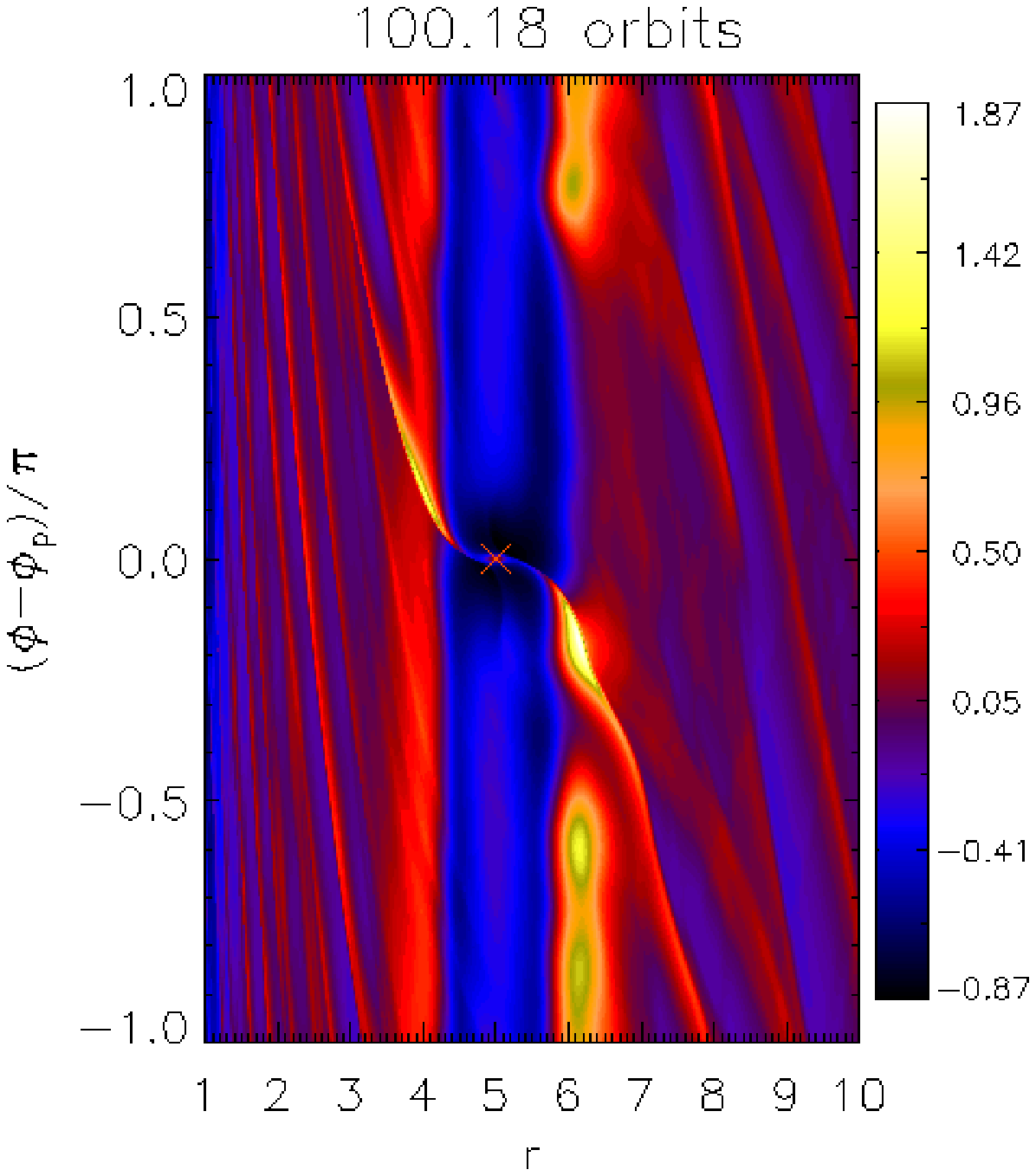}
\includegraphics[scale=.33,clip=true,trim=2.3cm 0cm 0cm 0cm]{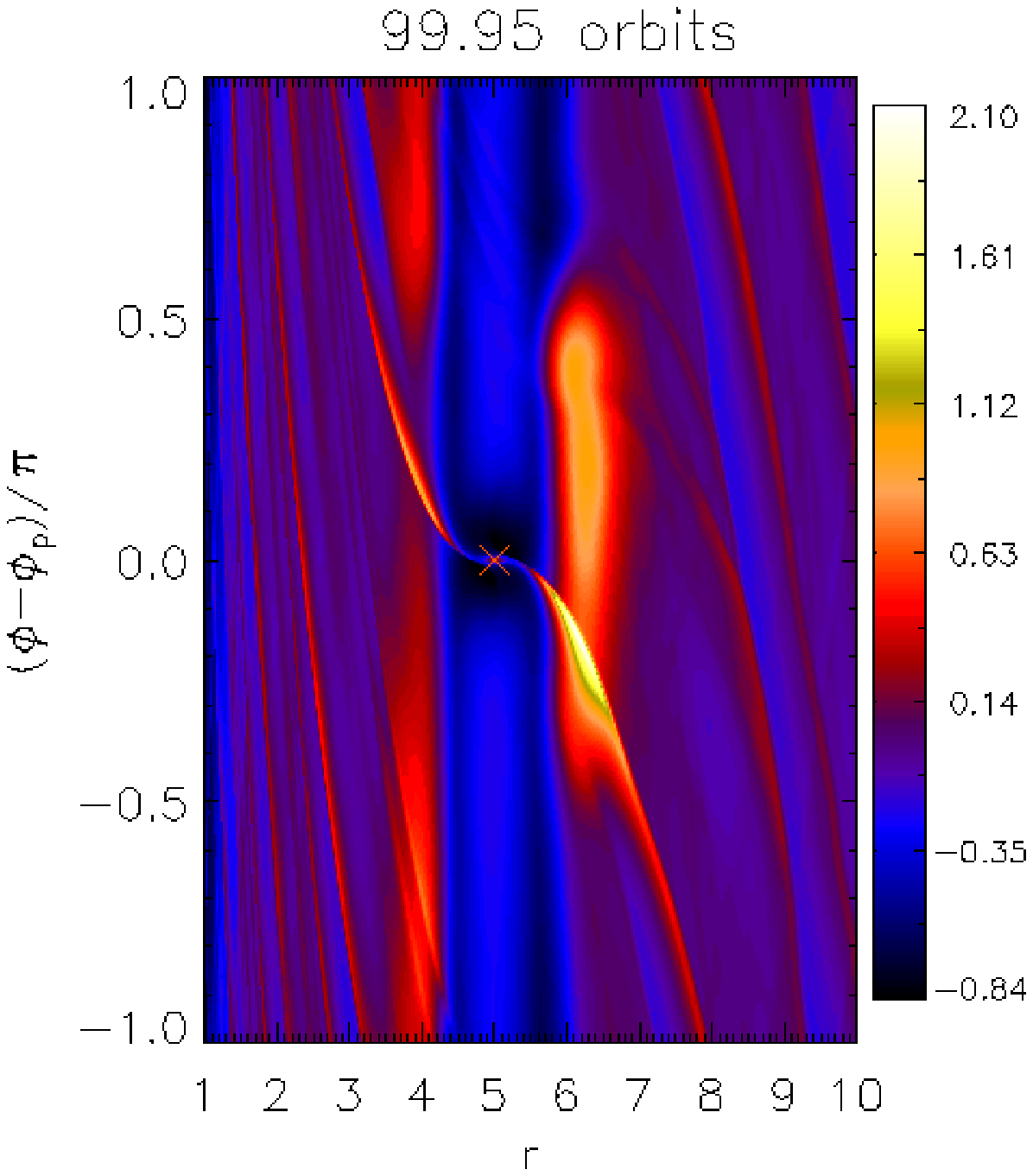}
\caption{Non-linear evolution of  vortex instabilities at the outer gap edge of a Saturn mass planet as a function of
  disc mass, parametrised by the minimum Toomre parameter $Q_o$, from left to right, of $Q_o=2.5,\,3.0,\,3.5,\,4.0,\,8.0$.
\label{compare_v_nonlin}}
\end{figure*}

We examine the evolution of vortices under the influence of
self-gravity. Fig. \ref{compare_v_nonlin} compares  surface density 
perturbations for a range of disc masses at $t=100P_0.$  Since the
vortices emerge at roughly $t=56P_0$, they have evolved for a  similar
time. Typical growth rates for vortex modes 
are  $\tau\sim 6P_0$, which implies the  vortices have evolved for
about $7\tau$, well into the non-linear regime.


The snapshot for $Q_o = 8$ shows  a single vortex, resulting from the
merging of the initial vortices. The vortex disturbance is largely
confined to within  a local scale-height of the gap edge.
 Such a result is typical for simulations with no self-gravity
\citep{valborro07}. The case with $Q_o = 4$ shows vortex merging taking place, as  
individual surface density maxima can still be identified in the
large vortex behind the shock. Notice there is a smaller vortex just
passing through the shock. When  $Q_o=4$, a single vortex  forms at
$t\simeq 110P_0$. 
Fig. \ref{compare_v_nonlin} shows  that increasing the disc mass delays
vortex merging. For $Q_o=3.5$, 5 vortices remain and for $Q_o=3$ and $Q_o=2.5$, 
7 vortices remain. They have not merged into a single vortex as happened when
$Q_o=8$. A  25\% increase in disc mass as  $Q_o=4\to3$ causes
merging to be  delayed by $50P_0$.This suggests that increased gravitational
interaction between vortices opposes merging.  

As $Q_o$ is lowered  
the inter-vortex distance increases. When  $Q_o=2.5$ their
azimuthal separation can be larger than the vortex itself. Also, vortices
become less elongated and more localised being symptomatic 
of  gravitational condensation. Trailing wakes 
from vortices also become more prominent as the vortices more strongly
perturb the disc via their  self-gravity. In fact, the planetary wake becomes less
identifiable among the vortex wakes. Notice  vortex wakes are
mostly identified with vortices upstream of the planetary wake, rather
than just  downstream. The vortex wakes appear to detach
from the vortex after passing through the planetary shock. 
There is a sharp contrast in gap structure between the $Q_o=2.5$ and $Q_o=8$
cases. The single vortex that results when 
 $Q_o=8$ is aligned along the outer gap edge, which is
still approximately identified as circle  $r=6.$
 However, when  $Q_o=2.5$ 
the vortices and  their wakes intersect the circle $r=6$ 
 making the radius of the outer gap edge less well-defined.

\begin{figure}
\centering
\includegraphics[width=0.99\linewidth]{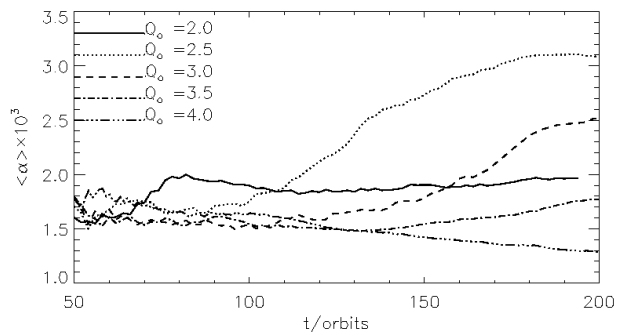}
\caption{The running time average of the 
 viscosity parameter $\alpha$ averaged over  the vortex region
 for a range of disc models.  
\label{alpha_visc}}
\end{figure}

An important feature of vortices is that the wakes
are associated with the excitation of density waves
which transport  angular momentum transport outwards
\citep[see e.g.][]{paardekooper10}. 
A measure of this is is made through the standard $\alpha$ viscosity
parameter. We define it
here through

\begin{align}
\alpha(r,\varphi) \equiv \Delta u_r\Delta u_\varphi/c_s^2(r)
\end{align}
where $\Delta$ denote deviation from azimuthal averaged values. 
As we are interested in the vortex region, $\alpha$ is spatially
averaged over the annulus $r\in [5.7,7.1]$ and its running-time average
$\langle \alpha \rangle$ is plotted in Fig. \ref{alpha_visc}.
The parameter $\langle \alpha\rangle $   associated with vortices
is $O(10^{-3})$, an order of magnitude larger than the imposed value 
associated with $\nu.$
When $Q_o=4$, $\langle\alpha\rangle$ decays steadily
 after an initial transient growth. The case
$Q_o=2$ is also  shown. In that case  $\langle\alpha\rangle \simeq 1.9\times10^{-3}$
 for $t\ga 80P_0.$ Interestingly for $Q_o=2.5$---3.5 there is growth 
in $\langle\alpha\rangle$ over several tens of orbits.

The behaviour of $\langle\alpha\rangle$ as a function of $Q_o$ is consistent with the
general picture of
vortex formation and evolution (Fig. \ref{compare_v_nonlin}).  
We  see below  that the decay in $\langle\alpha\rangle$ is associated with             
vortex merging leading to fewer vortices. This is the case for $Q_o=4$.
 Vortex merging happens more readily  
for lower disc masses, hence although multiple vortices  develop
 from the instability, this phase 
does not last long enough for the multiple-vortex configuration to
 significantly transport angular momentum. 
As we increase self-gravity when  $Q_o=3.5\to2.5$,
 vortices become less prone to merging. 
Hence, the multiple-vortex phase lasts longer.
 They have time to evolve into compact objects
that further perturb the disc.
 This is consistent with fact that $\langle\alpha\rangle$-growth is prolonged 
with increased disc mass, as merging is delayed.
 However, if self-gravity is too strong, such as when  $Q_o=2.0$, growth is
again limited, because the $m=2$ edge mode develops and hinders vortex evolution.

\subsection{ Long term evolution of gap edge vortices}
\begin{figure*}
\centering
\includegraphics[scale=.34]{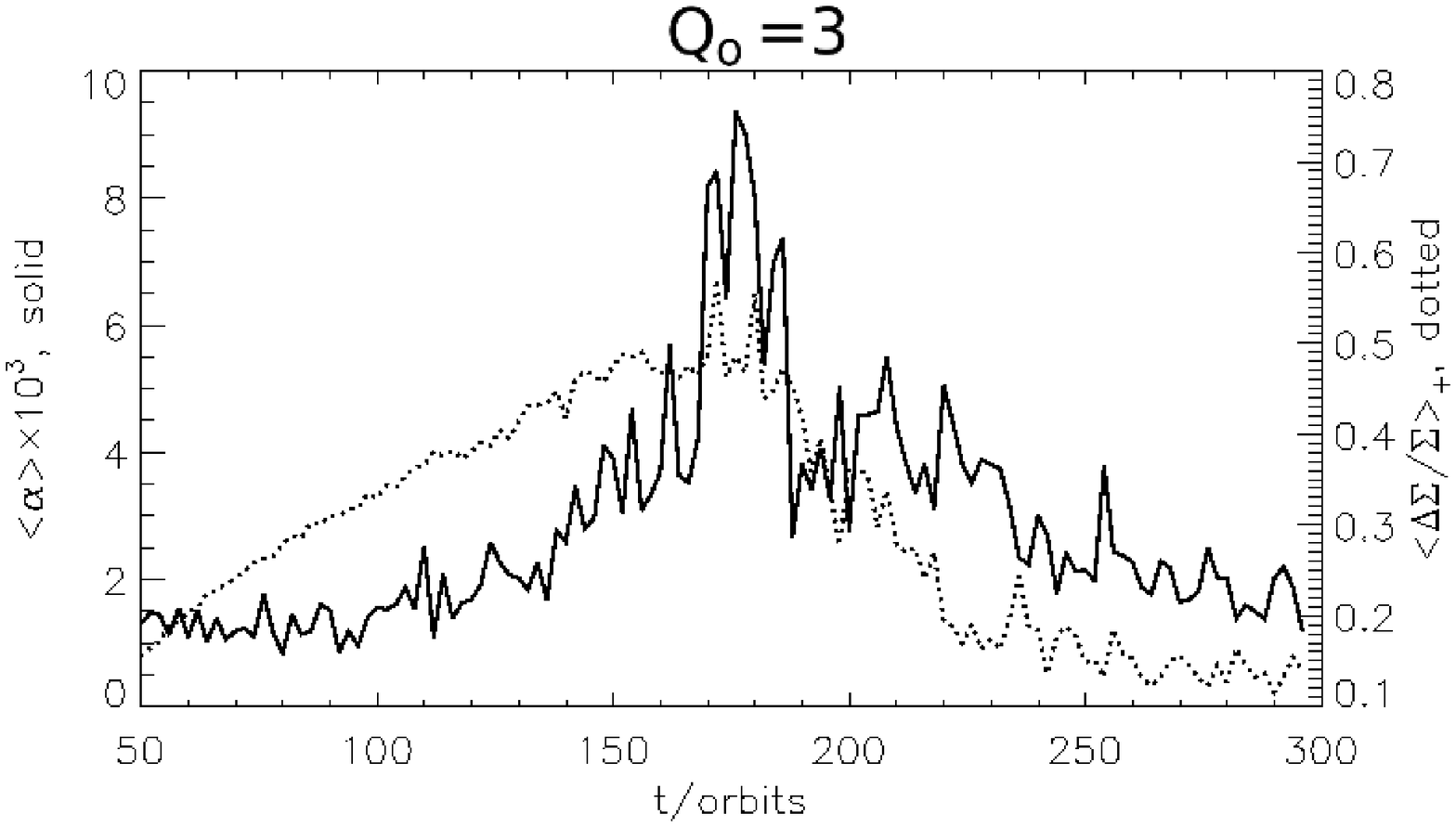}
\includegraphics[scale=0.32]{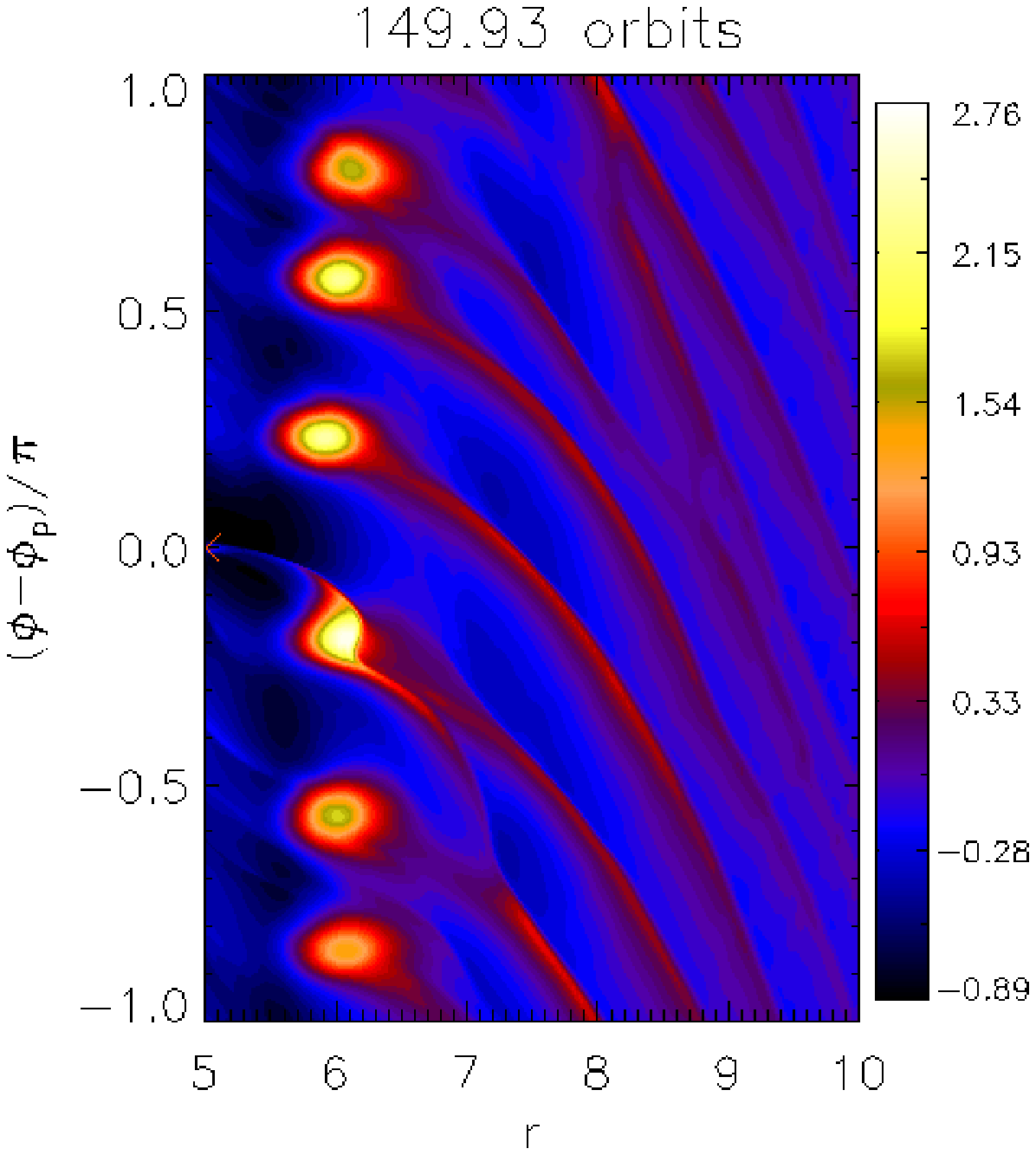}
\includegraphics[scale=0.32,clip=true,trim=2.3cm 0cm 0cm 0cm]{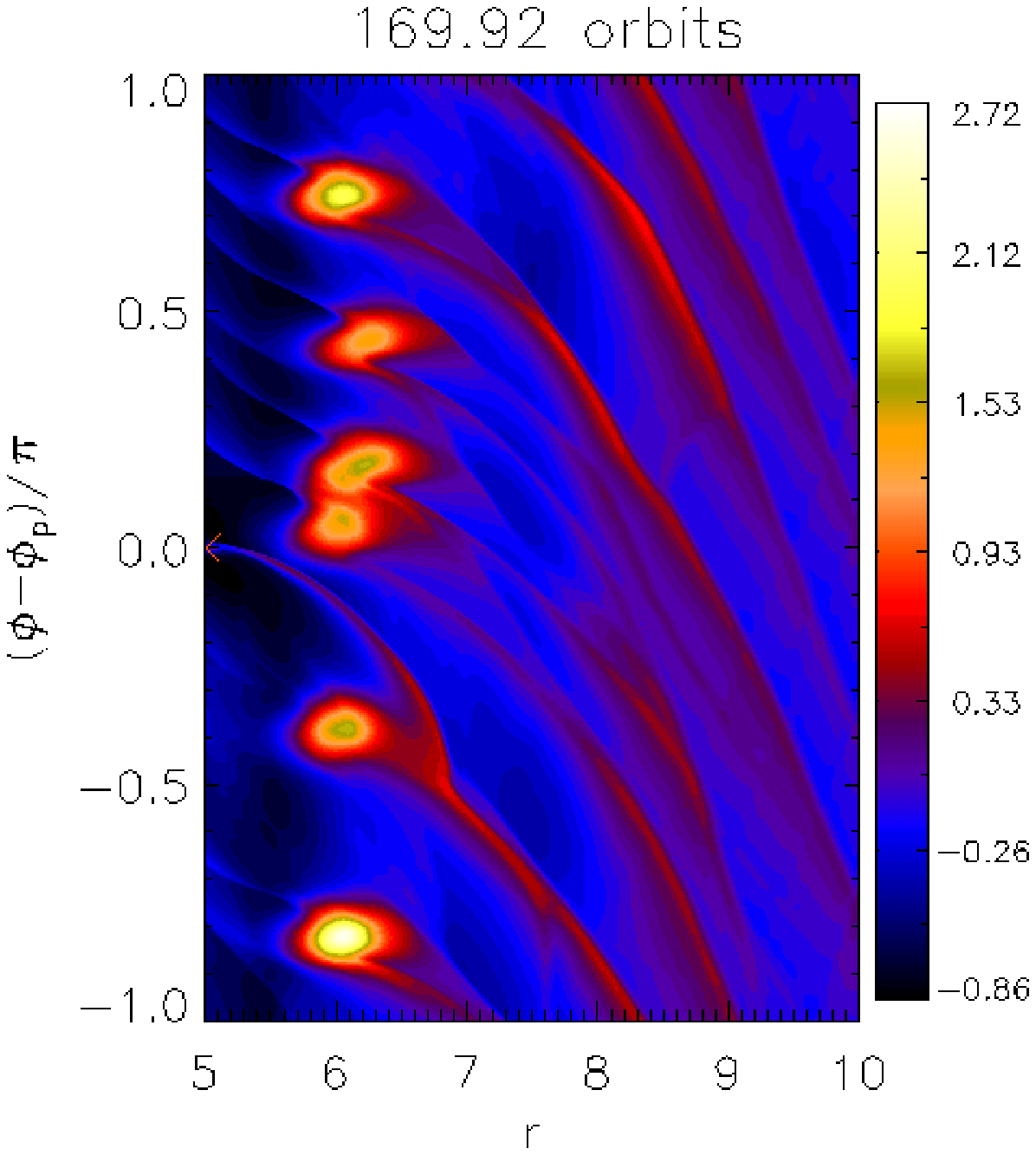}
\includegraphics[scale=0.32,clip=true,trim=2.3cm 0cm 0cm 0cm]{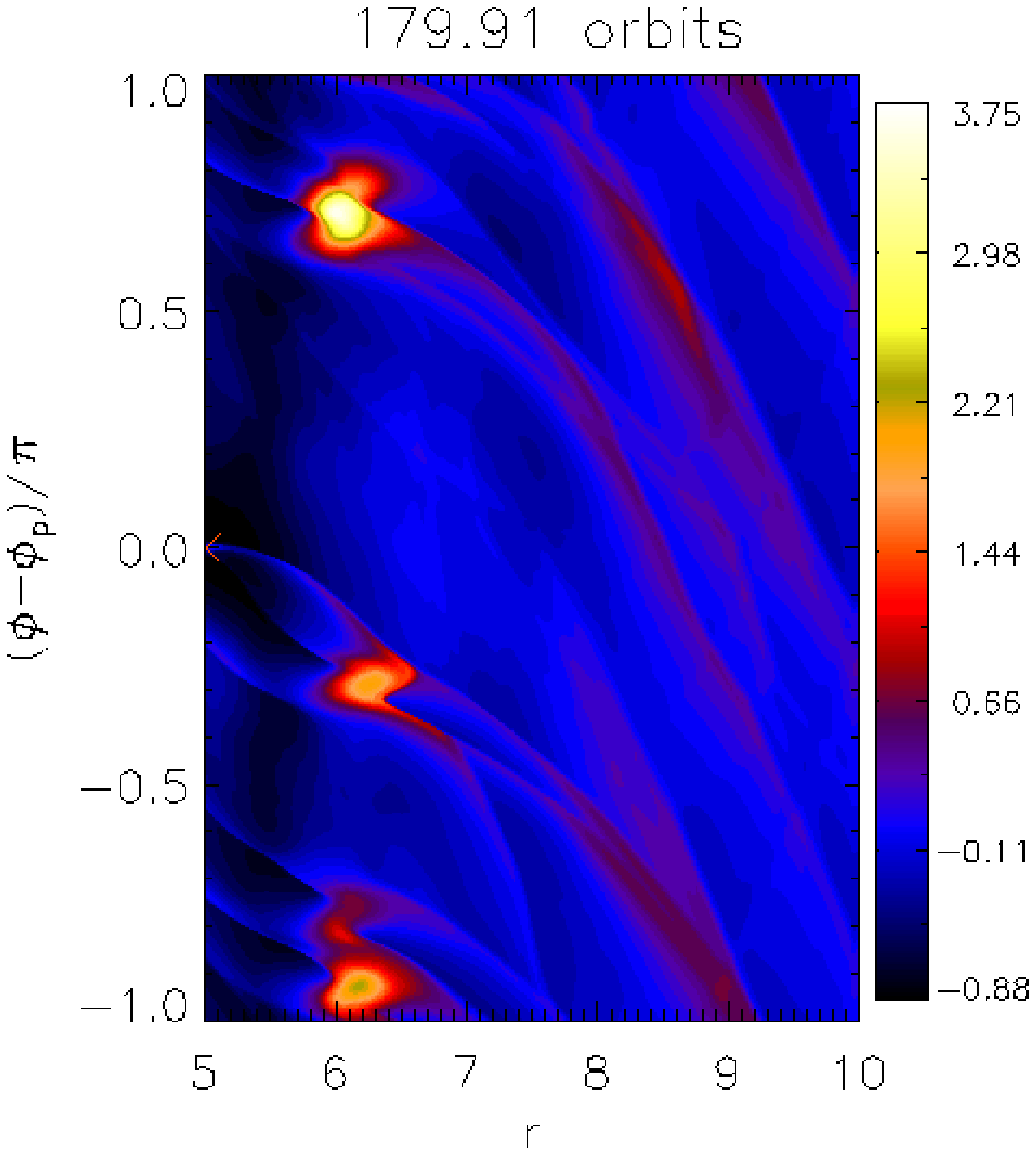}
\caption{Vortex evolution  for the disc with $Q_o=3$. Line plot: instantaneous
  $\alpha$ viscosities (solid) and average over-density
  (dotted). Contour plots:  relative surface density
  perturbations before and after vortex merging.
\label{Qm3_alpha}}
\end{figure*}

We consider the long term evolution of vortices in the disc with $Q_o=3.$  
Fig. \ref{Qm3_alpha} shows the  instantaneous $\alpha$ viscosity parameter, 
average over-density (defined in  regions where the relative
surface density perturbation is positive) and surface density
contour plots of the vortex region.
The viscosity parameter 
 $\alpha$ grows from $t=100P_0$ to $t=170P_0$ with $\mathrm{max}(\alpha) \simeq
8\times 10^{-3}$, which is 50 times larger than the contribution from the background
viscosity. At $t=150P_0$ there remains 6 distinct vortices, one of which just 
passing through the planetary wake. The multiple-vortex
configuration has been maintained for a further$\sim 50P_0$
since Fig. \ref{compare_v_nonlin}. 
With weak self-gravity, a single vortex would have formed through
merging. Delayed merging allows individual vortices to evolve and
become planet-like. The typical
over-density in a vortex is $\sim 1$ at $t=100P_0$ and increases to
$>1.5$ by $150P_0.$  As a consequence we expect increased
 disturbance in the surrounding disc. 
The phase of $\alpha$-growth correlates with a linear increase in the average
over-density in the vortex region. 

\begin{figure}
\centering
\includegraphics[width=0.49\linewidth]{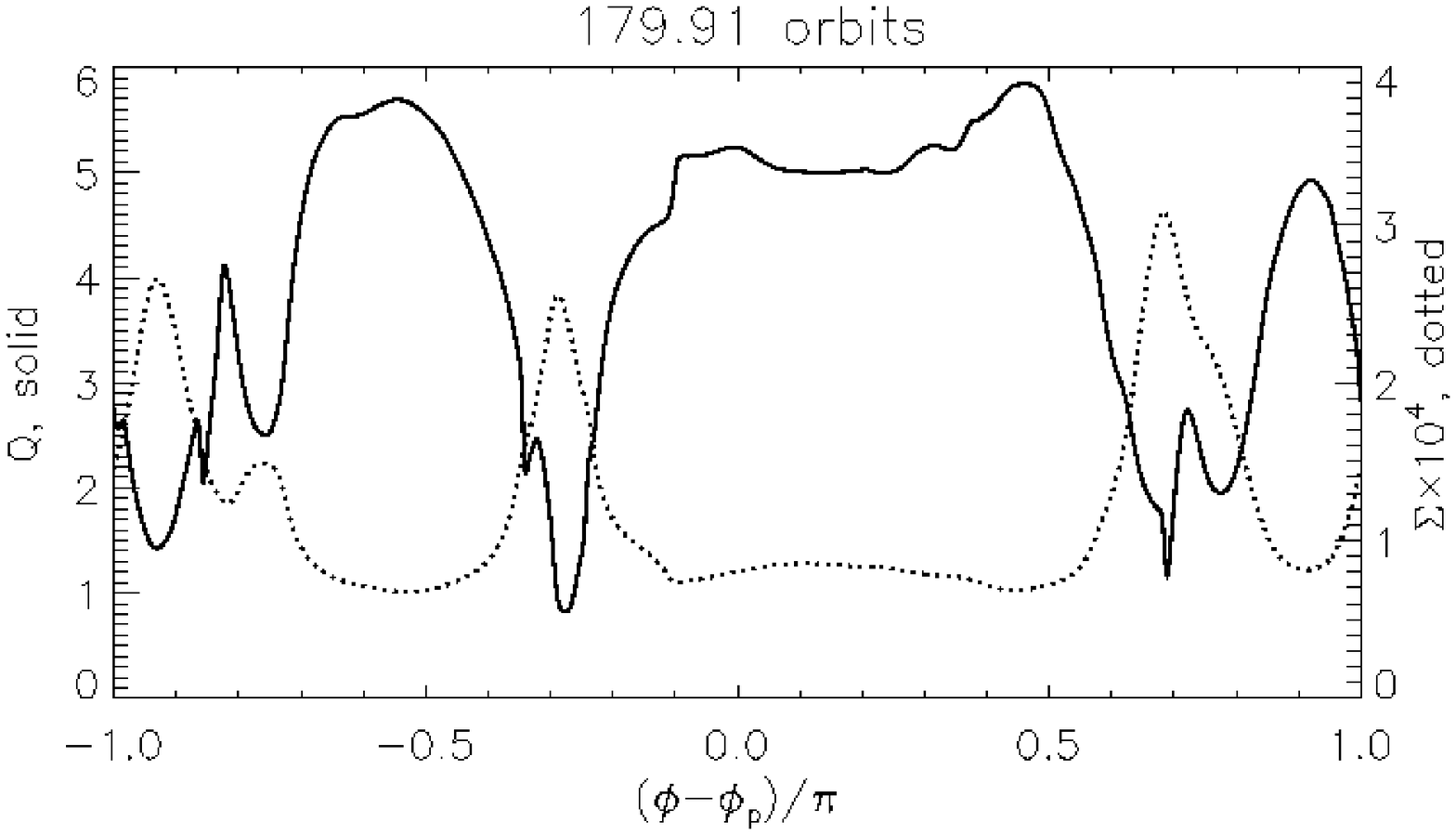}
\includegraphics[width=0.49\linewidth]{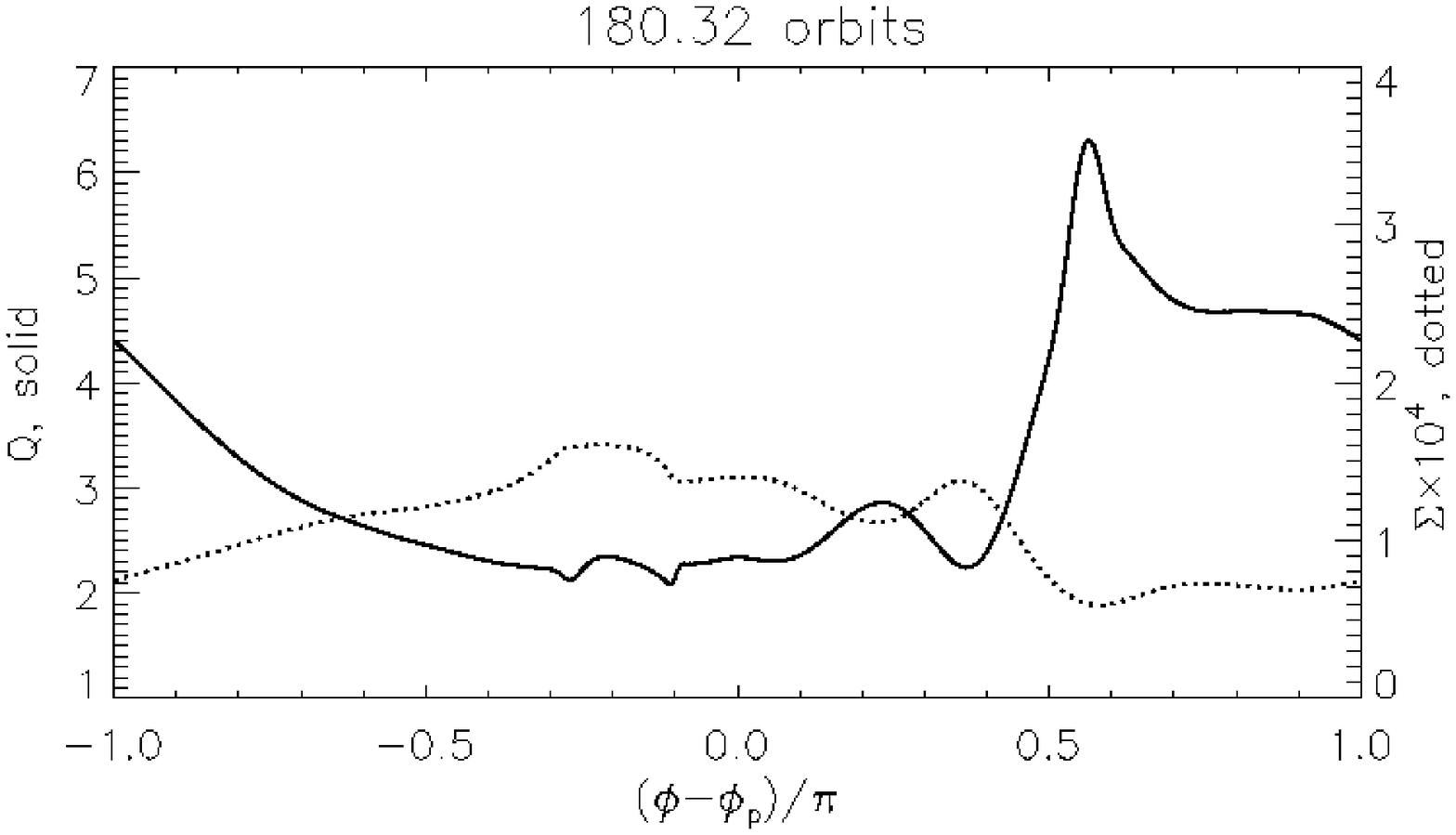}
\caption{Comparison of post-merging vortices for discs with  $Q_o=3$ (left) and
  $Q_o=4$ (right). The Toomre $Q$ (solid) and surface density (dotted)
  averaged over $r\in[6,6.5]$ is shown as a function of azimuth relative
  to the planet. 
\label{compare_merged_vortex}}
\end{figure}

At $t=170P_0$, Fig. \ref{Qm3_alpha} shows 6 vortices still remain, but the
pair just upstream of the planet is merging. The parameter $\alpha$ is a  maximum.
However, after a  burst of vortex merging events, 
$\alpha$ decreases rapidly. The snapshot at $t=180P_0$ shows a quieter
disc with only 3 vortices of similar size to those before
merging. This is unlike cases with weak self-gravity where a larger vortex results
from merging. Fig. \ref{compare_merged_vortex} compares
post-merging vortices in discs with  $Q_o=3$ and $Q_o=4.$
For $Q_o=3$, the post-merging vortices are localised in azimuth,
with  $Q\sim 1$ and the contour plot shows the densest vortex has
an over-density of $\simeq 3.75$. However,  for  the disc with  $Q_o=4$,
 a single vortex forms
that extends about half the total azimuth and has $Q>2.$
.

The mass of the disc with  $Q_o=3$ is $M_d=0.031M_*$, usually considered 
insufficient for gravitational collapse by itself. However, the
comparison above shows that addition of self-gravity to the
vortex instability induced by an embedded planet,
 may enable collapse into compact objects that
survive against shear.

\begin{figure}
   \centering
   \includegraphics[width=0.99\linewidth]{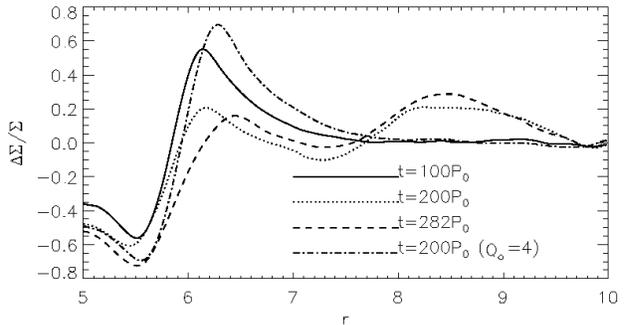}
 \caption{The  
gap profile for the $Q_o=3$ disc at different times 
 (solid, dotted and dashed) as shown through
	relative surface density perturbation. A snapshot of the
profile for the  disc 
with  $Q_o=4$ (dash-dot) is also 
	given for comparison. \label{v3_gap_evol}}
 \end{figure}

We found that vortices in the disc with $Q_o=3$ noticeably affect the gap structure. 
Fig. \ref{v3_gap_evol} shows several snapshots
of the gap structure from $t=100P_0,$ when there were
 multiple vortices, to the end of the simulation 
when a vortex pair remained.
 The gap profile for $Q_o=4$ is also shown for comparison. Neither  the single
vortex for the disc with  $Q_o=4$ nor multiple vortices with $Q_o=3$
affect the one-dimensional
gap profile  at  $t=100P_0$  because the disturbances only redistribute mass in the azimuthal direction.
 However,  the original bump at the outer
 edge is diminished after vortex  merging takes place in the disc with  $Q_o=3$ at $t=200P_0.$   
A surface density depression of  $\Delta\Sigma/\Sigma \simeq -0.1$ then
 develops at $r=7.3$  
and a bump develops at $r=8.5.$  These features last to the end of 
the simulation. 
 Self-gravitating vortices behave like planets
and gap opening is to be expected.
 Assuming vortices lie near the surface density maximum at
$r=6,$ the  creation of the surface density deficit 
for  $r\in[7,7.5]$ and the new surface density maximum  at $r=8.5$
could  be induced,  in a similar manner as  it would be by  a planet,
  through  the outward transport of angular momentum
by the density waves launched by the vortices. 
 The  material removed from the region of the deficit, which has gained angular momentum, ends up  contributing
 to the new surface density maximum at $r=8.5.$ 

\subsection{Anticyclonic vortices}

\begin{figure}
  \centering
  \subfigure[Gap vortices, $\Delta
  \Sigma/\Sigma$\label{v3_vortex}]{\includegraphics[width=0.32\linewidth]{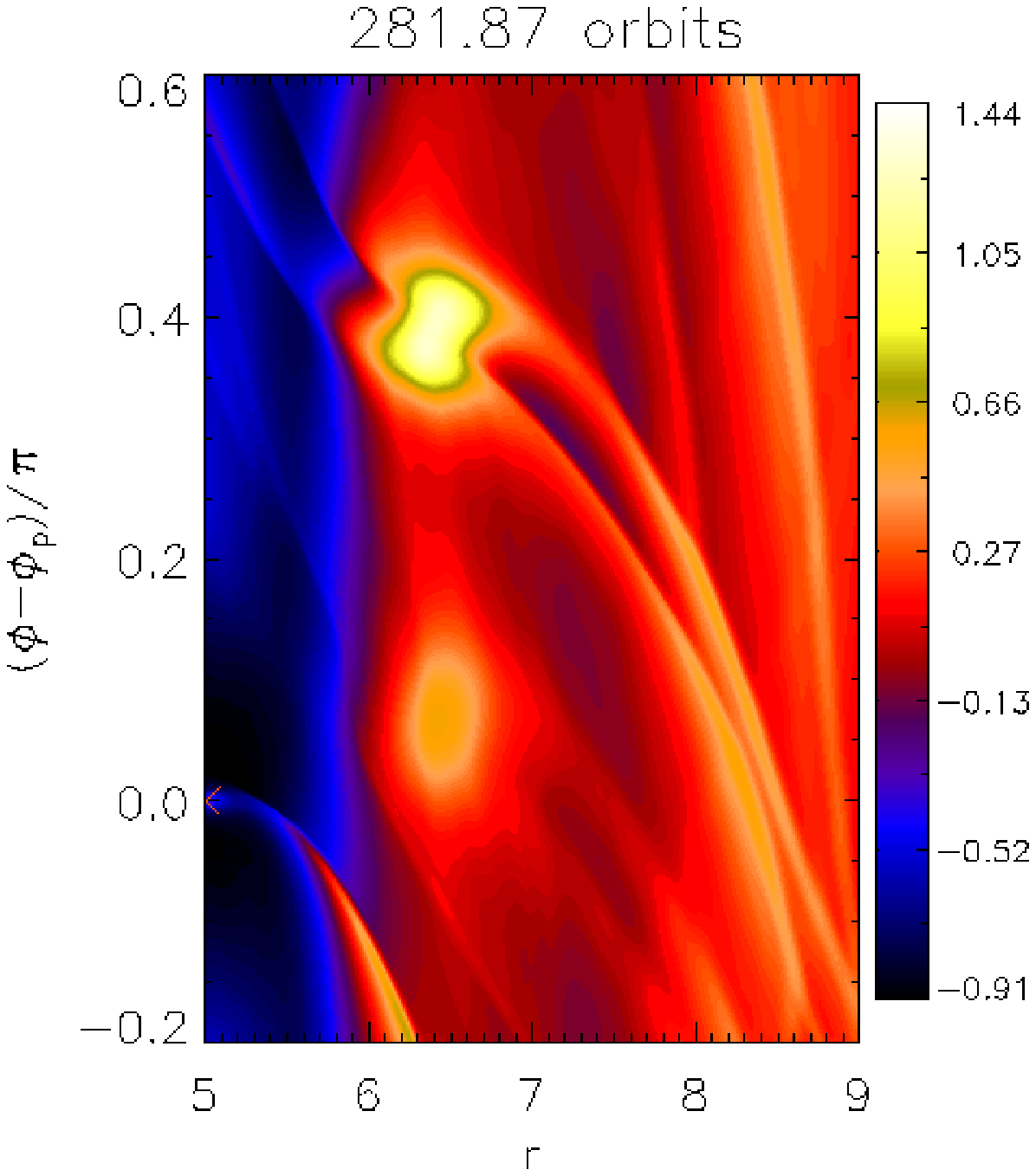}} 
  \subfigure[Gap vortex, $\eta$
  \label{v3_vorten}]{\includegraphics[width=0.32\linewidth]{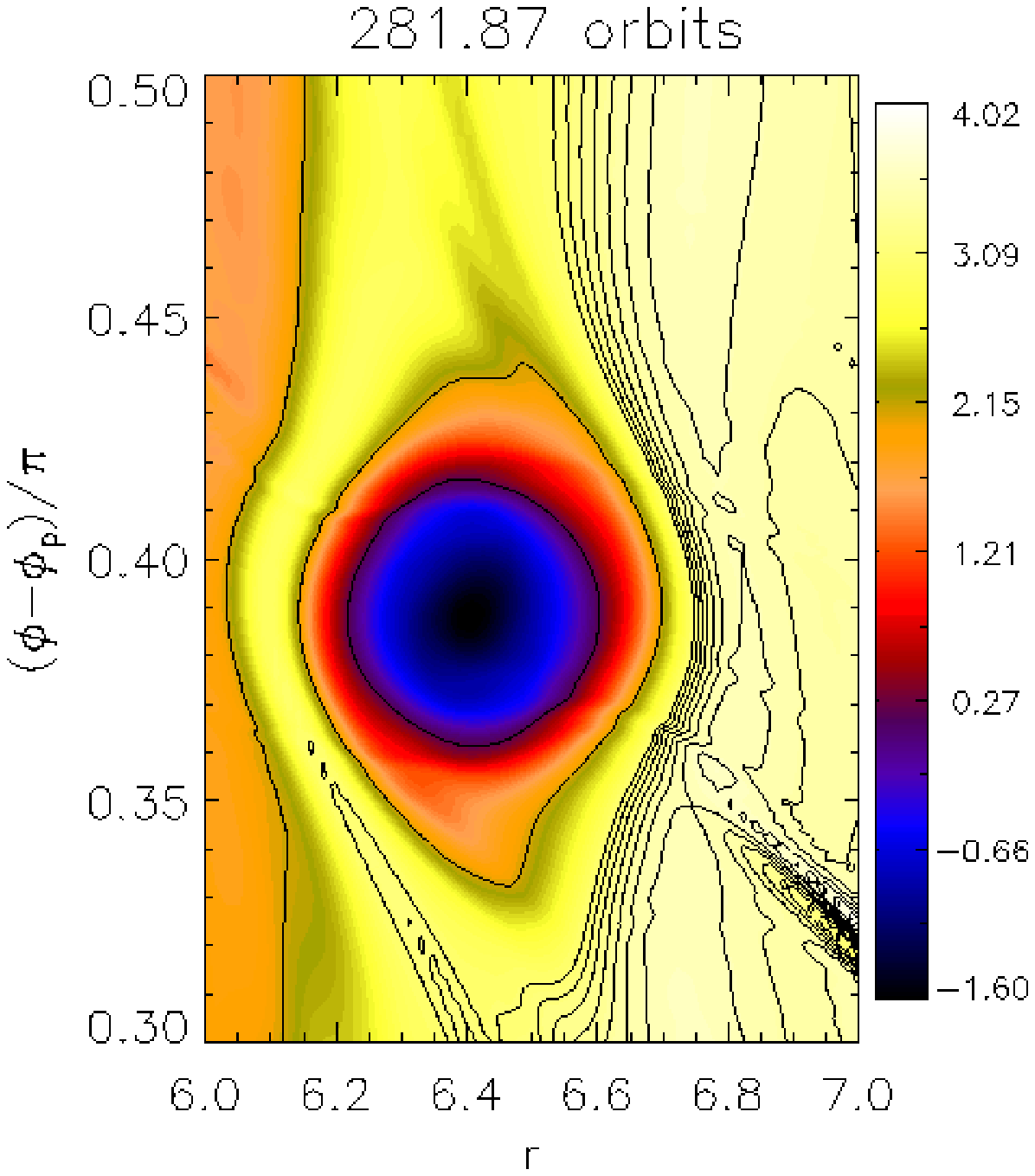}} 
  \subfigure[Kida-like vortices, $\Delta \Sigma/\Sigma$
  \label{kida_vortex}]{\includegraphics[width=0.32\linewidth]{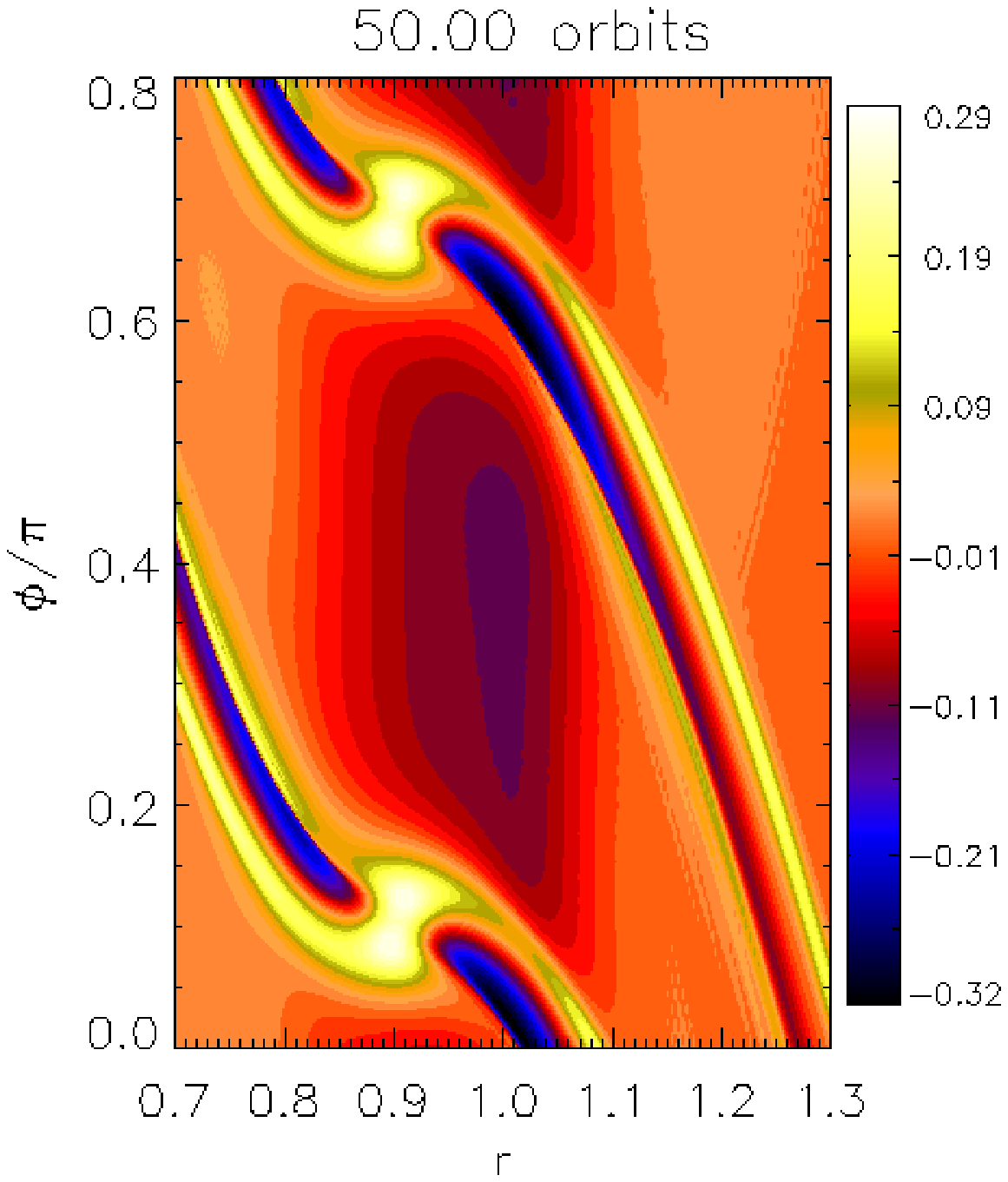}} 
\caption{The final vortex pair at the end of the  disc-planet
  simulation for $Q_o=3:$ (a) Relative surface density contour plot (b)Vortensity contours for the upper vortex. 
 Shown in (c) is the relative surface density for  a pair of  vortices formed by imposing
Kida vortex solutions as a perturbation in 
  a standard power-law disc. \label{Qm3_final_vortex}
}
\end{figure}

For the disc with $Q_o=3$, a vortex-pair forms at $t\simeq 230P_0$ and
lasts until the end of the simulation ($t\simeq 300P_0$) and is
reminiscent of co-orbital planets. The pair survives on a time-scale
beyond which merging occurs in the  simulations without self-gravity.
A snapshot is shown in Fig. \ref{v3_vortex}. Two blobs can be
identified along the gap edge, the upper vortex being more over-dense
than the lower one. They lie at a radius $r_v = 6.4$, corresponding to
local surface density maximum  in azimuthally
averaged  one-dimensional profiles, which is expected to be
neutral  for vortex migration \citep{paardekooper10}.

The upper vortex in Fig. \ref{v3_vortex} is different to  the pre-merging
vortices or those with weak self-gravity. It
has two spiral disturbances extending from the vortex to $(r,\phi -
\varphi_p) = (9,-0.2\pi)$, whereas the pre-merging vortices have one trailing
spiral.  Its vortensity field is shown in Fig. \ref{v3_vorten}.  The
vortex core has $\eta < 0$, whereas the final large vortex in the disc
with  $Q_o=4$
has $\eta >0$ in its core, though it is still a local minimum. 

From the vortensity field the region where $\eta < 0$ has a mass of 
$M_v\simeq 5.46\times10^{-5}M_*\simeq 0.18M_p$ and average radius
$\bar{r} \simeq 0.92H(r_v)$.
This is about 18 Earth masses if
$M_*=1M_s$. Denoting the mean square relative velocity (to
the vortex centre) as $\Delta v^2$, we found $GM_v/\bar{r}\Delta v^2
\sim 3.9$. This region is gravitationally bound. Although planets of
such mass are not expected to open gaps, the surface density
deficit in $r=[7,7.5]$ in Fig. \ref{v3_vortex}  indicates vortices may do
so. This  may be because  the process is assisted by the fact
that a vortex of size $H$ produces a  perturbation of a magnitude
similar what would be produced by   Saturn when  $h=0.05,$  even without self-gravity
\citep{paardekooper10}. We estimate by inspection that the region
with negative vorticity has semi-major and semi-minor axis of 
$a\sim 1.57H(r_v),\,b\simeq 0.55H(r_v)$, respectively, corresponding to
aspect-ratio of 2.9.   

Finally, Fig. \ref{kida_vortex} show $\Delta\Sigma/\Sigma$ for a pair of Kida-like vortices
placed in a disc. We note the similarity between the Kida-like vortex and the upper vortex in
Fig. \ref{v3_vortex}, particularly the double wake structure, the
tilted core and the surface density deficit just outside the
vortices. This is a surprising coincidence,
given that the simulation setups that produced them  were completely different (for more details see below). 

\section{Simulations of co-orbital Kida vortices}\label{hydro2}

In the  simulations above, we observed one of the effects of  self-gravity
is to  delay vortex
merging. This effect is significant when vortices develop into compact
structures. Furthermore, the similarity between
the final vortex in the disc with $Q_o=3$  and a Kida-like
vortex \citep{kida81} motivates us to consider the effect of self-gravity on the
interaction between two Kida vortices.

The  simulations presented here are
supplementary and focus on the interaction
between vortices without the perturbation of the planet.
Thus we isolate effects due to vortex-vortex interactions
and the influence of self-gravity on those without interference
from the planet.
In this way we obtain a better  understanding of some of the
results above.

\subsection{Numerical setup}
We use a standard power law disc with initial surface density profile
$\Sigma=\Sigma_0 r^{-1/2},$ where $\Sigma_0$ is a scaling constant  for
$r\in [0.25,2.5]$. The value of $\Sigma_0$ is chosen to provide 
a specified  $Q_\mathrm{Kep}$ at a specified radius as
before. The initial 
azimuthal velocity is chosen so that the disc is
in hydrostatic equilibrium. The initial radial velocity is zero. The disc is
locally isothermal with $h=0.05$. The viscosity is
$\nu=10^{-9}$ which is essentially the inviscid limit. No planets are
introduced, but in this section we use $(r_p,\varphi_p)$ to denote a vortex's centroid. 

To set up vortices in the disc,
we follow \cite{lesur09b} and introduce  velocity perturbations  
$(\dd u_r,\dd u_\varphi)$ that correspond to the  elliptical vortices of
\cite{kida81} in an incompressible shear flow. Details of the implementation
is described in Appendix \ref{kidasetup}.
Two vortex perturbations are imposed with initial angular
separation $\theta = \pi/2$. Table \ref{experiments} summarises our
numerical experiments. We consider switching self-gravity on or off, a
range of disc masses (here   specified   through $Q_1\equiv Q_\mathrm{Kep}(1)$) 
and initial radial separations  $X_0$ of the two
vortices (so that for one of the vortices, $r_p\to r_p + X_0$).    

We use the FARGO code with resolution $N_r\times N_\varphi = 800\times
2400$ giving $\simeq 17$ grid points per scale height.  The vortices
are typically of that size. Damping boundary conditions are applied
\citep{valborro06}. 

\begin{table}
  \centering
    \caption{Parameters for two-vortex interaction simulations.
     The first column gives the nomenclature for the simulations. 
     The  initial radial separation is $X_0$ and  $Q_1$ is the
    Keplerian Toomre parameter $Q_\mathrm{Kep}$ at unit radius.
     The fourth column indicates whether self-gravity was included.} 
  \begin{tabular}{lcccl}
    \hline
    Case & $X_0/H $ & $Q_1$ & self-gravity \\ 
    \hline\hline
    Fnsg  & 0  &  8.0 &  NO \\
    Fsg   & 0  &  8.0 &  YES \\
    M1  &  0 & 15.9 & YES \\
    M2  &  0 & 5.3 & YES \\
    S1  &  0.1 & 8.0 & YES \\
    S2  &  0.2 & 8.0 & YES \\
    S3  &  0.3 & 8.0 & YES \\
    S4  &  0.6 & 8.0 & YES \\
    \hline
  \end{tabular}
  \label{experiments}
\end{table}

\subsubsection{Vortex pair interactions}
\begin{figure}
  \centering
  \includegraphics[width=0.35\textwidth]{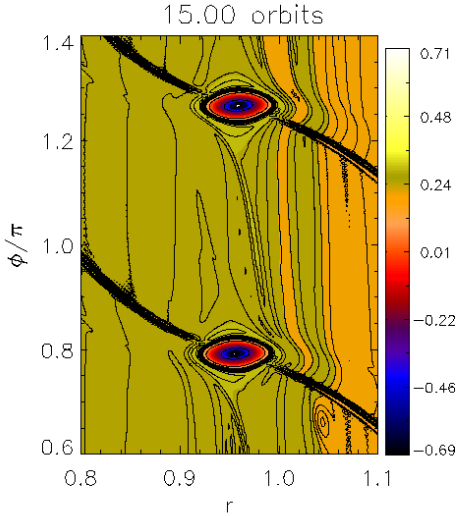}
  \caption{Vortices formed by imposing Kida vortex solutions as a 
    perturbation to a global disc. The vortensity field (scaled) is
    shown. 
    \label{nsg_vortices}}
\end{figure}

As an example of vortices formed by the above procedure,
Fig. \ref{nsg_vortices} shows  the vortensity field for the case Fnsg. 
The vortex pair circulates at $r\simeq 0.96$ and are localised in radius and azimuth.  
The vortex centroids have vorticity $-1.57$ (dimensionless units) in the non-rotating frame,
 close to the local Keplerian shear ($-1.5r_p^{-3/2}\simeq -1.59$). 
 The vortices are not
symmetric with respect to reflection in a co-orbital radius.
 The upper vortex has a long tail reaching
the outer part of the lower vortex rather than its centroid.
The wakes associated with each vortex are
similar to those induced by a planet and responsible for vortex
migration  \citep{paardekooper10}.  

Focusing on one vortex, the region with negative absolute vorticity has half
width in radius and azimuth of $\simeq 0.6H$ and $1.3H$
respectively.
 This region has mass $m_v \sim
1.2\times 10^{-5}M_*$, or about 4 Earth masses if $M_*=1M_s$.    
The estimate is not far from the assumption that the final vortex has
size $H$ with average density that of the unperturbed
disc at the location it was set up, which gives 
$m_v\sim 1.6\times10^{-5}$. If self-gravity is enabled, we can expect
gravitational interaction between vortices to behave like that between  two
co-orbital planets of at least a few Earth masses. 

\subsection{Simulation results}

\begin{figure*}
  \centering
\includegraphics[width=0.33\linewidth]{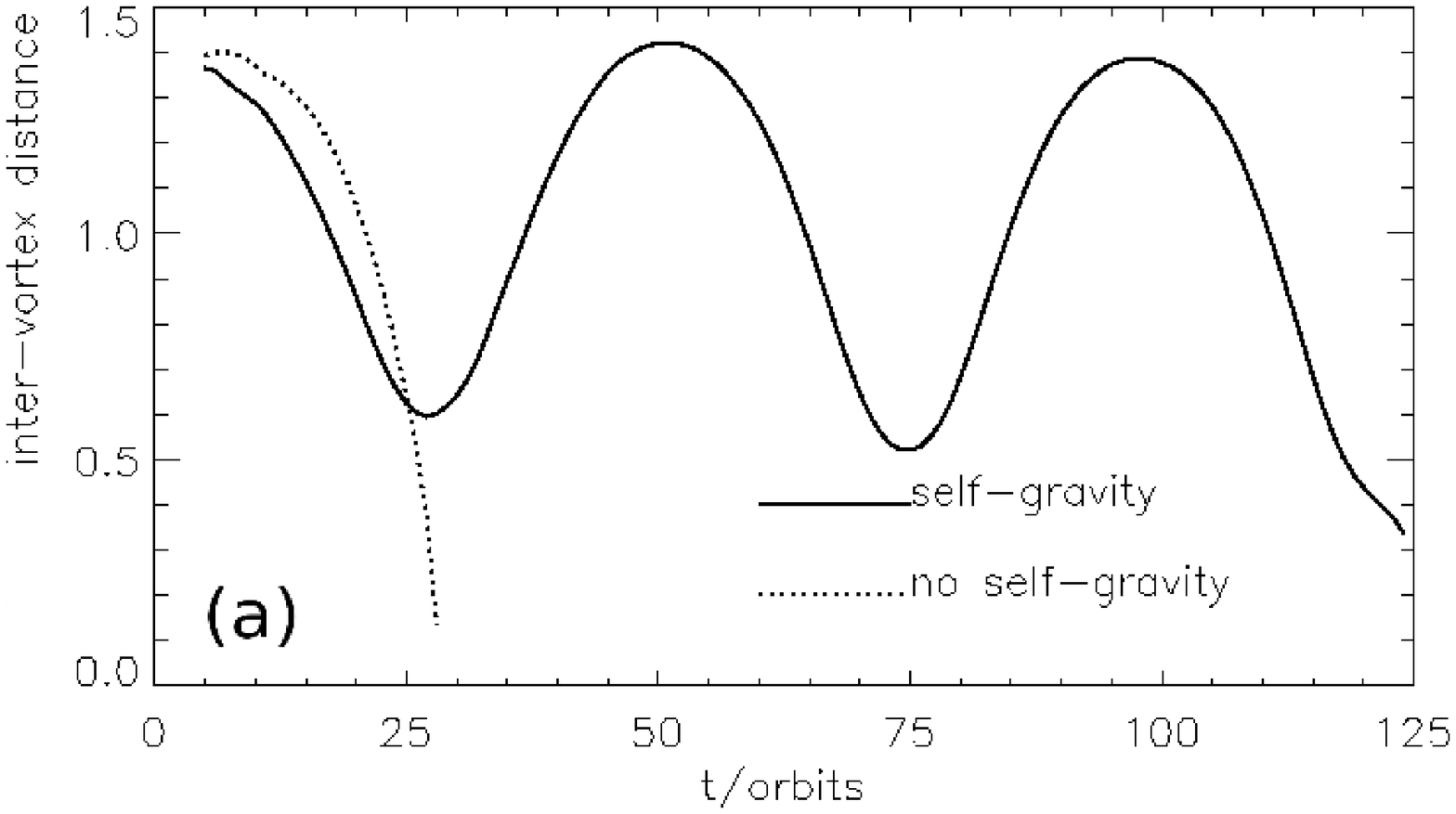}
\includegraphics[width=0.33\linewidth]{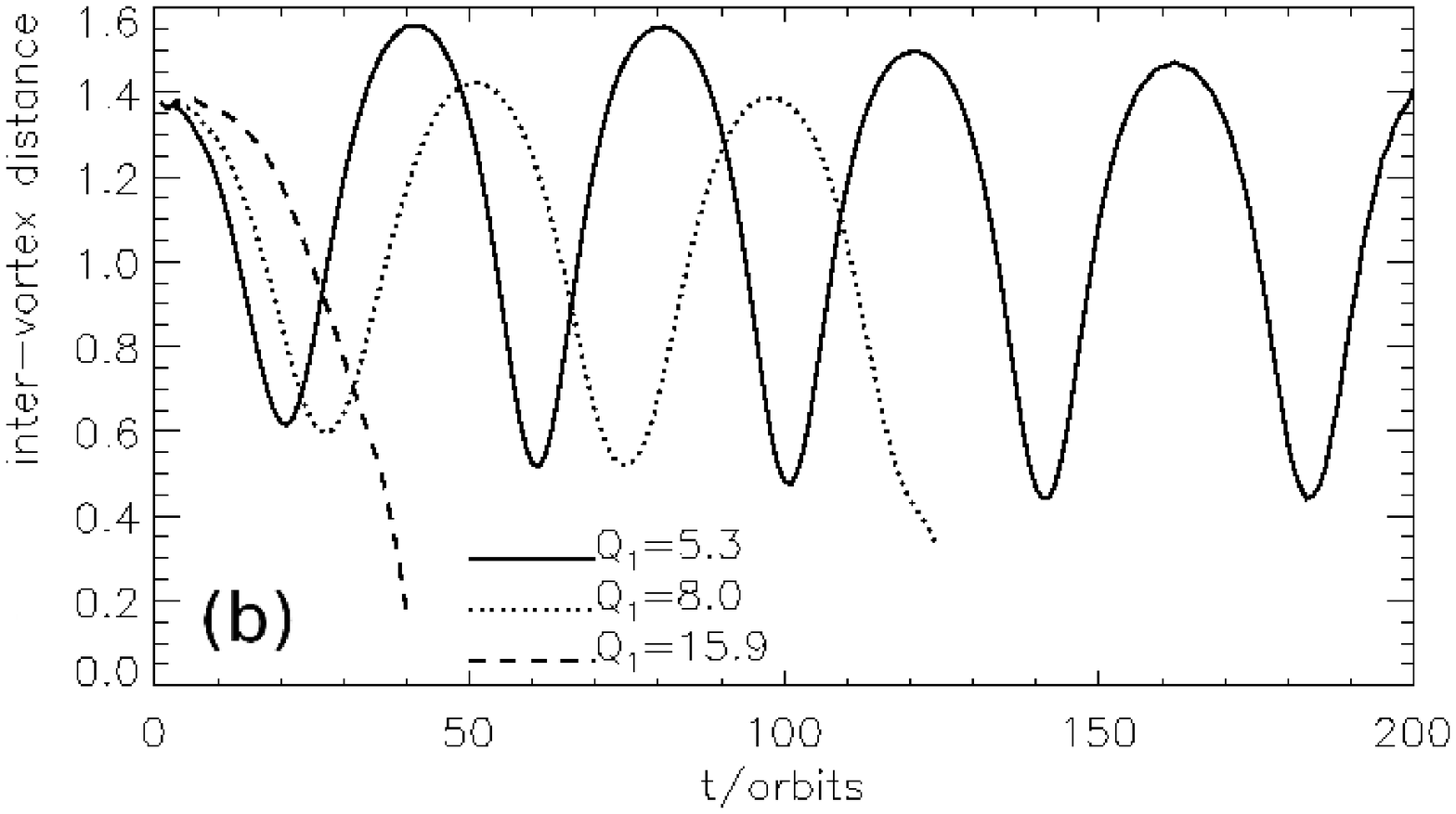}
\includegraphics[width=0.33\linewidth]{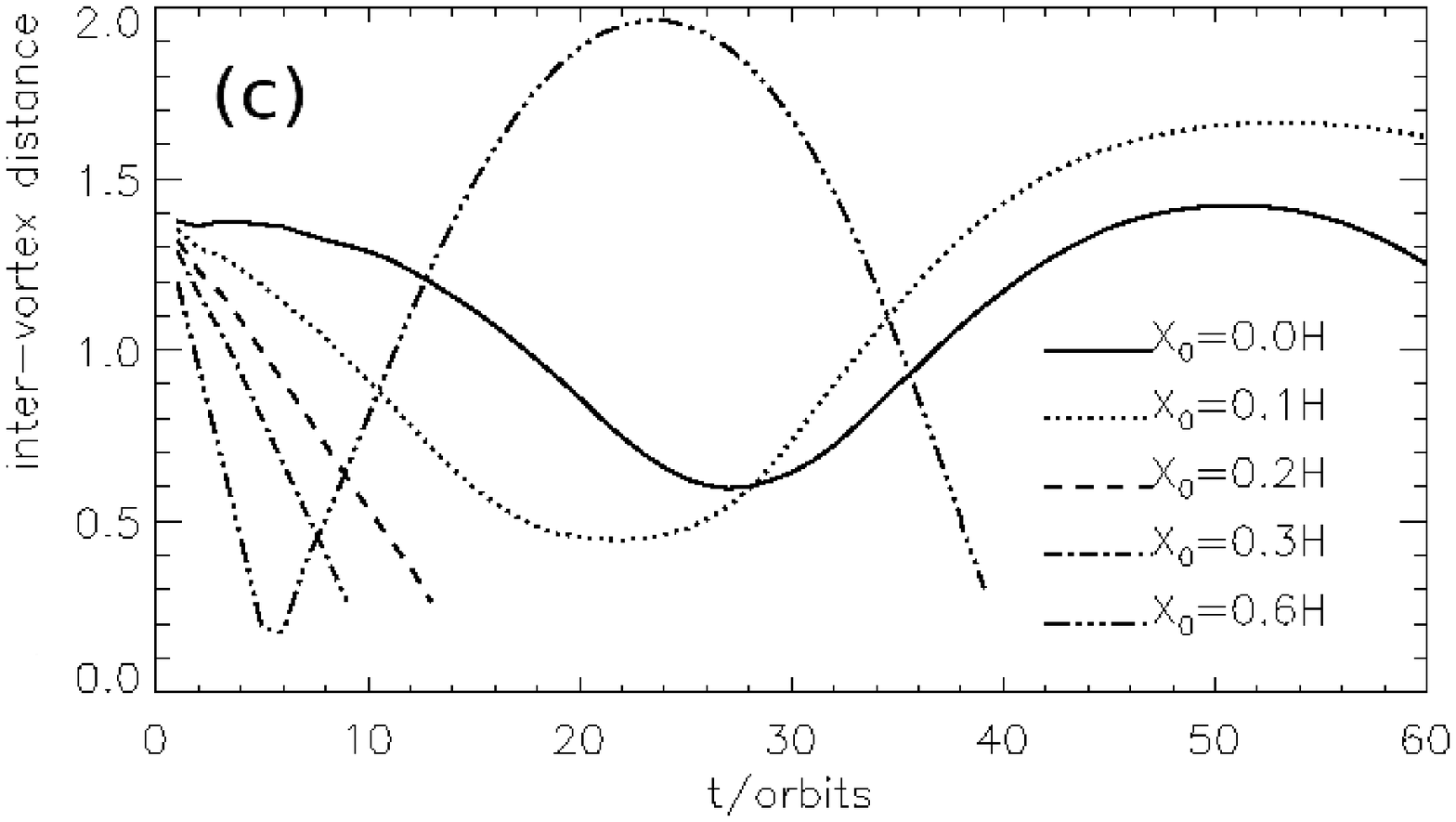}
  \caption{Inter-vortex distances for vortex pair simulations: (a) With and 
    without self-gravity for $Q_1=8$ ( simulations  Fsg and  Fnsg) (b) For  discs of different masses
    characterised  by $Q_1$, the Keplerian Toomre parameter at $r=1,$ self-gravity being included
    (simulations Fsg, M1, and M2).
    (c)Vortices input with  varying initial radial separation, $X_0,$ with self-gravity included
     for disc models with $Q_o=8$( simulations Fsg, S1,S2,S3, and S4).
    \label{kida_expts}}
\end{figure*}

We first compare cases Fsg and Fnsg. These are for the disc with $Q_1=8$ with
and without self-gravity respectively.  The distance between vortex centroids
as a function of time is shown in Fig. \ref{kida_expts}(a). It takes
almost 5 times as long for the self-gravitating vortices to merge.
Vortensity evolutionary plots show that
non self-gravitating vortices begin to merge at $t=28P_0,$  but
self-gravitating vortices at $t=126P_0.$  
For Fnsg, 
vortices  approach each other  within $\sim 23P_0$ of formation and  merge. 
Assuming they move relative to each other because of  Keplerian shear, 
the time taken to merge implies that 
their radial separation at vortex formation must be 
$\simeq 0.14H.$ This length-scale is resolved by about 2.5 grid cells
indicating that a non-zero initial separation is generated by grid effects, despite
the vortex perturbations being imposed at the same radius.  In a pair of lower resolution 
runs ($N_r=N_\varphi/3=400$),  vortex merging without self-gravity occurs at $t=23P_0$ whereas
the with self-gravity the vortex pair survive until the end of the simulation. These are consistent
if the initial radial separation is increased due to lowered resolution, giving stronger differential rotation
, hence earlier merging in the non-self-gravity run and increased impact parameter in terms of horseshoe orbits
in the self-gravitating run (see below). 

In the self-gravitating case Fsg, the inter-vortex distance
oscillates with period $50P_0$ and there are two close-encounters
before merging. 
This is analogous to the survival mode of  the vortex pair towards the end of the 
disc-planet simulation for $Q_o=3$. Given a  maximal separation of $\simeq 1.4$ and
that the vortices  are circulating  near unit radius, the maximal angular
separation is  $\simeq \pi/2$. 
 The minimal separation during the first two close encounters  is $\simeq 0.6$, or about $12H$
which implies  that the vortices are on tadpole
orbits.  
As the minimum separation is much larger than the typical vortex size,
 merging does not occur during the first two close encounters.

Next, in Fig. \ref{kida_expts}(b) we vary the
disc mass via
$Q_1$. The reference case has $Q_1=8$.  
Doubling $Q_1$ to $Q_1=15.9$ (case M1) weakens self-gravity and vortices merge within few tens
of $P_0.$  For $Q_1=5.3$ (case M2) the
oscillation period is $\simeq 40P_0$ and maximal
separation is 1.55, larger than for  $Q_1=8$, as is the first minimum
separation: the increased self-gravity has enhanced the mutual repulsion of
vortices. The M2 two-vortex configuration lasts until the end of the
simulation, but
there is a secular decrease in the minimum
separation (of $0.6$ at $t=20P_0$ and $0.45$ at $t=180P_0$) due to
vortex migration. We expect merging to occur eventually. Consistent with
the behaviour seen for  gap
edge vortices, increasing self-gravity delays merging.  

Our final experiment  varies the  initial
radial separation of the vortex perturbations.
Results are  shown in Fig. \ref{kida_expts}(c). Increasing $X_0$
to $0.1H$ from the reference case, the first minimum separation decreases to 0.45
from 0.6: vortices approach each other more closely, but still repel and
undergo co-orbital dynamics. The increased oscillation amplitudes  imply a larger tadpole
orbit. For $X_0=0.2H$ and $0.3H$, vortex separation decreases more rapidly
and become small enough for vortex merging. However, for $X_0=0.6H$,
vortices simply circulate past by each other and no merging occurs,
despite reaching a similar minimal separation  as when
$X_0=0.2H,\,0.3H$. This implies merging requires that  vortices should collide head
on. 

\subsection{Vortices as co-orbital planets}\label{vortices_planets}
The simulations presented above indicate that   self-gravitating vortex pairs behave 
like co-orbital planets. This means there exists an initial radial 
separation (or impact parameter) below which  vortices execute 
horseshoe orbits relative to each other. Analytical and numerical work indicates that vortices 
merge if their centroids are within $d\sim 3s$
of each other \citep[e.g.][]{zabusky79,melander88}, where $s$ is the
vortex size. We cannot assume the same critical $d/s$ apply to our
simulations because the situations  are very different, but it is
reasonable to expect merging if vortices can reach within some
critical distance of one another. 

The above results above can  be anticipated from existing treatments of co-orbital 
dynamics, which we discuss  here for completeness. The first is
\cite{murray00}'s  model of the co-orbital satellites of Saturn, the
Janus-Epimetheus system.  
The governing equation from \cite{murray00} gives a relationship between
two configurations,  the  `final'
configuration (subscript $f$) and the 
 `initial' configuration (subscript $i$) in the form
\begin{align}\label{murray_model}
  &\left(\frac{\Delta r_\mathrm{i}}{r_0}\right)^2 - \left(\frac{\Delta
      r_\mathrm{f}}{r_0}\right)^2 =
  -\frac{4}{3}q\left[H(\theta_\mathrm{i}) - H(\theta_\mathrm{f})\right],\\
  &H(\theta) = \left[\sin{(\theta/2)}\right]^{-1} - 2\cos{\theta} - 2 \notag
\end{align}
where $\Delta r$ is the radial separation of the satellites,
 $r_0$ the average orbital
radius (assumed fixed), $\theta$ their  angular 
separation and $q=\mathcal{M}/M_*,$  with $\mathcal{M}$ being the sum of
the satellite masses. It is assumed $q\ll1.$ 

Let us apply equation (\ref{murray_model}). We assume zero final radial
separation, $\Delta r_\mathrm{f}=0$ when the vortices are at their minimum
separation.  
Inserting $\theta_i = \pi/2,$
$\Delta r_\mathrm{i} = 7.1\times10^{-3}$ (corresponding to the initial
conditions deduced for Fnsg), $r_0=1$ and
$q=2.3\times10^{-5}$ (corresponding to the measured mass of the
 negative absolute vorticity region in Fsg), equation (\ref{murray_model}) gives    
$\theta_\mathrm{f} \simeq 0.41$ or a minimal angular separation of $\sim
8H$ so merging is not expected if vortices have size $H.$ 
This estimate is lower than the observed value of 0.6, but we
have used a lower limit on $q$. Inserting $q=1.2\times10^{-4}$ gives
$\theta_f = 0.6$ assuming all other parameters remain the same. This
implies the gravitational mass of the  vortex should be 20 Earth
masses. 

\begin{figure}
  \centering
\includegraphics[width=0.49\linewidth]{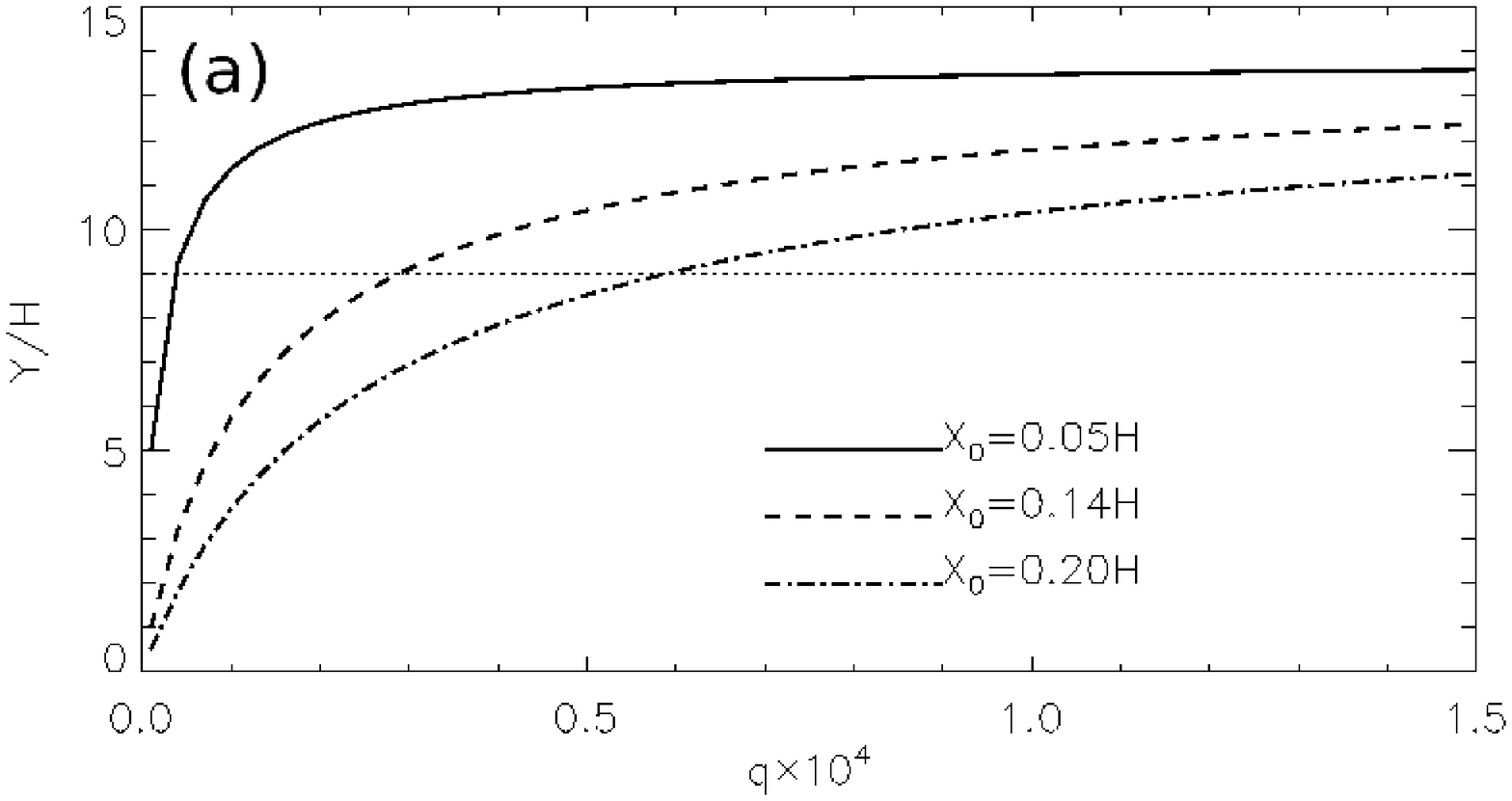}
\includegraphics[width=0.49\linewidth]{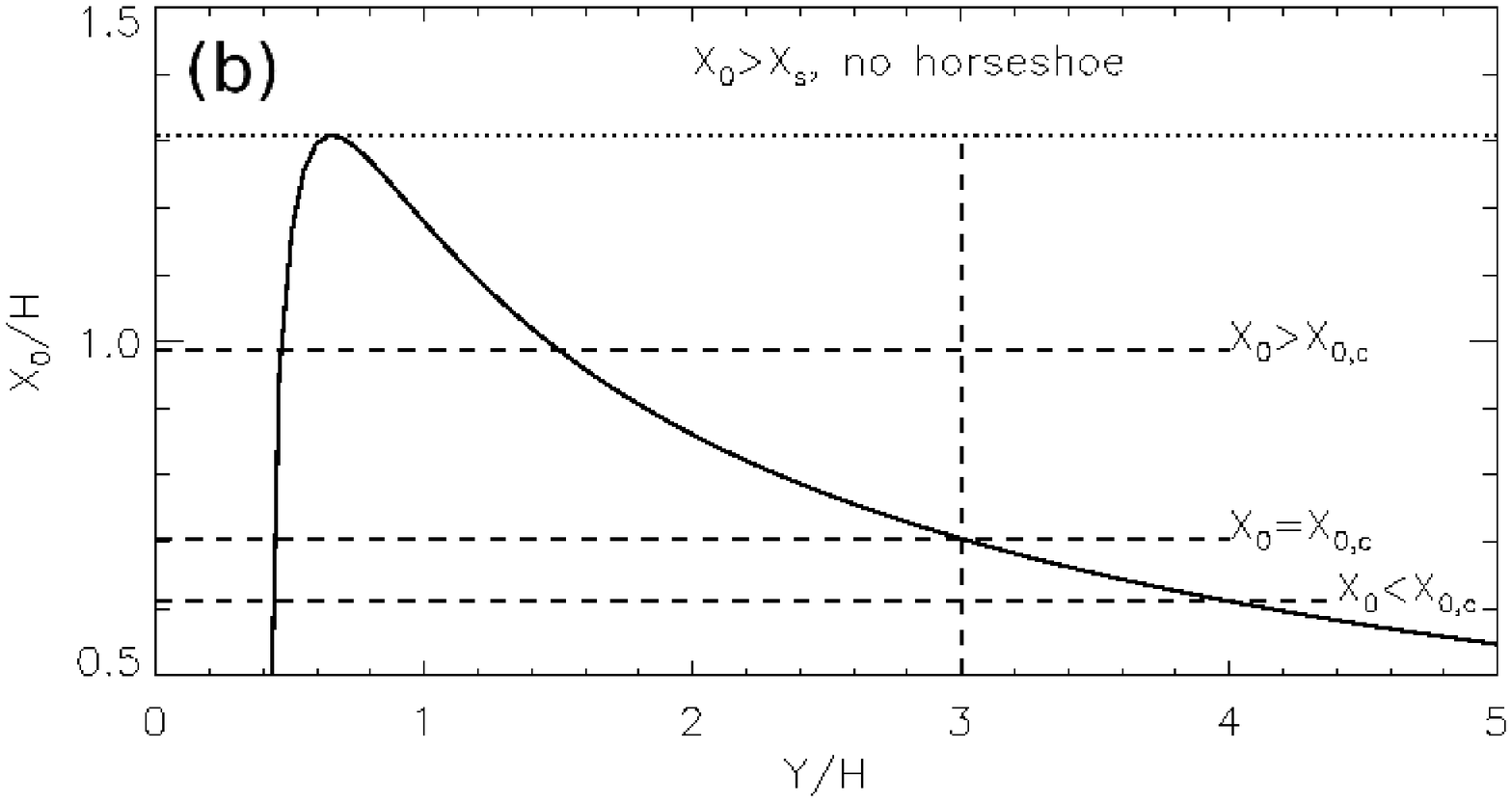}
\caption{Merging conditions based on
  \citeauthor{murray00}'s model of co-orbital satellites (a) and shearing
  sheet dynamics (b). In (a), $Y$ is the minimum inter-vortex
  separation and the horizontal line is a hypothetical critical
  separation below which merging occurs.  Thus merging is less likely
   for larger $q.$ Fig. \ref{merge_cond}(b)
  illustrates equation (\ref{min_sep2}): for a given $X_0$, the allowed values
  of minimum separation $Y$ lies between the intersection of the horizontal
  line $X_0/H =$ constant
  and the solid curve. 
   The vertical dashed line is a hypothetical critical
  separation, below which merging occurs.\label{merge_cond}}
\end{figure}

Equation (\ref{murray_model}) is illustrated in
Fig. \ref{merge_cond}(a).  For fixed $X_0$, as we increase $q,$
(e.g. as a by product of increasing the disc mass)
the minimal distance $Y$ increases, eventually
becoming too large for merging to occur.
For fixed $q,$  $Y$ decreases
as $X_0$ increases. This is  similar  to a test particle on a
horseshoe orbit in the restricted three body problem.
The larger its impact parameter
(while still in libration), the closer it approaches the secondary mass.

We may also view the horseshoe turns in a shearing sheet
approximation. In Appendix \ref{horseshoe} we derive
 the inequality
\begin{align}\label{min_sep2}
  \hat{x}_0 < \frac{1}{h}\left(\frac{8q}{3\hat{y}h} -
      \frac{q^2}{3\hat{y}^4h^4}\right)^{1/2},
\end{align}
where $\hat{x}_0 = X_0/H,\, \hat{y}=Y/H$. Given $X_0$,
equation( \ref{min_sep2})  can be inverted to give the range of possible $Y.$
This inequality is displayed in Fig. \ref{merge_cond}(b). There exists a
critical $X_0=X_s$ beyond which  
vortices do not execute horse-shoe turns and instead
circulate past  each other. There may also exist a value $X_{0,c}$ such
that   $X_{0,c}< X_s$ and  for $X_{0,c}<X_0<X_s$, vortices execute U-turns 
but the values of $Y$ attained are sufficiently small to allow merging. 
This happens  only when the vortices have small enough mass.
When $X_0<X_{0,c}$, $Y$  can be larger
than critical, so that merging does not occur.

These simple models give qualitatively the same results as vortex pair and
disc-planet simulations, and serve  to explain the effect of
self-gravity delaying merging by interpreting vortices as co-orbital
planets that execute mutual horseshoe turns, thereby imposing a
minimum inter-vortex separation (mainly in the azimuthal direction). 
If this minimum separation is still larger than a critical separation known to exist
for vortex merging, then merging cannot occur. Hence, multiple-vortex configurations can be sustained longer 
as the strength of self-gravity is increased. 


\section{Implications on vortex-induced migration}\label{migration}

Vortices at gap edges can lead to brief phases of rapid inwards migration. 
The case for Saturn and Jupiter mass planets 
in non-self-gravitating discs was investigated
by \cite{lin10}. In their simulations, gap edges become unstable to
 vortex modes leading to  vortex 
mergers  on dynamical time-scales, 
in turn resulting in a large-scale vortex circulating at  either gap edge.
Upon approaching the planet, the inner vortex can
 execute a horseshoe turn and  move  outwards 
across the gap, gaining angular momentum.
 Thus, the planet loses angular momentum and is 
scattered inwards.

\begin{figure}
   \centering
   \includegraphics[width=0.99\linewidth]{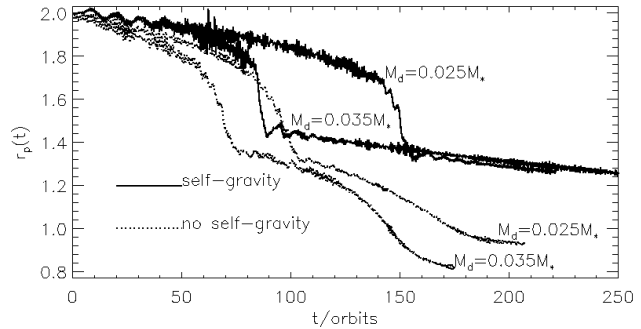}
   \caption{Vortex-induced migration with (solid) and without (dotted) self-gravity.
     \label{sgNsg}} 
\end{figure}

For comparison purposes, we repeated \citeauthor{lin10}'s simulations
with self-gravity. These
correspond to inviscid discs with initially uniform surface 
density. Fig. \ref{sgNsg} shows
the orbital migration of a Saturn-mass planet in discs with
total mass $M_d=0.035M_*$ and $0.025M_*$.  
In the $M_d=0.035M_*$ disc with self-gravity neglected,
 vortex-induced migration occurs at $t=60P_0$ whereas  in 
the self-gravitating case it is delayed to $t=85P_0.$
The delay is consistent
with both the stabilisation  of low $m$ modes, and the slower vortex merging induced  
by self-gravity. Furthermore, there could be gravitational influence of the outer gap edge vortices on the 
inner gap edge.

Indeed, although somewhat inhibited \footnote{This is also seen in disc-planet simulations in previous
sections, the inner gap edge is relatively more stable than outer gap edge}
, we found 
 the formation of an azimuthally extended coherent vortex at the inner gap edge
when  self-gravity was  included, just as in the non self-gravitating case..
However, it takes a longer time
for the inner vortex to build up, disrupt the co-orbital region and flow across the gap, 
therefore induced rapid  migration is delayed. For $M_d=0.025M_*$ it is delayed by
$\simeq 50P_0$, or twice of the higher mass disc, indicating stabilisation is more effective for the lower
disc mass. However, the extent of rapid migration is unaffected by self-gravity.
Notice also the increased oscillations when self-gravity is included.
This is because of the sustained multi-vortex
configuration at the outer gap edge (causing large oscillations in disc-planet torques), 
whereas without self-gravity these vortices would have merged.

After the first scattering event, migration stalls while the planet 
opens a gap at its  orbital radius and 
 vortex formation recurs. 
In the non self-gravitating case, a large inner
 vortex develops (taking about $15P_0$) 
and each time it passes by the planet,
 some vortex material splits off and flows across the planet's orbital radius, 
leading to inward  planet migration. 
With self gravity included,  the formation of a single vortex takes longer
 (about $35P_0$ in the higher mass disc) and it is narrower in radial extent than
the non self-gravitating case. 
We found that in this case, when the thinner vortex passes by the planet, little vortex material splits off from
the main vortex and flows across the gap. This  may be due to
the  self-gravity of the vortex. Hence,
there is a longer stalling period when self-gravity is included. 
Thus the net effect of self-gravity is to slow the migration in this 
example.

For the setup of \cite{lin10}, we do not observe a second fast migration episode 
within the simulation time-scale, 
but it may eventually occur. The total practical simulation time of 
a few hundred orbits is still very short compared to disc lifetimes.  
However, for the $Q_o=4$ disc model used in previous parts of this work, two episodes of 
rapid migration occur. Self-gravity does not change the physical nature of vortex-induced migration, 
but we comment that for discs of even larger mass
where vortex formation no longer occurs, being replaced by global spiral instabilities, 
the migration may be even faster than in the non self-gravitating case \citep{lin11}.

A detailed numerical study of the effect of self-gravity on vortex-induced migration 
and its  consequences for  disc structure
will be deferred to a future paper,
 but the simple experiments described here indicates that episodes of fast migration
punctuated by periods of slower migration are still expected as a consequence 
of vortex-induced migration.

\section{Conclusions}\label{conclusions}

We have studied the effect of self-gravity on vortex formation and evolution
 in protoplanetary discs. We specifically considered vortex production 
 through  instability associated with edges of  surface density dips or
gaps opened by a giant
planet. It was shown analytically that vortex forming modes are
stabilised by self-gravity through  its effect on the  linear mode
when the background remains fixed. This aspect has been
confirmed by linear calculations, which also showed that self-gravity
has a de-stabilising effect through its effect 
on  the background  state that is set up by  the perturbing action
of an embedded planet. Linear
calculations showed that  the vortex forming modes with the highest
growth rate  shift to higher $m$ 
with increasing disc mass or equivalently decreasing Toomre $Q$ value.

We performed hydrodynamic simulations that compare simulations
with and without  self-gravity for
 a range of disc masses. It was found that
more vortices form as the disc mass increased and minimum
$Q$  value decreased in accordance with linear calculations. However, for
sufficiently strong self-gravity, the vortex modes are
suppressed. In this case  global spiral 
modes develop instead. Self-gravity was found to delay vortex merging. 
In addition, multiple-vortex configurations can
be sustained for longer with  increasing disc mass, allowing vortices to evolve individually.

The nature of post-merging
vortices is also affected by self-gravity. With weak self-gravity ( i.e.. $M_d
\leq 0.024M_* $), a single vortex, extended in azimuth forms and circulates
at the outer gap edge. However, for $M_d=0.031M_*$ the final
configuration is a vortex pair at the outer gap edge, each localised
in azimuth. In this case a vortex, despite containing on the
order of $20M_{\oplus}$, can be gravitationally bound. 
The  effect of their gravitational influence on the rest of the disc is to
redistribute mass radially whereas for lower disc masses, 
redistribution is restricted  to the same azimuth.

We note that the internal flow in these self-gravitating, localised
vortices adjusts so that they are not destroyed by the background shear, 
taking on a structure similar to that of a Kida vortex \citep{kida81}. We found such
vortices form in discs much less massive than that required for direct disc fragmentation. Classical
gravitational instability would produce clumps of much larger mass than vortices in our models 
and would have difficulty surviving against shear beyond a few orbits. 
However, higher resolution calculations
in three dimensions that relax the assumption of a locally isothermal
equation of state are needed to assess the viability of the self-gravitational collapse of a 
vortex.

We performed supplementary simulations of Kida-like vortices to understand
the effect of self-gravity on inter-vortex interactions in a simpler
setting where it was not contaminated by the presence
of the embedded planet.  The effectiveness of merging as a function of disc 
mass may be understood from existing descriptions of co-orbital  horseshoe
dynamics. In essence it appears that pressure forces
do not play a role, so that in a self-gravitating disc,
vortex pairs  behave like co-orbital planets and
as a consequence  there exists a minimum inter-vortex
distance. Merging is  then avoided if this minimal separation
is still larger than a critical separation 
below which vortex merging occurs.

Finally, we briefly considered the consequence of self-gravity on vortex-induced migration. With self-gravity, 
the resistance to forming a single large vortex, which in non-self-gravitating
simulations is responsible   for scattering  the planet inwards, 
results in vortex induced migration being delayed. The  vortices
are less effective in scattering the planet because they are smaller and do not disrupt the co-orbital region as significantly
as their non self-gravitating  counterparts.
Thus, it can be said that in the regime of 
disc masses where vortices form and are significantly
affected by  self-gravity,  vortex-induced migration is slowed down.

\subsection{Outlook and caveats}
A possible issue in this work is that the introduction of a planet over 
short timescales of a few orbits may favour instability. However,  
\cite{valborro07} have considered non-self-gravitating discs with a pre-defined gap profile and found
similar growth rates for unstable modes as models without an initial gap.   
In addition, our linear calculations based on gap profiles attained in a 
quasi-steady state are in good agreement with simulations. 
This indicates the instability operates on such attained 
profiles and is not significantly affected by
how the gap is formed.

For the most part planetary migration has  been neglected in this work. 
The role of the embedded  planet was to create a  structured background  state 
from which vortices
develop. Hence, we expect our findings on  the effect of
self-gravity to be applicable to other types of  structured 
features  in a protoplanetary
disc that could  support vortex forming  instabilities.
Such a possibility is  the boundary region  
between  a dead zone  and active region of the disc \citep{lyra09}.

Although migration was briefly considered, 
we note that
there are many technical issues associated with a numerical study.  
These include
the treatment of the Roche lobe and the  time and the nature
 of the  release of the planet.
 The issue of the effects of the choice of  softening length
and disc viscosity should also  be investigated.
 A detailed parameter
survey is beyond the scope of this paper, but we hope to investigate migration
in more detail in the future. 

Vortices can trap dust particles due to their
association with pressure maxima. We have seen that increasing 
self-gravity leads to vortices of stronger density contrast 
(Fig. \ref{compare_merged_vortex}). We then anticipate self-gravitating 
vortices to be more effective at collecting solid particles. It would be interesting to consider
a two-component, gas and dust disc model to investigate the specific role of self-gravity
on dust trapping.

Another issue is that we adopted the two-dimensional approximation.
Clearly our studies should be extended to three dimensions.
It is known that Kida vortices with aspect ratio $\la 4$ are strongly
unstable to elliptic instability in three dimensional, unstructured 
non-self-gravitating discs \citep{lesur09b}. However, results
may change when vortices are produced by an instability arising from a 
a background structure and self-gravity is important. These should be investigated in 
the future. We point out that  recently
\cite{meheut10} demonstrated vortex formation in three-dimensional
 simulations without self-gravity  via the vortex forming  instability.
 Because the importance of self-gravity is
sensitive to the  disc thermodynamics and equation of state,
 it would be desirable to incorporate a realistic
 energy equation in future studies. 


\appendix
\section{Parameterisation of disc models for the vortex instability}\label{Qo_Md}
Disc models used to study the vortex instability are labelled by $Q_o$, 
which corresponds to values of $Q_p$ and disc-to-star mass ratio
given in Table  
\ref{Qo_Md_conversion}.  

\begin{table}
  \centering
  \caption{Relationship between the initial Keplerian Toomre stability
    parameter at the outer boundary, $Q_o$, at the planet's initial
    orbital radius, $Q_p$, and the disc-to-star mass ratio, for disc models used  in disc-planet interactions. }
  \begin{tabular}{ccc}
    \hline
    $Q_o$ & $Q_p$ & $M_d/M_*$ \\
    \hline\hline
    1.5 & 2.62 & 0.063 \\
    2.0 & 3.49 & 0.047 \\
    2.5 & 4.36 & 0.038 \\
    3.0 & 5.23 & 0.031 \\
    3.5 & 6.11 & 0.027 \\
    4.0 & 6.98 & 0.024 \\
    8.0 & 14.0 & 0.012 \\
    \hline
  \end{tabular}
  \label{Qo_Md_conversion}
\end{table}

\section{Artificial vortices in an accretion disc}\label{kidasetup}
The artificial Kida-like vortices used in \S\ref{hydro2} are setup as follows.
Consider a small patch of the disc, whose centre $(r_p, \varphi_p)$
rotates at angular speed $\Omega_p$ about the primary and set up local
Cartesian co-ordinates $x=r_p(\varphi-\varphi_p),\,y=r_p-r$. In the $(x,y)$
frame, there exists a Kida vortex solution for incompressible flow,
whose velocity field

$(u_x, u_y)$ is
\begin{align}  
   u_x = \frac{3\Omega_p\zeta y}{2(\zeta -1)},\quad
   u_y = -\frac{3\Omega_p x}{2\zeta(\zeta-1)}
\end{align}
inside the vortex core.  The ratio of the vortex
semi-major to semi-minor axis being  $\zeta=a/b$ is a free
parameter.  This velocity field is such that the
 vorticity $\omega$ is constant in the rotating frame. The elliptical boundary of the vortex
is defined such that
 $\omega =-3\Omega_p(1+\zeta^2)/(2\zeta(\zeta-1)) = \omega_v - 3\Omega_p/2 $ inside the
boundary  of the vortex and $\omega = - 3\Omega_p/2$ outside. The quantity
 $\omega_v= -3\Omega_p(1+\zeta)/(2\zeta(\zeta-1))$ is then
the vorticity of the vortex core relative to the background.
Being negative, this corresponds to an anticyclonic vortex.
In order to introduce perturbations corresponding
to Kida vortices,  we impose perturbations $\dd u_r \equiv
-v_y,\,\dd u_\varphi \equiv  v_x$ inside a specified elliptical boundary with an  exponential decay
outside. The boundary is fixed  by specifying  $\zeta = 8$
and its  semi-major axis $a=H(r_p)$ where the  reference radius is
$r_p=1$.

\section{Mutual horseshoe turns in the shearing sheet}\label{horseshoe}
We describe the gravitational interactions between two vortices in the
shearing sheet. It is assumed they behave like point masses and that
pressure forces may be neglected. We indicate below why this is a
reasonable assumption.

Consider a local Cartesian co-ordinate system $(x,y)$ that co-rotates
with a small patch of fluid  with angular velocity  $\Omega$ about the
primary, at a distance $r_p$. We have
$x=r-r_p,\,y=r_p(\varphi - \Omega 
t)$. Let $(x_j,y_j)$ denote the co-ordinates of the  centroid of  the $j^\mathrm{th}$
vortex  and $m_j$ be its  mass $(j = 1 , 2).$  Defining
$X\equiv x_2 - x_1,\,Y \equiv y_2 - y_1$ and $\mathcal{M} \equiv
m_2+m_1$, the equations of motion give 
\begin{align}
 & \ddot{X} - 2\Omega\dot{Y} = 3\Omega^2X - \frac{G\mathcal{M}X}{R^3},\\
 & \ddot{Y} + 2\Omega\dot{X} =  - \frac{G\mathcal{M}Y}{R^3},
\end{align}
where $R^2\equiv X^2 + Y^2$. 
These equations imply the constancy  of the Jacobi constant
\begin{align}
  J \equiv \frac{1}{2}\left( \dot{Y}^2 + \dot{X}^2\right) -
  \frac{3}{2}\Omega^2 X^2 - \frac{G\mathcal{M}}{R}.
\end{align}
  Let the  initial conditions be $X=
X_0,\,Y=\infty,\,\dot X=0,\,\dot Y = -3\Omega X_0/2.$ We assume the point of
closest approach occurs when  $X=\dot{Y}=0$. Equating $J$  at the initial time  and
at the time of closest approach we obtain
\begin{align}
  -\frac{3}{8} \Omega^2 X_0^2 = \frac{1}{2}\dot{X}^2 - \frac{G\mathcal{M}}{Y}
\label{Jacobi}
\end{align}
at the time of closest approach.
Since the vortices are then  at minimum separation, $\ddot{Y}>0$.
 The $y$  component of the equation of motion then  implies
\begin{align}
  \frac{1}{2}\dot{X}^2 > \frac{1}{8}\left(\frac{G\mathcal{M}}{\Omega Y^2}\right)^2.
\end{align}
Substituting $\dot X$ from  (\ref{Jacobi})  the inequality becomes 
\begin{align}\label{min_sep1}
\frac{3}{8}\hat{X}_0^2  < \frac{q}{\hat{Y}} - \frac{q^2}{8\hat{Y}^4}  
\end{align}
at minimum separation, where $\hat{X}_0 = X_0/r_p$, $\hat{Y} =
Y/r_p$, $q=\mathcal{M}/M_*$ and we have assumed $\Omega^2 =
GM_*/r_p^3$.

Eq. \ref{min_sep1} is useful for the case there vortices are just able
to undergo U turns. For fixed $q$, the function
\begin{align}\label{fy}
  f(\hat{Y};q) = \frac{q}{\hat{Y}} - \frac{q^2}{8\hat{Y}^4}
\end{align}
has a maximum value at $\hat{Y} = (q/2)^{1/3},$
corresponding to the
maximum conceivable  initial separation $X_0=X_s,$ where
\begin{align}\label{xs}
  X_s = 2^{2/3}q^{1/3}r_p.
\end{align}
If $X_0>X_s$ then equation (\ref{min_sep1})  cannot be satisfied and there
can be no horseshoe turns. For initial separations  $X_0<X_s$, (\ref{min_sep1})
implies that  the
minimal inter-vortex distance must exceed  $q^{1/3}/2$ (so that $f>0$). 
Now for sufficiently large $q,$ $q^{1/3}/2$ will be larger
than the critical separation  for merging, 
so merging is avoided during the  encounter.  

It is interesting to compare equation (\ref{xs})  to the estimate of the horseshoe
half-width $x_s$ of  \cite{paardekooper09}. They found $x_s=
1.68(q/h)^{1/2}r_p$ based on hydrodynamic simulations for low mass
planets. Equating $x_s$ and $X_s$
with $h=0.05$ we find $q=8.9\times10^{-5}.$
This should give the minimum $q$ for which pressure effects
could be ignored.
 Inserting
this value in  \cite{murray00}'s model of co-orbital satellites
gives a minimal separation of $0.58$ (see the estimate in
\S\ref{vortices_planets}), close to simulation results.   
Hence, if the vortex-pair interaction is purely gravitational, a
single vortex behaves in a similar way to  $\sim 15$ Earth masses, i.e. a low-mass
protoplanet.

Considering a vortex size of order $H$, the vortex-to-star mass ratio
is $q\sim \pi H^2\Sigma/M_*\simeq h^3/Q$. For a self-gravitating vortex where
$Q\sim 1$ we have $q\simeq h^3=1.25\times10^{-4}$ for $h=0.05$,
slightly exceeding the threshold value above.
Hence we expect the pressureless treatment of self-gravitating
vortex-vortex interactions to be acceptable for the purpose 
of explaining the resisted-merging of self-gravitating vortices. 






\end{document}